\documentclass[12pt]{article}
\usepackage{xcolor}
\usepackage[utf8]{inputenc}
\usepackage{textcomp}
\usepackage[numbers,sort&compress]{natbib}
\usepackage[left=2.5cm,right=2.5cm,top=2.5cm,bottom=3cm]{geometry}
\usepackage{tikz} 
\usepackage{url,hyperref}
\usepackage{amssymb,amsfonts,mathrsfs,dsfont,yfonts,bbm}
\usepackage{xcolor}
\usepackage{braket}
\usepackage{physics}
\usepackage{longtable,pdflscape}
\usetikzlibrary{calc,decorations.markings, shapes.misc, decorations.pathmorphing, tqft}
\tikzset{cross/.style={cross out, draw, 
	minimum size=2*(#1-\pgflinewidth), 
	inner sep=0pt, outer sep=0pt}}
\tikzset{every picture/.style={line width=0.75pt}}
\usepackage{color,colortbl}
\usepackage{comment}

\newtheorem{theorem}{Theorem}

\newcommand{\cclass}{C}
\newcommand{\cclasses}[1]{\mathrm{Cl}(#1)}
\newcommand{\irrep}{R}
\newcommand{\irreps}[1]{\mathrm{Irr}(#1)}
\newcommand{\fusionalg}[1]{R(#1)}
\newcommand{\FGhandle}{\Theta}

\makeatletter
\@addtoreset{equation}{section}
\makeatother

\def\one{{\hbox{ 1\kern-.8mm l}}}
\def\zero{{\hbox{ 0\kern-1.5mm 0}}}

\def\mC{ \mathbb{C}}

 \def\cZ{{\cal Z}}

\def\Sym{ \hbox{Sym} } 
 
\def\Aut{ {\rm Aut} }

\newcommand{\diag}{ {\rm Diag} }

\newtheorem{lemma}{Lemma}

\newtheorem{proposition}{Proposition}[section]

\newcommand{\be}{\begin{equation}}
\newcommand{\ee}{\end{equation}}
\newcommand{\beq}{\begin{equation}}
\newcommand{\eeq}{\end{equation}}
\newcommand{\bea}{\begin{eqnarray}\displaystyle}
\newcommand{\eea}{\end{eqnarray}}

\newcommand{\DZ}{\mathds{Z}}

\newcommand*{\boxcoloro}{orange}
\makeatletter
\newcommand{\boxedo}[1]{\textcolor{\boxcoloro}{%
\tikz[baseline={([yshift=-1ex]current bounding box.center)}] \node [rectangle, minimum width=1ex,rounded corners,draw] {\normalcolor\m@th$\displaystyle#1$};}}
 \makeatother
\newcommand*{\boxcolorr}{red}
\makeatletter
\newcommand{\boxedr}[1]{\textcolor{\boxcolorr}{%
\tikz[baseline={([yshift=-1ex]current bounding box.center)}] \node [rectangle, minimum width=1ex,rounded corners,draw] {\normalcolor\m@th$\displaystyle#1$};}}
 \makeatother
\newcommand*{\boxcolorb}{blue}
\makeatletter
\newcommand{\boxedb}[1]{\textcolor{\boxcolorb}{%
\tikz[baseline={([yshift=-1ex]current bounding box.center)}] \node [rectangle, minimum width=1ex,rounded corners,draw] {\normalcolor\m@th$\displaystyle#1$};}}
 \makeatother
\newcommand*{\boxcolorg}{green}
\makeatletter
\newcommand{\boxedg}[1]{\textcolor{\boxcolorg}{%
\tikz[baseline={([yshift=-1ex]current bounding box.center)}] \node [rectangle, minimum width=1ex,rounded corners,draw] {\normalcolor\m@th$\displaystyle#1$};}}
 \makeatother
 \newcommand*{\boxcolorp}{purple}
\makeatletter
\newcommand{\boxedp}[1]{\textcolor{\boxcolorp}{%
\tikz[baseline={([yshift=-1ex]current bounding box.center)}] \node [rectangle, minimum width=1ex,rounded corners,draw] {\normalcolor\m@th$\displaystyle#1$};}}
 \makeatother
  \newcommand*{\boxcolorc}{cyan}
\makeatletter
\newcommand{\boxedc}[1]{\textcolor{\boxcolorc}{%
\tikz[baseline={([yshift=-1ex]current bounding box.center)}] \node [rectangle, minimum width=1ex,rounded corners,draw] {\normalcolor\m@th$\displaystyle#1$};}}
 \makeatother
  \newcommand*{\boxcolory}{yellow}
\makeatletter
\newcommand{\boxedy}[1]{\textcolor{\boxcolory}{%
\tikz[baseline={([yshift=-1ex]current bounding box.center)}] \node [rectangle, minimum width=1ex,rounded corners,draw] {\normalcolor\m@th$\displaystyle#1$};}}
 \makeatother

\usepackage{amsmath}	

\title{Row-Column duality and combinatorial topological strings}

\author{\normalsize Adrian Padellaro$^{a,\diamondsuit}$, Rajath Radhakrishnan$^{a,b,\clubsuit}$, Sanjaye Ramgoolam$^{a,c,\heartsuit}$}

\date{%
	{\normalsize \textit{${}^a$Centre for Theoretical Physics, School of Physical and Chemical Sciences, \\ Queen Mary University of London, London E1 4NS, United Kingdom.\\
		${}^b$ International Centre for Theoretical Physics, Strada Costiera 11, 34151 Trieste, Italy.\\
		${}^c$ National Institute for Theoretical Physics, School of Physics and Centre for Theoretical Physics,  University of the Witwatersrand, Wits, 2050, South Africa \\[2ex]}}
	\normalsize \today
}
\makeatletter
\def\blfootnote{\xdef\@thefnmark{}\@footnotetext}
\makeatother

\begin{document}

\begin{flushright}
QMUL-PH-23-07\\
\end{flushright}

{\let\newpage\relax\maketitle}

\maketitle
\blfootnote{$^{\diamondsuit}$a.k.s.padellaro@qmul.ac.uk, $^{\clubsuit}$ rradhakr@ictp.it, $^{\heartsuit}$s.ramgoolam@qmul.ac.uk}
\begin{abstract}
	Integrality properties of partial sums over irreducible representations, along columns of character tables of finite groups, were recently derived using combinatorial topological string theories (CTST). These CTST were  based on Dijkgraaf-Witten theories of flat $G$-bundles for finite groups $G$ in two dimensions, denoted $G$-TQFTs. We define analogous combinatorial topological strings related to  two dimensional TQFTs based on fusion coefficients of finite groups. These TQFTs  are denoted as  $R(G)$-TQFTs and allow  analogous integrality results to be derived for partial row sums of characters over conjugacy classes along fixed rows. This relation between the $G$-TQFTs and  $R(G)$-TQFTs defines a row-column duality for character tables, which provides a physical framework for exploring the mathematical analogies between rows and columns of character tables.  These constructive proofs of integrality are complemented with the proof of similar and complementary results using the more traditional Galois theoretic framework for integrality properties of character tables. The partial row and column sums are used to define generalised partitions of the integer row and column sums, which are of interest in combinatorial representation theory.
\end{abstract}
\newpage
\tableofcontents

\newpage

\section{Introduction}

Quantum field theories have led to the discovery of new mathematical structures, and have given us deep insights into several areas of mathematics. Some examples include the study of low-dimensional topology using Chern-Simons theory \cite{Witten:1988hf,reshetikhin1990ribbon,reshetikhin1991invariants}, the discovery of mirror symmetry from string theory \cite{hori2003mirror}, the discovery of vertex operator algebras from conformal field theory \cite{goddard1986kac, frenkel1989vertex, borcherds1992monstrous} and many more \cite{bah2022panorama}.

In this paper we continue this tradition by studying the implications of string-theoretic constructions for finite group representation theory. $G$-TQFTs are two-dimensional topological field theories (TQFTs) based on a finite group $G$ \cite{DijkgraafWitten, Witten:1991we, Freed:1991bn, FukumaKawai}. The partition functions compute topological invariants of flat $G$-bundles over two-manifolds.  Using representation theoretic Fourier transforms over  the centres of the group algebras $\cZ ( \mC ( G ) ) $ these partition functions are also expressible in terms of representation theory data. In combinatorial representation theory, properties of group representations are studied by introducing combinatorial objects (see \cite{ReviewCombRepTheory} for an overview of combinatorial representation theory). Recently \cite{IDFCTS}, combinatorial topological string theories ($G$-CTSTs) based on $G$-TQFTs were used to obtain  constructions of representation theoretic data.
Finite algorithms based on combinatorial amplitudes in the $G$-CTST were given for computing the integers
\begin{equation}
	\frac{\abs{G}}{d_R},
\end{equation}
where $d_R$ is the dimension of irreducible representation $R$ of $G$. Algorithms for constructing characters from $G$-CTST amplitudes were given, and related to the Burnside algorithm for computing characters. 

These results were developed further in  \cite{GCTST2} where combinatorial  observables in $G$-CTST, constructed from the combinatorics of flat $G$-bundles on surfaces of arbitrary genus and one boundary labelled by a fixed conjugacy class $\cclass$,  were  used to derive  integrality properties of partial sums  of characters of $\cclass$  over irreducible representations of $G$. These are partial sums along columns of the character table of $G$. For example, Proposition 3.4.1-II \cite{GCTST2}  states that sums over irreducible characters of a given conjugacy class,  for representations of fixed dimension,  are integers. It follows immediately that the sum $n_\cclass$, over irreducible characters along a column $\cclass$ of the character table, is an integer (see Proposition 2.1 and a positive integrality property for normalized characters in Proposition 2.2 in this paper).  It is a known fact in representation theory that the sum $n_\irrep$, of irreducible characters of a group $G$ along a row $\irrep$, is a positive integer and it is an open problem to find a combinatorial interpretation of this positive integer \cite[Problem 12]{StanleyPositivityProblems}. 

This raises the question of whether there is a CTST/TQFT approach to study the sums along rows of character tables of finite groups $G$.  In this paper, we give a positive answer to this question  by introducing a related family of CTSTs, based on fusion algebras $\fusionalg{G}$ where results complementary to those coming from $G$-CTSTs can be derived, with the role of columns switched with that of rows. We therefore refer to the relation between $G$-CTSTs and $\fusionalg{G}$-CTSTs as a row-column duality. The integrality property of the partial column sums for a column  corresponding to conjugacy class $\cclass$,   restricted by fixing the dimension of the irreps being summed, was generalised to  obtain integrality for restrictions involving additional conjugacy classes which have the property that they have integer characters for all rows (e.g. Propositions 3.4.1-III and 3.4.1-IV in \cite{GCTST2}). These integrality properties were obtained from  the counting of flat $G$-bundles  on surfaces with one boundary labelled by $\cclass$ and  multiple additional boundaries labelled by these additional conjugacy classes. Following the theme of row-column duality we will obtain analogous results for partial row sums. 

The algebra structure of $\cZ ( \mC ( G ) ) $ is defined using the product of conjugacy classes of $G$ and its decomposition into conjugacy classes. Similarly, the fusion algebra $R(G)$ is defined using the tensor product of representations of $G$ and its decomposition into irreducible representations. Both the product of conjugacy classes as well as the tensor product of representations play a crucial role in the fusion rules of line operators in three-dimensional G-TQFTs \cite{DijkgraafWitten}. In \cite{aradherzog}, the Arad-Herzog conjecture in finite group theory \cite{arad2006products} was studied in the context of three-dimensional topological quantum field theories (TQFTs). It was shown that this conjecture implies special fusion rules for line operators in these TQFTs. These special fusion rules were then independently proved giving evidence for the validity of the conjecture. In closely related work \cite{abc}, special fusion rules of line operators in three-dimensional TQFTs based on finite and Lie groups were studied, relating them to the representation theory of these groups. 

The more traditional approach to integrality properties of finite group character tables uses Galois theory and analogies between rows and columns have been investigated in this framework \cite{Navarro}. In this paper, we will complement the constructive TQFT/CTST approaches to properties of character tables with 
the Galois theoretic approaches. We reproduce the integrality properties of the partial row/column sums using Galois theory. We also derive some new results motivated by these TQFT/CTST investigations. 

 Every character table has at least one integer row and one integer column :  namely, the trivial representation, and the the identity element. We study, using the Galois theory methods,  the conditions under which the number of integer rows equals the number of integer columns. We refine this analysis to study the number of rows and columns of the character table belonging to a field extension of $\mathds{Q}$. We capture the number theoretic properties of the rows and columns in two graphs, one defined using rows of the character table and the other using columns. We find sufficient conditions under which the graphs are isomorphic.

The plan of the paper is as follows. We start with a review of the construction of a $G$-CTST in section \ref{sec: CTST}, which is followed by a review of the relevant integrality results in \cite{GCTST2}. Following this, we introduce the $\fusionalg{G}$-CTST and illustrate how it is used to prove dual integrality results. A $G$-TQFT is defined through a semi-simple commutative Frobenius algebra with a basis labelled by conjugacy classes with a product on the classes inherited from group multiplication, and an appropriate Frobenius pairing.  A $\fusionalg{G}$-TQFT is defined through a basis of elements labelled by irreducible representations (irreps)  with a  product given by a tensor product multiplicities of irreps  and an appropriate Frobenius pairing (A closely related TQFT with the same product but different pairing is defined in \cite{Kock}). The idempotent (projector) basis for $G$-TQFTs is naturally labelled by irreducible representations, while the idempotent basis for a $\fusionalg{G}$-TQFT is naturally labelled by conjugacy classes. The properties of the idempotent basis for $\fusionalg{G}$-TQFT imply that fusion data can be used to construct the characters of $G$. This is analogous to Burnside's algorithm for constructing characters from structure constants of the centre of the group algebra.
Section \ref{sec:implications} builds on section \ref{sec: CTST}, where we prove refined integrality results for character sums using CTST. The refined integrality results are used in the next subsection, where we give generalized integer partitions of the integers $n_\irrep$ and $n_\cclass$. These consist of the expression of $n_\irrep $ or $n_{\cclass}$ as a sum of zeroes with a multiplicity along with a positive integer equipped with a partition and the negative of another positive integer equipped with a partition. These expressions are obtained by inspecting the column $\cclass$   adding up to  $n_{ \cclass}$ in terms of level sets of a function  defined by considering all the  columns other than $ \cclass$ which are all integers. For the row $\irrep$ the generalised partition is obrained from the row-column dual of the construction used for $\cclass$. As mentioned, Galois theory is a powerful toolbox for studying number and field theoretic properties of characters. Therefore, we have dedicated section \ref{sec: Galois Theory} to the Galois theory actions on conjugacy classes and representations of finite groups, and their consequences of interest in relation to CTST. We start with a review of the background and then prove a set of generalizations of the previously mentioned integrality results obtained from $G$-CTST and $R(G)$-CTST. In section \ref{sec: int rows and int cols} we use Galois theory to study the number of integer rows and integer columns of character tables. In section \ref{ConstRCD} we describe an algorithm which starts from the $G$-TQFT and arrives at the  $\fusionalg{G}$-TQFT. The same algorithm applied to the   $\fusionalg{G}$-TQFT produces the $G$-TQFT. The problem of deducing the integrality results in section \ref{sec: CTST} from this construction of the duality involves some challenges which we discuss.

In appendix \ref{apx: fusion tqft} we prove that the fusion algebra defines a TQFT by explicitly exhibiting the Frobenius algebra structure. Appendix \ref{apx: inverse vandermonde lemma} contains a proof of the Vandermonde matrix lemma used to prove the refined integrality results in section \ref{sec:implications}. We have put some additional constructive formulas and algorithms based on the $\fusionalg{G}$-CTST in appendix \ref{apx: additional formulas}. Appendix \ref{sec:Galrowcoltable} contains a table of groups of order up to 100, the number of integer rows and columns in the character table of these groups, and the isomorphism class of their Galois groups. We also compute this data for sporadic groups. We explore an interesting connection between Harada's conjecture and row-column duality in appendix \ref{sec:Harada}.

\section{Row-Column Duality}\label{sec: CTST}
In this section we will review the combinatorial construction of partition functions in a Dijkgraaf-Witten theory based on finite group $G$, following the presentation in \cite{IDFCTS}. Closely related discussions have appeared in \cite{Couch:2021wsm}\cite{Banerjee:2022pmw} motivated by the investigation of low-dimensional models of wormhole physics \cite{Marolf:2020xie}.   We follow this with a review the derivation from \cite{GCTST2} of integrality properties of partial sums along a column of the character table of $G$. 
We then give a combinatorial description of a dual TQFT, based on the fusion algebra $\fusionalg{G}$, which gives integrality results for partial sums along rows rather than columns.

\subsection{$G$-CTST}
The Dijkgraaf-Witten theory for a finite group $G$ can be described as follows.
Let $\mathbb{C}(G)$ be the group algebra of a finite group $G$ and $\mathcal{Z}(\mathbb{C}(G))$ the subalgebra of central elements which commute with all $\mathbb{C}(G) $ , also known as the center of $\mathbb{C}(G)$. The center of a finite group algebra has two useful bases. The first one is based on the set
\begin{equation}
	\cclasses{G} = \{\cclass_1, \dots, \cclass_K\},
\end{equation}
of distinct conjugacy classes of $G$, which can be used to label a set of basis elements of the center since
\begin{equation}
	\abs{\cclasses{G}} = \dim \mathcal{Z}(\mathbb{C}(G)).
\end{equation}
The sums over elements of each conjugacy class are central elements 
\begin{equation}
	T_{\cclass} = \sum_{g \in \cclass}  g \, . 
\end{equation}
The set
\begin{equation}
	\{T_{\cclass_1}, \dots, T_{\cclass_K}\},
\end{equation}
form a basis of the center.

The second basis is labelled by irreducible representations of $G$. It is a standard result  in representation theory  that
\begin{equation}
	\abs{\cclasses{G}} = \abs{\irreps{G}},
\end{equation}
where
\begin{equation}
	\irreps{G} = \{\irrep_1, \dots, \irrep_K\},
\end{equation}
is a complete set of non-isomorphic irreducible representations of $G$.
This implies that
\begin{equation}
	\abs{\irreps{G}} = \dim \mathcal{Z}(\mathbb{C}(G)),
\end{equation}
and in particular there exists a basis for $ \mathcal{Z}(\mathbb{C}(G))$ labelled by irreducible representations of $G$.
The representation basis is constructed using irreducible characters $\chi^{\irrep}$ of $G$
\begin{equation}
	P_{{\irrep}} = \frac{d_\irrep}{\abs{G}} \sum_{g \in G} \chi^{\irrep}(g^{-1})g = \frac{d_\irrep}{\abs{G}} \sum_{g \in G} \overline { ( \chi^\irrep ( g )  ) }   g,
\end{equation}
where $\overline { \chi^\irrep ( g) ) } $ is the complex conjugate of $\chi^\irrep( g )$ and $d_\irrep = \chi^\irrep(1)$ the dimension of $\irrep$. The two bases are related by
\bea 
P_\irrep = { d_\irrep \over |G| } \sum_{ \cclass \in \cclasses{G}} \overline { ( \chi^\irrep_{ \cclass }   ) } T_{ \cclass }  \label{eq: rep change of basis}
\eea 
and 
\bea 
T_{ \cclass} = \sum_{ \irrep \in \irreps{G} }  { \chi^\irrep ( T_{ \cclass } ) \over d_\irrep  } P_\irrep, \label{eq: cc change of basis}
\eea
where
\begin{equation}
	\chi^{\irrep}_\cclass \equiv \chi^{\irrep}(g) \qq{for any $g \in \cclass$.}
\end{equation}

Note that the structure constants $f^{\cclass_3}_{\cclass_1 \cclass_2}$ in the conjugacy class  basis, defined by
\begin{equation}\label{ClassAlgInteg} 
	T_{\cclass_1}T_{\cclass_2} = \sum_{\cclass_3 \in \cclasses{G}} f^{\cclass_3}_{\cclass_1 \cclass_2} T_{\cclass_3},
\end{equation}
are integers. They have an expression in terms of characters
\begin{equation}
	f^{\cclass_3}_{\cclass_1 \cclass_2} = \frac{\abs{\cclass_1} \abs{\cclass_2}}{\abs{G}} \sum_{\irrep \in \irreps{G}} \frac{1}{d_\irrep} \chi^\irrep_{\cclass_1}\chi^\irrep_{\cclass_2} \overline{\chi^\irrep_{\cclass_3}}.
\end{equation}
The structure constants in the representation basis are diagonal,
\begin{equation}
	P_{\irrep_1} P_{\irrep_2} = \delta_{\irrep_1 \irrep_2} P_{\irrep_1}. \label{eq: flat rep basis product}
\end{equation}
so that the projectors form an orthogonal set and further satisfy the completeness property 
\begin{equation}
	\quad \sum_{\irrep \in \irreps{G}} P_\irrep = 1.
\end{equation}

To define the partition functions in a Dijkgraaf-Witten theory we introduce the delta function on $G$
\begin{equation}
	\delta(g) = \begin{cases} 1 \qq{if $g$ is the identity,}\\ 0 \qq{otherwise.}\end{cases}
\end{equation}
In the representation basis we have
\begin{equation}
	\frac{1}{\abs{G}}\delta(P_\irrep) = \frac{d_\irrep}{\abs{G}^2} \sum_{g \in G} \overline { ( \chi^\irrep ( g )  ) }   \delta(g)= \frac{d_\irrep^2}{\abs{G}^2}. \label{eq: flat rep basis delta}
\end{equation}
We use the following short-cut to arrive at the genus $h$ partition functions. Define the handle creation operator
\begin{equation}
	\Pi = \sum_{\irrep \in \irreps{G}} \frac{\abs{G}^2}{d_\irrep^2} P_{\irrep}.
\end{equation}
Because
\begin{equation}
	T_{\cclass} P_{\irrep} = \abs{C}\frac{\chi^{\irrep}_\cclass}{d_\irrep}P_{\irrep},
\end{equation}
the eigenvalues of the integer matrix $f_{\cclass}$ of structure constants are given by normalized characters. In particular, using \eqref{eq: rep change of basis} we have
\begin{equation}
	\Pi = \sum_{\irrep \in \irreps{G}} \frac{\abs{G}}{d_\irrep} \sum_{ \cclass \in \cclasses{G}} \overline { ( \chi^\irrep_{ \cclass }   ) } T_{ \cclass } =  \sum_{ \cclass \in \cclasses{G}} \frac{\abs{G} }{\abs{\cclass}} \overline{\tr(f_{\cclass})} T_{\cclass} = \sum_{ \cclass \in \cclasses{G}} \Sym \cclass \,  {\tr(f_{\cclass})} T_{\cclass},
\end{equation}
where
\begin{equation}
	\Sym \cclass = \frac{\abs{G}}{\abs{C}}, \label{eq: def sym C}
\end{equation}
is the size of the centralizer of $g \in \cclass$.
Therefore, it is possible to express $\Pi$ as an integer linear combination of conjugacy class basis elements.

Inserting powers of the handle creation operator into the delta function gives
\begin{equation}
	\frac{1}{\abs{G}}\delta(\Pi^h) = \sum_{\irrep \in \irreps{G}}  \qty(\frac{\abs{G}}{d_\irrep})^{2h-2}.
\end{equation}
This is the partition function of a manifold of genus $h$ with no boundaries,
\begin{equation}
	Z^G_{h} = \frac{1}{\abs{G}}\delta(\Pi^h) = \vcenter{\hbox{
			\tikzset{every picture/.style={line width=0.75pt}} 
			
			\begin{tikzpicture}[x=0.75pt,y=0.75pt,yscale=-1,xscale=1]
				
				\draw   (50,119) .. controls (50,96.36) and (108.52,78) .. (180.71,78) .. controls (252.9,78) and (311.42,96.36) .. (311.42,119) .. controls (311.42,141.64) and (252.9,160) .. (180.71,160) .. controls (108.52,160) and (50,141.64) .. (50,119) -- cycle ;
				\draw    (80,120) .. controls (89.99,130.02) and (100.39,130.02) .. (110,120) ;
				\draw    (84.79,124.42) .. controls (92.39,115.62) and (96.39,114.42) .. (105.19,123.62) ;
				\draw    (130,112.75) .. controls (139.99,122.77) and (150.39,122.77) .. (160,112.75) ;
				\draw    (134.79,117.17) .. controls (142.39,108.37) and (146.39,107.17) .. (155.19,116.37) ;
				\draw    (250,112.75) .. controls (259.99,122.77) and (270.39,122.77) .. (280,112.75) ;
				\draw    (254.79,117.17) .. controls (262.39,108.37) and (266.39,107.17) .. (275.19,116.37) ;
				\draw  [dash pattern={on 0.84pt off 2.51pt}]  (170,114) .. controls (189.59,108.02) and (215.59,108.22) .. (240,113) ;
	\end{tikzpicture}}}
\end{equation}
In general, we are interested in partition functions of manifolds with boundaries labelled by conjugacy classes. For example, the genus zero partition function with $b$ boundaries labeled by conjugacy classes $C_1, \dots, C_b$ is equal to
\begin{equation}
	Z^G_{h=0, C_1; \dots;C_b} = \frac{1}{\abs{G}}\delta(T_{C_1} \dots T_{C_b}) = \vcenter{\hbox{

			\tikzset{every picture/.style={line width=0.75pt}} 
			
			\begin{tikzpicture}[x=0.75pt,y=0.75pt,yscale=-1,xscale=1]
				
				\draw   (400,100) .. controls (400,72.39) and (429.1,50) .. (465,50) .. controls (500.9,50) and (530,72.39) .. (530,100) .. controls (530,127.61) and (500.9,150) .. (465,150) .. controls (429.1,150) and (400,127.61) .. (400,100) -- cycle ;
				\draw [line width=0.75]    (400,100) .. controls (414.33,120.24) and (516.08,120.24) .. (530,100) ;
				\draw [line width=0.75]  [dash pattern={on 4.5pt off 4.5pt}]  (400,100) .. controls (415.58,80.49) and (515.83,79.99) .. (530,100) ;
				\draw    (412.36,46.66) .. controls (428.78,66.83) and (432.75,72.5) .. (430,80) ;
				\draw    (430,40) .. controls (431.81,56.46) and (436.21,66.86) .. (440,70) ;
				\draw   (410,40) .. controls (410,34.48) and (414.48,30) .. (420,30) .. controls (425.52,30) and (430,34.48) .. (430,40) .. controls (430,45.52) and (425.52,50) .. (420,50) .. controls (414.48,50) and (410,45.52) .. (410,40) -- cycle ;
				\draw    (537.64,56.66) .. controls (521.22,76.83) and (517.25,82.5) .. (520,90) ;
				\draw    (520,50) .. controls (518.19,66.46) and (513.79,76.86) .. (510,80) ;
				\draw   (540,50) .. controls (540,44.48) and (535.52,40) .. (530,40) .. controls (524.48,40) and (520,44.48) .. (520,50) .. controls (520,55.52) and (524.48,60) .. (530,60) .. controls (535.52,60) and (540,55.52) .. (540,50) -- cycle ;
				\draw  [dash pattern={on 4.5pt off 4.5pt}]  (470.17,36.76) .. controls (477.11,61.79) and (450.36,71.54) .. (440,70) ;
				\draw  [dash pattern={on 4.5pt off 4.5pt}]  (488.9,38.85) .. controls (475.61,54.79) and (480.86,73.54) .. (510,80) ;
				\draw  [dash pattern={on 4.5pt off 4.5pt}] (471.09,29.78) .. controls (473.61,24.89) and (479.63,22.94) .. (484.55,25.45) .. controls (489.47,27.95) and (491.42,33.95) .. (488.9,38.85) .. controls (486.39,43.75) and (480.36,45.69) .. (475.45,43.19) .. controls (470.53,40.69) and (468.58,34.68) .. (471.09,29.78) -- cycle ;
				
				\draw (388,18) node [anchor=north west][inner sep=0.75pt]    {$C_{1}$};
				\draw (529,19) node [anchor=north west][inner sep=0.75pt]    {$C_{b}$};
		\end{tikzpicture}}}
\end{equation}
This can be re-expressed in terms of characters of $G$ using \eqref{eq: cc change of basis}, \eqref{eq: flat rep basis product} and \eqref{eq: flat rep basis delta}
\begin{align}
	Z^G_{h=0, \cclass_1; \dots;\cclass_b} &=  \sum_{\irrep \in \irreps{G}} \frac{d_R^2}{\abs{G}^2} \frac{\chi^\irrep(T_{\cclass_1})}{d_\irrep} \dots \frac{\chi^\irrep(T_{\cclass_b})}{d_\irrep} \\
	&= \sum_{\irrep \in \irreps{G}} \frac{d_R^2}{\abs{G}^2} \abs{\cclass_1}\frac{\chi^\irrep_{\cclass_1}}{d_\irrep} \dots \abs{\cclass_b}\frac{\chi^\irrep_{\cclass_b}}{d_\irrep}.
\end{align}
To compute genus $h$ partition functions with $b$ boundaries we insert handle creation operators
\begin{align}
	Z^G_{h; \cclass_1; \dots; \cclass_b} &= \frac{1}{\abs{G}}\delta(\Pi^h T_{\cclass_1} \dots T_{\cclass_b}) = \sum_{\irrep \in \irreps{{G}}} \qty(\frac{\abs{G}}{d_\irrep})^{2h-2}\frac{\chi^\irrep(T_{\cclass_1})}{d_\irrep} \dots \frac{\chi^\irrep(T_{\cclass_b})}{d_\irrep} \\
	&=\vcenter{\hbox{%
			\tikzset{every picture/.style={line width=0.75pt}} 
			\begin{tikzpicture}[x=0.75pt,y=0.75pt,yscale=-1,xscale=1]
				\draw   (40,321) .. controls (40,298.36) and (98.52,280) .. (170.71,280) .. controls (242.9,280) and (301.42,298.36) .. (301.42,321) .. controls (301.42,343.64) and (242.9,362) .. (170.71,362) .. controls (98.52,362) and (40,343.64) .. (40,321) -- cycle ;
				\draw    (70,322) .. controls (79.99,332.02) and (90.39,332.02) .. (100,322) ;
				\draw    (74.79,326.42) .. controls (82.39,317.62) and (86.39,316.42) .. (95.19,325.62) ;
				\draw    (120,314.75) .. controls (129.99,324.77) and (140.39,324.77) .. (150,314.75) ;
				\draw    (124.79,319.17) .. controls (132.39,310.37) and (136.39,309.17) .. (145.19,318.37) ;
				\draw    (240,314.75) .. controls (249.99,324.77) and (260.39,324.77) .. (270,314.75) ;
				\draw    (244.79,319.17) .. controls (252.39,310.37) and (256.39,309.17) .. (265.19,318.37) ;
				\draw  [dash pattern={on 0.84pt off 2.51pt}]  (160,316) .. controls (179.59,310.02) and (205.59,310.22) .. (230,315) ;
				\draw    (69.36,276.66) .. controls (85.78,296.83) and (89.75,302.5) .. (87,310) ;
				\draw    (87,270) .. controls (88.81,286.46) and (93.21,296.86) .. (97,300) ;
				\draw   (67,270) .. controls (67,264.48) and (71.48,260) .. (77,260) .. controls (82.52,260) and (87,264.48) .. (87,270) .. controls (87,275.52) and (82.52,280) .. (77,280) .. controls (71.48,280) and (67,275.52) .. (67,270) -- cycle ;
				\draw    (257.64,266.66) .. controls (241.22,286.83) and (237.25,292.5) .. (240,300) ;
				\draw    (240,260) .. controls (238.19,276.46) and (233.79,286.86) .. (230,290) ;
				\draw   (260,260) .. controls (260,254.48) and (255.52,250) .. (250,250) .. controls (244.48,250) and (240,254.48) .. (240,260) .. controls (240,265.52) and (244.48,270) .. (250,270) .. controls (255.52,270) and (260,265.52) .. (260,260) -- cycle ;
				\draw    (120.28,262.34) .. controls (126.81,287.51) and (128.05,294.33) .. (122.43,300) ;
				\draw    (139.09,263.63) .. controls (133.88,279.35) and (133.55,290.64) .. (135.69,295.07) ;
				\draw   (120.91,255.3) .. controls (123.21,250.28) and (129.14,248.08) .. (134.16,250.38) .. controls (139.18,252.68) and (141.39,258.61) .. (139.09,263.63) .. controls (136.79,268.65) and (130.85,270.86) .. (125.83,268.56) .. controls (120.81,266.26) and (118.61,260.33) .. (120.91,255.3) -- cycle ;
				\draw  [dash pattern={on 4.5pt off 4.5pt}]  (180.28,252.87) .. controls (186.81,278.05) and (181.97,293.64) .. (160,300) ;
				\draw  [dash pattern={on 4.5pt off 4.5pt}]  (199.09,254.17) .. controls (193.88,269.88) and (197.3,295.31) .. (220,300) ;
				\draw  [dash pattern={on 4.5pt off 4.5pt}] (180.91,245.84) .. controls (183.21,240.82) and (189.14,238.61) .. (194.16,240.91) .. controls (199.18,243.21) and (201.39,249.15) .. (199.09,254.17) .. controls (196.79,259.19) and (190.85,261.39) .. (185.83,259.09) .. controls (180.81,256.79) and (178.61,250.86) .. (180.91,245.84) -- cycle ;
				%
				\draw (45,248) node [anchor=north west][inner sep=0.75pt]    {$C_{1}$};
				\draw (259,229) node [anchor=north west][inner sep=0.75pt]    {$C_{b}$};
				\draw (116,228) node [anchor=north west][inner sep=0.75pt]    {$C_{2}$};
	\end{tikzpicture}%
	}}
\end{align}

\subsection{Integral partial column sums from $G$-CTST} \label{subsec: integral row sums}
We now review relevant integrality results from Section 3.4 of \cite{GCTST2}, based on the above construction of $G$-CTST.

Let $D \in \cclasses{G}$ be a conjugacy class such that $\chi_D^R$ is an integer for all $\irrep \in \irreps{G}$. That is, $D$ defines an integer column in the character table of $G$. Note that this implies that
\begin{equation}
	\frac{\chi^\irrep(T_D)}{d_\irrep}=\abs{D}\frac{\chi_D^\irrep}{d_\irrep}
\end{equation}
is rational for all $\irrep \in \irreps{G}$, which will be important in what follows.
Fix an irreducible representation $S \in \irreps{G}$ and define the set
\begin{equation}\label{eq:[S,D]}
\begin{aligned}
&[(S, D)] = \qty{\irrep \in \irreps{G} \qq{s.t. } \frac{\chi^\irrep(T_D)}{d_\irrep} = \frac{\chi^S(T_D)}{d_S}}\\
&~~~~~~~~= \qty{\irrep \in \irreps{G} \qq{s.t. } \abs{D}\frac{\chi^\irrep_D}{d_\irrep} = \abs{D}\frac{\chi^S_D}{d_S}} \\
		&~~~~~~~~= \qty{\irrep \in \irreps{G} \qq{s.t. } \frac{\chi^\irrep_D}{d_\irrep} = \frac{\chi^S_D}{d_S}}.	
\end{aligned}
\end{equation}
It was proved  in  \cite{GCTST2} (Proposition 3.4.1-II)  that 
\begin{equation} 
	\sum_{\irrep \in [(S,D)]} \abs{C}\frac{\chi_\cclass^\irrep}{d_\irrep} \in \mathbb{Z}, \label{eq: row sum thm 1}
\end{equation}
for every $\cclass \in \cclasses{G}$.

The proof uses the family of partition functions
\begin{equation}
	Z^{G}_{h=1; D^b, C} = \sum_{\irrep \in \irreps{G}} \qty(\frac{\chi^\irrep(T_D)}{d_R})^b  \frac{\chi^\irrep(T_\cclass)}{d_\irrep} = \vcenter{\hbox{

			\tikzset{every picture/.style={line width=0.75pt}} 
			
			\begin{tikzpicture}[x=0.75pt,y=0.75pt,yscale=-1,xscale=1]
				
				\draw   (394.58,320) .. controls (394.58,295.15) and (429.53,275) .. (472.63,275) .. controls (515.74,275) and (550.69,295.15) .. (550.69,320) .. controls (550.69,344.85) and (515.74,365) .. (472.63,365) .. controls (429.53,365) and (394.58,344.85) .. (394.58,320) -- cycle ;
				\draw    (440,315.35) .. controls (459.47,334.88) and (479.74,334.88) .. (498.45,315.35) ;
				\draw    (449.34,323.97) .. controls (464.15,306.83) and (471.94,304.49) .. (489.09,322.41) ;
				\draw    (402.36,287.66) .. controls (418.78,307.83) and (422.75,313.5) .. (420,321) ;
				\draw    (420,281) .. controls (421.81,297.46) and (426.21,307.86) .. (430,311) ;
				\draw   (400,281) .. controls (400,275.48) and (404.48,271) .. (410,271) .. controls (415.52,271) and (420,275.48) .. (420,281) .. controls (420,286.52) and (415.52,291) .. (410,291) .. controls (404.48,291) and (400,286.52) .. (400,281) -- cycle ;
				\draw    (440.28,263.34) .. controls (446.81,288.51) and (448.05,295.33) .. (442.43,301) ;
				\draw    (459.09,264.63) .. controls (453.88,280.35) and (453.55,291.64) .. (455.69,296.07) ;
				\draw   (440.91,256.3) .. controls (443.21,251.28) and (449.14,249.08) .. (454.16,251.38) .. controls (459.18,253.68) and (461.39,259.61) .. (459.09,264.63) .. controls (456.79,269.65) and (450.85,271.86) .. (445.83,269.56) .. controls (440.81,267.26) and (438.61,261.33) .. (440.91,256.3) -- cycle ;
				\draw    (537.64,276.66) .. controls (521.22,296.83) and (517.25,302.5) .. (520,310) ;
				\draw    (520,270) .. controls (518.19,286.46) and (513.79,296.86) .. (510,300) ;
				\draw   (540,270) .. controls (540,264.48) and (535.52,260) .. (530,260) .. controls (524.48,260) and (520,264.48) .. (520,270) .. controls (520,275.52) and (524.48,280) .. (530,280) .. controls (535.52,280) and (540,275.52) .. (540,270) -- cycle ;
				\draw  [dash pattern={on 4.5pt off 4.5pt}]  (475,257.45) .. controls (481.52,282.62) and (470.43,287.95) .. (467.71,294.57) ;
				\draw  [dash pattern={on 4.5pt off 4.5pt}]  (493.8,258.74) .. controls (488.59,274.46) and (492.43,289.95) .. (497.71,294.57) ;
				\draw  [dash pattern={on 4.5pt off 4.5pt}] (475.62,250.41) .. controls (477.92,245.39) and (483.86,243.18) .. (488.88,245.48) .. controls (493.9,247.78) and (496.1,253.72) .. (493.8,258.74) .. controls (491.5,263.76) and (485.57,265.97) .. (480.55,263.67) .. controls (475.53,261.37) and (473.32,255.43) .. (475.62,250.41) -- cycle ;
				\draw  [dash pattern={on 4.5pt off 4.5pt}]  (475,257.45) .. controls (481.52,282.62) and (470.43,287.95) .. (467.71,294.57) ;
				\draw  [dash pattern={on 4.5pt off 4.5pt}]  (493.8,258.74) .. controls (488.59,274.46) and (492.43,289.95) .. (497.71,294.57) ;
				\draw  [dash pattern={on 4.5pt off 4.5pt}] (475.62,250.41) .. controls (477.92,245.39) and (483.86,243.18) .. (488.88,245.48) .. controls (493.9,247.78) and (496.1,253.72) .. (493.8,258.74) .. controls (491.5,263.76) and (485.57,265.97) .. (480.55,263.67) .. controls (475.53,261.37) and (473.32,255.43) .. (475.62,250.41) -- cycle ;
				\draw    (539.36,361.59) .. controls (522.7,341.62) and (517.88,336.64) .. (510,337.9) ;
				\draw    (549.26,345.55) .. controls (533.45,340.63) and (524.08,334.32) .. (521.73,330) ;
				\draw   (545.44,365.18) .. controls (550.86,366.24) and (556.11,362.7) .. (557.17,357.28) .. controls (558.22,351.86) and (554.68,346.61) .. (549.26,345.55) .. controls (543.84,344.49) and (538.59,348.03) .. (537.54,353.45) .. controls (536.48,358.88) and (540.02,364.13) .. (545.44,365.18) -- cycle ;
				
				\draw (378,259) node [anchor=north west][inner sep=0.75pt]    {$D$};
				\draw (436,229) node [anchor=north west][inner sep=0.75pt]    {$D$};
				\draw (539,239) node [anchor=north west][inner sep=0.75pt]    {$D$};
				\draw (554,339) node [anchor=north west][inner sep=0.75pt]    {$C$};
	\end{tikzpicture}}}
\end{equation}
for $b$ a non-negative integer. The sum over irreducible representations is organized in terms of level sets of the normalized characters $\frac{\chi_D^{\irrep}}{d_{R}}$ 
\begin{equation}
	Z^{G}_{h=1; D^b, C} = \sum_{\irrep'} \qty(\frac{\chi^{\irrep'}(T_D)}{d_{R'}})^b \sum_{\irrep \in [(\irrep',D)]}\frac{\chi^\irrep(T_\cclass)}{d_\irrep}, \label{eq: review int result}
\end{equation}
where the first sum is over a set $\{\irrep'_1, \dots, \irrep'_k\}$ of irreducible representations such that
\begin{equation}
	\frac{\chi_D^{\irrep'_1}}{d_{\irrep'_1}}, \dots, \frac{\chi_D^{\irrep'_k}}{d_{\irrep'_k}},
\end{equation}
are all distinct. Now we restrict to the partition functions with $b=1,\dots,k$. We defined \eqref{eq:[S,D]} as the set of irreps having a fixed normalized character
 $ |D| { \chi^{R}_D \over d_R }$ which implies  that the conjugacy class $D$ determines a set partition of the set of irreps $\irreps{G} $ : 
\begin{equation}
	[(\irrep_1', D)] \cup [(\irrep_2', D)] \cup {\dots} \cup [(\irrep_k', D)] = \irreps{G}.
\end{equation}
Equation \eqref{eq: review int result} can be understood as a matrix equation
\begin{equation}
	y = V x  ~~~ ; ~~~ \hbox{ with indices explicit } y_b = \sum_{ R'} V_{ bR'} x_{R'} 
\end{equation}
where
\begin{align}
	&y_b = Z^{G}_{h=1; D^b, C}, \\
	&x_{\irrep'} =  \sum_{\irrep \in [(\irrep',D)]} \frac{\chi^\irrep(T_\cclass)}{d_\irrep} = \sum_{\irrep \in [(\irrep',D)]} \abs{C} \frac{\chi_\cclass^\irrep}{d_\irrep} \\
	&V_{b\irrep'} =  \qty( \frac{\chi^{\irrep'}(T_D)}{d_{R'}})^b = \qty(\abs{D} \frac{\chi_D^{\irrep'}}{d_{R'}})^b.
\end{align}
Since we assumed that $\chi_D^\irrep$ is an integer for all $\irrep$, the matrix $V$ is a $k \times k$ Vandermonde matrix with rational entries. Consequently, the inverse matrix $V^{-1}$ is a rational matrix as well and we have
\begin{equation}
	x = V^{-1}y.
\end{equation}
The r.h.s. of this equation is rational vector, while the l.h.s. is a vector of normalized characters which are known to be  algebraic integers. An algebraic integer is rational if and only if it is an integer and therefore
\begin{equation}
	x_{S} = \sum_{\irrep \in [(S,D)]}  \frac{\chi^\irrep(T_\cclass)}{d_\irrep} =\sum_{\irrep \in [(S,D)]} \abs{C} \frac{\chi_\cclass^\irrep}{d_\irrep} \in \mathbb{Z}.
\end{equation}
which is the result \eqref{eq: row sum thm 1}.

Integrality can also be shown for sums over irreducible representations with fixed dimension. Define the sets
\begin{align}
	[d_{\irrep'}] &= \{\irrep \in \irreps{G} \qq{s.t. } \frac{\abs{G}^2}{d_\irrep^2} = \frac{\abs{G}^2}{d_{\irrep'}^2}  \} \\
	&=\{\irrep \in \irreps{G} \qq{s.t. } d_\irrep = d_{\irrep'}\}.
\end{align}
A slight variation of the above construction, using insertions of handle creation operators, was used in \cite{GCTST2} to prove that (see  \cite[Proposition 3.4.1-II]{GCTST2})
\begin{equation}
	\sum_{\irrep \in [d_{\irrep'}]}  \chi^\irrep_\cclass \in \mathbb{Z}. \label{eq: TQFT dim sum is integer}
\end{equation}
Since the sum of characters along any column is equivalent to a sum of partial sums over irreducible representations of fixed dimension the following proposition follows.
\begin{proposition}\label{prop: sum over all irreps is integer}
Column sums of character tables are integers
\begin{equation}
	\sum_{{\irrep} \in \irreps{G}} \chi_\cclass^{\irrep} \in \mathbb{Z}, \quad \forall \cclass \in \cclasses{G}.
\end{equation}
\end{proposition}

Interestingly, the above sum is not necessarily positive, but a simple G-CTST proof gives the following proposition.
\begin{proposition}\label{prop: sum of normalized chars is pos int}
	Let $\mathbb{Z}^+$ be the set of positive integers, then
	\begin{equation}
		\sum_{{\irrep} \in \irreps{G}} \frac{\chi^{\irrep}(T_\cclass)}{d_\irrep} \in \mathbb{Z}^+, \quad \forall \cclass \in \cclasses{G}.
	\end{equation}
\end{proposition}
\noindent \textbf{Proof of Proposition \ref{prop: sum of normalized chars is pos int}}
Consider the amplitude
\begin{equation}
	Z^G_{h=1;\cclass} = { 1 \over |G|} \delta ( \Pi ~ T_\cclass ) = \sum_{R \in \irreps{G}}\frac{\chi^{\irrep}(T_\cclass)}{d_\irrep}.
\end{equation}
the l.h.s. is a positive rational number for all $\cclass \in \cclasses{G}$, while the r.h.s. is a sum of algebraic integers. Therefore, it is a positive integer, which proves the proposition.

The  equation \eqref{eq: row sum thm 1} involves sums over sets of  irreducible representations (irreps)  which are level sets of a function of irreps defined by a conjugacy class $S$. A natural question to ask is if similar integrality results exist for sums over level sets of conjugacy classes. In other words, are there similar integrality results to the above obtained by exchanging the role of rows and columns of character tables. Analogies between rows and columns have motivated a number of investigations on number-theoretic properties of characters \cite{Navarro}. 
In the next section we will introduce a new TQFT based on fusion algebras. A similar algebra was defined in \cite{Kock}, but with different Frobenius structure. We will find that the one used in the next section is the natural setting for deriving integrality results in the context of CTST which are dual under an exchange of rows and columns to the above integrality results. This motivates the notion of a physical row-column duality in the context of CTST which provides a physical framework for thinking about mathematical analogies based on  the exchange of rows and columns of character tables.

\subsection{$\fusionalg{G}$-CTST} 

In this section we give a  formulae for the partition functions in the $\fusionalg{G}$-TQFT in terms of fusion coefficients (tensor product multiplicities) of irreps of $G$. This is analogous to formulae in the 
the previous subsection for $G$-TQFT with fusion coefficients replacing class algebra structure constants in the central algebra $\cZ ( \mC ( G ) )$. To define a TQFT, it is useful to appeal to the language of Frobenius algebras. In appendix \ref{apx: fusion tqft} we show that the set of distinct (isomorphism classes) of irreducible representations of $G$ can be given the structure of a commutative Frobenius algebra, using fusion coefficients, and therefore defines $R(G)$-TQFT. The appendix contains expressions for genus $h$ partition functions with $n$ boundaries, derived using the Frobenius structure. In this section we show that equivalent expressions can be computed using a handle creation operator $\FGhandle$, counit $\varepsilon$ and fusion of irreducible representations.

The algebra underlying the $\fusionalg{G}$-TQFT is a vector space of dimension
\begin{equation}
	\dim \fusionalg{G} = \abs{\irreps{G}},
\end{equation}
where  $\irreps{G}$ is a complete set of non-isomorphic irreducible representations of $G$.
It has a basis
\begin{equation}
	\{a_{\irrep_1}, \dots, a_{\irrep_K}\},
\end{equation}
where the structure constants
\begin{equation}
	N_{\irrep_1 \irrep_2}^{\irrep_3} = \frac{1}{\abs{G}} \sum_{g \in G} \chi^{\irrep_1}(g)\chi^{\irrep_2}(g)\overline{\chi^{\irrep_3}(g)}.
\end{equation}
are fusion coefficients and
\begin{equation}
	a_{\irrep_1} a_{\irrep_2} = \sum_{\irrep_3 \in \irreps{G}} N_{\irrep_1 \irrep_2}^{\irrep_3} a_{\irrep_3}. \label{eq: fusion product}
\end{equation}
The trivial representation, denoted $R_0$, corresponds to the identity in this algebra: $a_{R_0}a_R = a_R$.

The fusion algebra has a basis of  orthogonal idempotents (projectors) 
\bea\label{projector}  
A_{ \cclass } = { 1 \over \Sym \cclass } \sum_{ \irrep }  {\chi^{ \overline \irrep }_\cclass }  a_{ \irrep},
\eea
where  $\Sym\, \cclass = \abs{G}/\abs{C}$ as in \eqref{eq: def sym C}. The orthogonal idempotent property is expressed in 
\bea\label{orthoproj}  
	A_{\cclass_1} A_{\cclass_2} = \delta_{\cclass_1 \cclass_2} A_{\cclass_1} 
\eea
This proved as follows 
\bea 
&& A_{ \cclass_1} A_{ \cclass_2} = { 1 \over \Sym \cclass_1 } . { 1\over \Sym \cclass_2 } \sum_{ R } \chi^{ \bar R }_{ \cclass_1 } a_R \sum_{ S } \chi_{ \cclass_2}^{ \bar S } a_S \cr 
&&  = { 1 \over \Sym \cclass_1 } . { 1\over \Sym \cclass_2 } \sum_{ R , S } \chi^{ \bar R }_{ \cclass_1} 
\chi_{ \cclass_2}^{ \bar S } a_R . a_S \cr 
&& = { 1\over \Sym \cclass_1} { 1 \over \Sym \cclass_2} \sum_{ R , S , T }
\chi^{ \bar R }_{ \cclass_1}  \chi^{ \bar S }_{ \cclass_2} N_{ RS}^T a_T \cr 
&& = { 1 \over \Sym \cclass_1 ~ \Sym \cclass_2 } \sum_{ R , S , T }  \chi^{ \bar R }_{ \cclass_1} \chi^{ \bar S }_{ \cclass_2} \sum_{ \cclass } \chi^{ R}_\cclass \chi^S_{\cclass} \chi^{ \bar T}_{ \cclass } . { |\cclass| \over |G| } . a_T \cr 
&& = { 1 \over \Sym \cclass_1 ~ \Sym \cclass_2 } \sum_{ T, C} \delta_{ \cclass_1 , \cclass   } \delta_{ \cclass_2 , \cclass } \Sym \cclass_1 ~ \Sym \cclass_2 . { |\cclass| \over |G| } . \chi^{ \bar T }_{ \cclass  } a_T \cr 
&& = { 1 \over \Sym \cclass_1 } \sum_{ T } \chi^{ \bar T }_{ \cclass_1} a_T  . \delta_{ \cclass_1 , \cclass_2 } \cr 
&& = A_{ \cclass_1} \delta_{ \cclass_1 ,  \cclass_2 }.
\eea
where we have used an expression for the fusion coefficient as a sum over characters, along with character orthogonality relations. The  equation \eqref{orthoproj}  also shows that the structure constants of the multiplication are diagonal in the idempotent basis.
 
The orthogonal  projectors  obey the completeness relation 
\begin{equation}
	\sum_{\cclass} A_{\cclass} = a_{\irrep_0},
\end{equation}
where $\irrep_0$ is the trivial representation since
\bea 
&& \sum_{ \cclass }  A_\cclass = \sum_{ \cclass } { 1 \over \Sym \cclass } \sum_{ R } \chi^{ \bar R }_\cclass a_R \cr 
&&  = \sum_{ R } \sum_{ \cclass } { |\cclass| \chi^{ \bar R }_\cclass \over |G| } a_R \cr 
&& = \sum_{ R } \sum_{ g } { \chi^\irrep (g^{-1} ) \over |G| } ~ a_R \cr 
&& = \sum_{ R  }  \sum_{ g } { \chi^\irrep (g^{-1} ) \over |G| } \chi^{ R_0} ( g )  ~ a_R \cr 
&& = \sum_R \delta^{ R , R_0 } a_{R_0}  = a_{ R_0}.
\eea

The idempotents  are also  eigenvectors of the operation of multiplication by  $a_\irrep$ : 
\begin{equation}
	a_\irrep A_\cclass = \chi^\irrep_\cclass A_\cclass, \label{eq: fusion eigenvectors} \, . 
\end{equation}
This is derived as follows 
\bea 
a_R A_\cclass  && = { 1 \over \Sym \cclass } \sum_{ S } \chi^{ \bar S }_\cclass a_R a_S \cr 
&& = { 1 \over \Sym \cclass} \sum_{ S , T } \chi^{ \bar S }_\cclass N_{ RS}^T a_T \cr 
&& = { 1 \over \Sym \cclass } \sum_{ S , T } \chi^{ \bar S }_\cclass \sum_{ \cclass'} \chi^\irrep_{ \cclass'} \chi^{S}_{ \cclass'} \chi^{ \bar T }_{ \cclass'} { |\cclass'|\over |G| }   a_T \cr 
&& =  { 1\over \Sym \cclass } \sum_{ \cclass', T }
\left ( \sum_{ S } \chi^{\bar S }_{ \cclass }  \chi^S_{ \cclass'} \right ) \chi^\irrep_{ \cclass'} \chi^{\bar T}_{ \cclass'} . { |\cclass'| \over |G| } . a_T \cr 
&& = { 1\over \Sym C } \sum_{ \cclass', T } \left ( \delta_{ \cclass , \cclass'} \Sym \cclass' \right ) 
\chi^\irrep_{ \cclass'} \chi^{\bar T}_{ \cclass' } . {  |\cclass'| \over |G| } . a_T \cr 
&& = \sum_{ T } { 1 \over \Sym \cclass } \chi^\irrep_\cclass \chi^{ \bar T}_{ \cclass } a_T \cr 
&& = \chi^\irrep_{ \cclass }  { 1 \over \Sym \cclass } \sum_{ T } \chi^{ \bar T }_\cclass a_T  \cr 
&& = \chi^\irrep_\cclass A_\cclass.
\eea
Comparing \eqref{eq: fusion eigenvectors} and \eqref{eq: fusion product}, we see that the matrix $N_{\irrep}$ corresponding to left multiplication of $a_\irrep$ has eigenvalues corresponding to characters $\chi^\irrep_\cclass$ for $\cclass \in \cclasses{G}$. Therefore, the fusion algebra gives an algorithm analogous to the Burnside's algorithm \cite{burnside1911theory,dixon1967high,schneider1990dixon}, which constructs characters of $G$ from the structure constant matrices $f_\cclass$ of the centre of the group algebra. As discussed in section 3.1  of \cite{IDFCTS},  $G$-CTST amplitudes on genus one surfaces  with boundaries provide the combinatorial data from multiplication of conjugacy classes in $ \cZ ( \mC ( G ) ) $  which enters the Burnside construction of the normalized characters $ { \chi^R ( T_C ) \over d_R }$. Similarly the corresponding  amplitudes in $R(G)$-CTST give the positive integer fusion data which can be used to construct the characters $\chi^R ( T_C) $.

To define the partition functions of the fusion TQFT we introduce a counit on $\fusionalg{G}$ (see \eqref{eq: fusion counit})
\begin{equation}
	\varepsilon(a_{\irrep}) = \begin{cases}
		1 \qq{if $\irrep = \irrep_0$,} \\
		0 \qq{otherwise.}
	\end{cases}
\end{equation}
In the idempotent basis we have
\bea 
\varepsilon ( A_\cclass  ) = { 1 \over \Sym \cclass } \sum_{ R } \chi^{ \bar R  }_\cclass  \varepsilon (  a_{ R}  ) 
= { 1 \over \Sym \cclass } \chi^{ R_0}_\cclass = { 1 \over \Sym \cclass } \label{eq: counit cc basis}
\eea
Just as in the $G$-TQFT, it is possible to construct a handle creation operator $\FGhandle$ in the fusion algebra such that
\begin{equation}
	Z_h^{\fusionalg{G}} = \varepsilon(\FGhandle^h) = \vcenter{\hbox{
			\tikzset{every picture/.style={line width=0.75pt}} 
			
			\begin{tikzpicture}[x=0.75pt,y=0.75pt,yscale=-1,xscale=1]
				
				\draw   (50,119) .. controls (50,96.36) and (108.52,78) .. (180.71,78) .. controls (252.9,78) and (311.42,96.36) .. (311.42,119) .. controls (311.42,141.64) and (252.9,160) .. (180.71,160) .. controls (108.52,160) and (50,141.64) .. (50,119) -- cycle ;
				\draw    (80,120) .. controls (89.99,130.02) and (100.39,130.02) .. (110,120) ;
				\draw    (84.79,124.42) .. controls (92.39,115.62) and (96.39,114.42) .. (105.19,123.62) ;
				\draw    (130,112.75) .. controls (139.99,122.77) and (150.39,122.77) .. (160,112.75) ;
				\draw    (134.79,117.17) .. controls (142.39,108.37) and (146.39,107.17) .. (155.19,116.37) ;
				\draw    (250,112.75) .. controls (259.99,122.77) and (270.39,122.77) .. (280,112.75) ;
				\draw    (254.79,117.17) .. controls (262.39,108.37) and (266.39,107.17) .. (275.19,116.37) ;
				\draw  [dash pattern={on 0.84pt off 2.51pt}]  (170,114) .. controls (189.59,108.02) and (215.59,108.22) .. (240,113) ;
	\end{tikzpicture}}}
\end{equation}
We can define the handle creation operator
\bea 
\FGhandle = \sum_{ S \in \irreps{G} } N_{ S \bar S}^R a_R 
\eea
which reflects the fact that a torus with a hole  (labelled with state $R$) is obtained by gluing two boundaries from a sphere with three boundaries. The gluing operation is associated with the sum over $S$. 
This can be used to obtain an expression for $\Theta $  as a sum over the projectors $A_{ \cclass}$. It is useful to observe 
\bea\label{identNchi}  
 \sum_{ S }  N_{ S \bar S}^R  &=&  { 1 \over |G| } \sum_{ S }  \sum_{ g \in G } \chi^S ( g ) \chi^{ \bar S } ( g ) \chi^{\bar R } ( g) \cr 
&  =& \sum_{ C \in \cclasses{G}  } \chi^{\bar R}_C ( g )  
\eea
where the second line is obtained by using the orthogonality of characters to simplify the sum over $S$. It then follows, using the formula \eqref{projector}  for the projector $A_{ \cclass }$ that 
\bea 
\FGhandle = \sum_\cclass  ( \Sym \cclass ) A_\cclass.
\eea
From \eqref{eq: counit cc basis} we have
\bea 
Z_h^{\fusionalg{G}} = \epsilon ( \Theta^h ) =\sum_{ \cclass }  \epsilon  ( ( \Sym \cclass)^h A_K ) = \sum_{ \cclass } ( \Sym \cclass )^{h-1},
\eea
which agrees with \eqref{eq: fusion genus h} derived from the Frobenius structure.
Genus $h$ partition functions with $n$ boundaries are given by
\begin{equation}
	Z^{\fusionalg{G}}_{h,\irrep_1, \dots, \irrep_n} = \varepsilon(\FGhandle^h a_{\irrep_1} \dots a_{\irrep_n}) =\vcenter{\hbox{%
			\tikzset{every picture/.style={line width=0.75pt}} 
			\begin{tikzpicture}[x=0.75pt,y=0.75pt,yscale=-1,xscale=1]
				\draw   (40,321) .. controls (40,298.36) and (98.52,280) .. (170.71,280) .. controls (242.9,280) and (301.42,298.36) .. (301.42,321) .. controls (301.42,343.64) and (242.9,362) .. (170.71,362) .. controls (98.52,362) and (40,343.64) .. (40,321) -- cycle ;
				\draw    (70,322) .. controls (79.99,332.02) and (90.39,332.02) .. (100,322) ;
				\draw    (74.79,326.42) .. controls (82.39,317.62) and (86.39,316.42) .. (95.19,325.62) ;
				\draw    (120,314.75) .. controls (129.99,324.77) and (140.39,324.77) .. (150,314.75) ;
				\draw    (124.79,319.17) .. controls (132.39,310.37) and (136.39,309.17) .. (145.19,318.37) ;
				\draw    (240,314.75) .. controls (249.99,324.77) and (260.39,324.77) .. (270,314.75) ;
				\draw    (244.79,319.17) .. controls (252.39,310.37) and (256.39,309.17) .. (265.19,318.37) ;
				\draw  [dash pattern={on 0.84pt off 2.51pt}]  (160,316) .. controls (179.59,310.02) and (205.59,310.22) .. (230,315) ;
				\draw    (69.36,276.66) .. controls (85.78,296.83) and (89.75,302.5) .. (87,310) ;
				\draw    (87,270) .. controls (88.81,286.46) and (93.21,296.86) .. (97,300) ;
				\draw   (67,270) .. controls (67,264.48) and (71.48,260) .. (77,260) .. controls (82.52,260) and (87,264.48) .. (87,270) .. controls (87,275.52) and (82.52,280) .. (77,280) .. controls (71.48,280) and (67,275.52) .. (67,270) -- cycle ;
				\draw    (257.64,266.66) .. controls (241.22,286.83) and (237.25,292.5) .. (240,300) ;
				\draw    (240,260) .. controls (238.19,276.46) and (233.79,286.86) .. (230,290) ;
				\draw   (260,260) .. controls (260,254.48) and (255.52,250) .. (250,250) .. controls (244.48,250) and (240,254.48) .. (240,260) .. controls (240,265.52) and (244.48,270) .. (250,270) .. controls (255.52,270) and (260,265.52) .. (260,260) -- cycle ;
				\draw    (120.28,262.34) .. controls (126.81,287.51) and (128.05,294.33) .. (122.43,300) ;
				\draw    (139.09,263.63) .. controls (133.88,279.35) and (133.55,290.64) .. (135.69,295.07) ;
				\draw   (120.91,255.3) .. controls (123.21,250.28) and (129.14,248.08) .. (134.16,250.38) .. controls (139.18,252.68) and (141.39,258.61) .. (139.09,263.63) .. controls (136.79,268.65) and (130.85,270.86) .. (125.83,268.56) .. controls (120.81,266.26) and (118.61,260.33) .. (120.91,255.3) -- cycle ;
				\draw  [dash pattern={on 4.5pt off 4.5pt}]  (180.28,252.87) .. controls (186.81,278.05) and (181.97,293.64) .. (160,300) ;
				\draw  [dash pattern={on 4.5pt off 4.5pt}]  (199.09,254.17) .. controls (193.88,269.88) and (197.3,295.31) .. (220,300) ;
				\draw  [dash pattern={on 4.5pt off 4.5pt}] (180.91,245.84) .. controls (183.21,240.82) and (189.14,238.61) .. (194.16,240.91) .. controls (199.18,243.21) and (201.39,249.15) .. (199.09,254.17) .. controls (196.79,259.19) and (190.85,261.39) .. (185.83,259.09) .. controls (180.81,256.79) and (178.61,250.86) .. (180.91,245.84) -- cycle ;
				%
				\draw (45,248) node [anchor=north west][inner sep=0.75pt]    {$R_{1}$};
				\draw (259,229) node [anchor=north west][inner sep=0.75pt]    {$R_{n}$};
				\draw (116,228) node [anchor=north west][inner sep=0.75pt]    {$R_{2}$};
			\end{tikzpicture}%
	}}
\end{equation}
From \eqref{eq: fusion eigenvectors} we have
\begin{equation}\label{RGhrepspart} 
	\begin{aligned}		
		\varepsilon(\FGhandle^h a_{\irrep_1} \dots a_{\irrep_n}) 
		&= \sum_{\cclass \in \cclasses{G}} (\Sym C)^h \varepsilon(A_\cclass a_{\irrep_1} \dots a_{\irrep_n} ) \\
		&= \sum_{\cclass \in \cclasses{G}} (\Sym C)^h \chi_\cclass^{\irrep_1} \dots \chi_\cclass^{\irrep_n} \varepsilon(A_\cclass) \\
		&=\sum_{\cclass \in \cclasses{G}} (\Sym C)^{h-1} \chi_\cclass^{\irrep_1} \dots \chi_\cclass^{\irrep_n},		
	\end{aligned}
\end{equation}
which agree with \eqref{eq: fusion genus h n boundaries} derived from the Frobenius structure.

\subsection{Integral partial row sums from $\fusionalg{G}$-CTST} \label{subsec: integral column sums}
The $\fusionalg{G}$-CTST will allow us to derive integrality results for  sums of characters along rows. 
 The proof mirrors that of section \ref{subsec: integral row sums}.  The input of the integrality properties of multiplication in $\cZ ( \mC ( G )  ) $ in the basis of class central elements $T_C$ \eqref{ClassAlgInteg} exploited in  section \ref{subsec: integral row sums}  will now  be replaced by the integrality properties of fusion coefficients which play a role as building blocks for amplitudes  in $R(G)$-TQFT. 

\begin{theorem}\label{thm: unrefined partial row sums}
Let $S \in \irreps{G}$ be an irreducible representation such that $\chi_\cclass^S$ is an integer for all $\cclass \in \cclasses{G}$. This defines an integer row in the character table of $G$. Now, fix a conjugacy class $D$ and define the set
\begin{equation}
	\langle (S,D) \rangle = \qty{\cclass \in \cclasses{G} \qq{s.t. } \chi^S_\cclass = \chi^S_D}.
\end{equation}
Now consider any $ R \in \irreps{G}$ with $ R \ne S $.  We will show that
\begin{equation}
	\sum_{\cclass \in \langle (S,D) \rangle} \chi_\cclass^{\irrep} \in \mathbb{Z}. 
\end{equation}
\end{theorem}
\noindent \textbf{Proof of Theorem \ref{thm: unrefined partial row sums}}
Consider the vector of partition functions
\begin{equation}
	Z^{\fusionalg{G}}_{h=1; S^b, \irrep} = \sum_{\cclass \in \cclasses{G}} \qty(\chi_\cclass^S)^b \chi_\cclass^\irrep = \vcenter{\hbox{

			\tikzset{every picture/.style={line width=0.75pt}} 
			
			\begin{tikzpicture}[x=0.75pt,y=0.75pt,yscale=-1,xscale=1]
				
				\draw   (394.58,320) .. controls (394.58,295.15) and (429.53,275) .. (472.63,275) .. controls (515.74,275) and (550.69,295.15) .. (550.69,320) .. controls (550.69,344.85) and (515.74,365) .. (472.63,365) .. controls (429.53,365) and (394.58,344.85) .. (394.58,320) -- cycle ;
				\draw    (440,315.35) .. controls (459.47,334.88) and (479.74,334.88) .. (498.45,315.35) ;
				\draw    (449.34,323.97) .. controls (464.15,306.83) and (471.94,304.49) .. (489.09,322.41) ;
				\draw    (402.36,287.66) .. controls (418.78,307.83) and (422.75,313.5) .. (420,321) ;
				\draw    (420,281) .. controls (421.81,297.46) and (426.21,307.86) .. (430,311) ;
				\draw   (400,281) .. controls (400,275.48) and (404.48,271) .. (410,271) .. controls (415.52,271) and (420,275.48) .. (420,281) .. controls (420,286.52) and (415.52,291) .. (410,291) .. controls (404.48,291) and (400,286.52) .. (400,281) -- cycle ;
				\draw    (440.28,263.34) .. controls (446.81,288.51) and (448.05,295.33) .. (442.43,301) ;
				\draw    (459.09,264.63) .. controls (453.88,280.35) and (453.55,291.64) .. (455.69,296.07) ;
				\draw   (440.91,256.3) .. controls (443.21,251.28) and (449.14,249.08) .. (454.16,251.38) .. controls (459.18,253.68) and (461.39,259.61) .. (459.09,264.63) .. controls (456.79,269.65) and (450.85,271.86) .. (445.83,269.56) .. controls (440.81,267.26) and (438.61,261.33) .. (440.91,256.3) -- cycle ;
				\draw    (537.64,276.66) .. controls (521.22,296.83) and (517.25,302.5) .. (520,310) ;
				\draw    (520,270) .. controls (518.19,286.46) and (513.79,296.86) .. (510,300) ;
				\draw   (540,270) .. controls (540,264.48) and (535.52,260) .. (530,260) .. controls (524.48,260) and (520,264.48) .. (520,270) .. controls (520,275.52) and (524.48,280) .. (530,280) .. controls (535.52,280) and (540,275.52) .. (540,270) -- cycle ;
				\draw  [dash pattern={on 4.5pt off 4.5pt}]  (475,257.45) .. controls (481.52,282.62) and (470.43,287.95) .. (467.71,294.57) ;
				\draw  [dash pattern={on 4.5pt off 4.5pt}]  (493.8,258.74) .. controls (488.59,274.46) and (492.43,289.95) .. (497.71,294.57) ;
				\draw  [dash pattern={on 4.5pt off 4.5pt}] (475.62,250.41) .. controls (477.92,245.39) and (483.86,243.18) .. (488.88,245.48) .. controls (493.9,247.78) and (496.1,253.72) .. (493.8,258.74) .. controls (491.5,263.76) and (485.57,265.97) .. (480.55,263.67) .. controls (475.53,261.37) and (473.32,255.43) .. (475.62,250.41) -- cycle ;
				\draw  [dash pattern={on 4.5pt off 4.5pt}]  (475,257.45) .. controls (481.52,282.62) and (470.43,287.95) .. (467.71,294.57) ;
				\draw  [dash pattern={on 4.5pt off 4.5pt}]  (493.8,258.74) .. controls (488.59,274.46) and (492.43,289.95) .. (497.71,294.57) ;
				\draw  [dash pattern={on 4.5pt off 4.5pt}] (475.62,250.41) .. controls (477.92,245.39) and (483.86,243.18) .. (488.88,245.48) .. controls (493.9,247.78) and (496.1,253.72) .. (493.8,258.74) .. controls (491.5,263.76) and (485.57,265.97) .. (480.55,263.67) .. controls (475.53,261.37) and (473.32,255.43) .. (475.62,250.41) -- cycle ;
				\draw    (539.36,361.59) .. controls (522.7,341.62) and (517.88,336.64) .. (510,337.9) ;
				\draw    (549.26,345.55) .. controls (533.45,340.63) and (524.08,334.32) .. (521.73,330) ;
				\draw   (545.44,365.18) .. controls (550.86,366.24) and (556.11,362.7) .. (557.17,357.28) .. controls (558.22,351.86) and (554.68,346.61) .. (549.26,345.55) .. controls (543.84,344.49) and (538.59,348.03) .. (537.54,353.45) .. controls (536.48,358.88) and (540.02,364.13) .. (545.44,365.18) -- cycle ;
				
				\draw (378,259) node [anchor=north west][inner sep=0.75pt]    {$S$};
				\draw (436,229) node [anchor=north west][inner sep=0.75pt]    {$S$};
				\draw (539,239) node [anchor=north west][inner sep=0.75pt]    {$S$};
				\draw (554,339) node [anchor=north west][inner sep=0.75pt]    {$R$};
	\end{tikzpicture}}}
\end{equation}
for $b$ a non-negative integer. We organize the sum into level sets
\begin{equation}
	Z^{\fusionalg{G}}_{h=1; S^b, \irrep} = \sum_{\cclass'}  \qty(\chi_{\cclass'}^S)^b \sum_{\cclass \in \langle (S,\cclass') \rangle} \chi_\cclass^\irrep, \label{eq: cc level sets sum}
\end{equation}
where the first sum is over a set $\{\cclass'_1, \dots, \cclass_k'\}$ of conjugacy classes such that
\begin{equation}
	\chi_{\cclass'_1}^S, \dots, \chi_{\cclass'_k}^S
\end{equation}
are all distinct and
\begin{equation}
	\langle (S,\cclass'_1) \rangle \cup 	\langle (S,\cclass'_2) \rangle  \cup \dots \cup 	\langle (S,\cclass'_k) \rangle = \cclasses{G}.
\end{equation}
Now we restrict to the partition functions with $b=1,\dots,k$. Equation \eqref{eq: cc level sets sum} is a matrix equation
\begin{equation}
	y = Vx,
\end{equation}
where
\begin{align}
	&y_b = Z^{\fusionalg{G}}_{h=1; S^b, \irrep} \\
	&x_{\cclass'} = \sum_{\cclass \in \langle (S,\cclass') \rangle} \chi_\cclass^\irrep \\
	&V_{b\cclass'} =  \qty(\chi_{\cclass'}^S)^b.
\end{align}
The matrix $V$ is an integer $k \times k$ Vandermonde matrix and its inverse $V^{-1}$ is a rational matrix. We have
\begin{equation}
	x = V^{-1}y.
\end{equation}
The l.h.s. is a vector of algebraic integers and the r.h.s. is a vector of rational numbers. Therefore
\begin{equation}
	x_{D} = \sum_{\cclass \in \langle (S,D) \rangle} \chi_\cclass^\irrep \in \mathbb{Z},
\end{equation}
for all $\irrep \in \irreps{G}$, which concludes the proof of Theorem \ref{thm: unrefined partial row sums}.

\begin{proposition}
A useful special case to consider is $S=R_0$, the trivial representation, and any $D \in \cclasses{G}$.
Then
\begin{equation}
	\langle (R_0,D) \rangle = \cclasses{G},
\end{equation}
and it follows that
\begin{equation}
	\sum_{\cclass \in \cclasses{G}} \chi_\cclass^\irrep \in \mathbb{Z}, \label{eq: proof row sum is integer}
\end{equation}
for any $\irrep \in \irreps{G}$.
\end{proposition}
In fact, the $\fusionalg{G}$-CTST readily gives the following positivity property.
\begin{proposition}\label{prop: col sum of characters is positive integer}
	The sum
	\begin{equation}
		\sum_{\cclass \in \cclasses{G}} \chi_\cclass^\irrep \in \mathbb{Z}^+, \label{eq: proof row sum is pos integer}
	\end{equation}
	for any $\irrep \in \irreps{G}$.
\end{proposition}
\noindent \textbf{Proof of Proposition \ref{prop: col sum of characters is positive integer}}
Consider the partition function the expressions for $Z^{\fusionalg{G}}_{h=1,\irrep}$ obtained by specialising 
\eqref{RGhrepspart} to 
\bea
	Z^{\fusionalg{G}}_{h=1,\irrep} && = \epsilon ( \Theta a_{\bar R} ) =  \sum_{\cclass \in \cclasses{G}} \chi_\cclass^\irrep.
	    \label{eq: pos integer char sum}
\eea
We also have 
\bea 
	Z^{\fusionalg{G}}_{h=1,\irrep}    = \sum_{S \in \irreps{G}} N^R_{S \bar S }  
\eea
using the identity \eqref{identNchi}. Since the fusion coefficients $N_{ S \bar S}^R$ are positive, the proposition follows.

\section{Row-Column Duality: refined integrality properties of rows/columns and generalised partitions } \label{sec:implications}
In this section we will extend the techniques used in the previous section to prove refinements of the character integrality results in \cite[Section 3.4]{GCTST2}. In particular, we relax the assumption that entire rows/columns are integer. As we review in section \ref{sec: Galois Theory}, these results can also be proven using Galois theory. Here, we give a combinatorial topological string theory derivation and interpretation of these results. We end this section by highlighting a new partition structure in the integer row and column sums that emerges from our considerations.

\subsection{Refined integral column sums from $G$-CTST}\label{subsec: integer level set sums}
In this subsection we will use the following lemma, which we prove in Appendix \ref{apx: inverse vandermonde lemma}.
\begin{lemma}
Let $Z$ be a diagonalizable $K \times K$ matrix of non-negative integers with eigenvalues $\lambda_1 \dots, \lambda_K$. Let the set of distinct eigenvalues be $\{z_1, \dots, z_k\}$. Note that $k \leq K$ generally.
Define the $k \times k$ Vandermonde matrix
\begin{equation}
	V_{b,i} = z_i^b.
\end{equation}
In general, the eigenvalues $z_1, \dots, z_k$ are algebraic integers.
But suppose $z_j$ is rational, then the inverse Vandermonde $V^{-1}$ has rational entries in the $j$th row:
\begin{equation}
	(V^{-1})_{j,b} \in \mathbb{Q} \quad \forall \, b \in \{1,\dots,k\}. \label{eq: inv vand lemma}
\end{equation}
We are interested in the special case of $z_j$ being an integer.
\end{lemma}
The first character integrality result, which we will now prove, is the following.
\begin{theorem} \label{thm: refined GCTST sum}
Let $D$ be a conjugacy class and $S$ an irreducible representation of $G$ such that
\bea
	\frac{\chi^S(T_D)}{d_S} = \abs{D}\frac{\chi^S_D}{d_S} \in \mathbb{Z}.
\eea
Define the subset of $\irreps{G}$
\begin{align}
	[(S, D)] &=  \qty{\irrep \in \irreps{G} \qq{s.t. } \frac{\chi^\irrep(T_D)}{d_\irrep} = \frac{\chi^S(T_D)}{d_S}}. \\
	&= \qty{\irrep \in \irreps{G} \qq{s.t. } \abs{D}\frac{\chi^\irrep_D}{d_\irrep} = \abs{D}\frac{\chi^S_D}{d_S}}.
\end{align}
Then the sum
\bea
\sum_{\irrep \in [(S,D)]}  \frac{\chi^\irrep(T_\cclass)}{d_\irrep} = \sum_{\irrep \in [(S,D)]} \abs{C}\frac{\chi^\irrep_\cclass}{d_\irrep} \in \mathbb{Z}, \quad \forall \cclass \in \cclasses{G}. \label{eq: level sets sum is integer}
\eea
is an integer for all conjugacy classes. Note that, compared to \eqref{eq: row sum thm 1},  we have dropped the assumption that $\chi^\irrep_D$ is an integer for all $\irrep \in \irreps{G}$.
\end{theorem}

\noindent \textbf{Proof of Theorem \ref{thm: refined GCTST sum}} (Constructive)
To prove the above result, consider the $l$-dimensional integer vector $y = (y_1, \dots, y_l)$ of amplitudes
\bea
y_b(D,C) = Z_{h=1,D^b,\cclass}^G = \sum_\irrep \qty(\frac{\chi^\irrep(T_D)}{d_\irrep})^{b} \frac{\chi^\irrep(T_{\cclass})}{d_{\irrep}}, \quad b=1,\dots,l.
\eea
where $l$ is a non-negative integer. We re-write the sum as a sum over level sets
\bea
y_b(D,C)  = \sum_{\irrep'} \qty(\frac{\chi^{\irrep'}(T_D)}{d_{\irrep'}})^{b} \sum_{\irrep \in [(\irrep',D)]}  \frac{\chi^\irrep(T_{C})}{d_{\irrep}}, \label{eq: level set amplitude}
\eea
where the first sum is over a set of irreducible representations $\irrep'_1, \dots, \irrep'_k$ such that
\begin{equation}
	\frac{\chi^{\irrep'_1}(T_D)}{d_{\irrep'_1}}, \dots, \frac{\chi^{\irrep'_k}(T_D)}{d_{\irrep'_k}},
\end{equation}
are all distinct and
\begin{equation}
	[(\irrep'_1, D)] \cup [(\irrep'_2, D)] \cup \dots \cup [(\irrep'_k, D)]= \irreps{G}.
\end{equation}
Now we restrict to the partition functions with $b=1,\dots,k$. Define the vector
\bea
x_{\irrep'}(D,\cclass) = \sum_{\irrep \in [(\irrep', D)]}  \frac{\chi^\irrep_{C}(T_C)}{d_{\irrep}},
\eea
and the $k \times k$ Vandermonde matrix
\bea
V_{b,R'}(D) = \qty(\abs{D}\frac{\chi^{\irrep'}_D}{d_{\irrep'}})^{b} = \qty(\frac{\chi^{\irrep'}(T_D)}{d_{\irrep'}})^{b}.
\eea
Note that
\begin{equation}
	T_{\cclass} P_{\irrep} =\frac{\chi^{\irrep}(T_\cclass)}{d_\irrep}P_{\irrep} =  \abs{C}\frac{\chi^{\irrep}_\cclass}{d_\irrep}P_{\irrep},
\end{equation}
and therefore normalized characters are the eigenvalues of the integer structure constant matrix $f_{\cclass}$, as required to use the inverse Vandermonde lemma.
Now equation \eqref{eq: level set amplitude} reads
\bea
y(D,C) = V(D)x(D,\cclass).
\eea
Since $V(D)$ is a Vandermonde matrix of distinct normalized characters it is invertible and
\bea
x(D,\cclass) = V^{-1}(D)y(D,\cclass).
\eea
In terms of the inverse Vandermonde, equation \eqref{eq: level sets sum is integer} reads,
\bea
x_{S}(D,C) = \sum_{b=1}^K (V^{-1})_{Sb}(D) y_{b}(D,C). \label{eq: vander monde used GCTST}
\eea
By lemma \eqref{eq: inv vand lemma}, $(V^{-1})_{Sb}(D)$ is rational for all $b=1,\dots,k$. Since $y_b$ is an integer vector, the r.h.s is rational. The l.h.s is a sum of algebraic integers. A sum of algebraic integers that is also rational is necessarily an integer. This concludes the constructive proof of Theorem \ref{thm: refined GCTST sum}. In section \ref{sec: Galois Theory} we give a Galois theoretic proof of the same proposition.

The sum in \eqref{eq: level sets sum is integer} was specified by a single pair $(S,D)$. We now prove a refinement of the above proposition.
\begin{theorem}\label{thm: twice refined GCTST sum}
	
Consider two pairs $(S_1,D_1), (S_2,D_2)$ with the property
\begin{equation}
	\frac{\chi^{S_1}(T_{D_1})}{d_{S_1}}, \quad \frac{\chi^{S_2}(T_{D_2})}{d_{S_2}} \in \mathbb{Z}.
\end{equation}
As we will prove, 
\begin{equation}
	\sum_{\irrep \in [(S_1,D_1)] \cap [(S_2,D_2)]} \frac{\chi^{\irrep}(T_{\cclass})}{d_{\irrep}} \in \mathbb{Z}, \quad \forall \cclass \in \cclasses{G},
\end{equation}
where the sum is over the set
\begin{equation}
	[(S_1,D_1)] \cap [(S_2,D_2)]	= \qty{\irrep \in \irreps{G} \qq{s.t. } \frac{\chi^{\irrep}(T_{D_1})}{d_\irrep} = \frac{\chi^{S_1}(T_{D_1})}{d_{S_1}} \qq{and } \frac{\chi^\irrep(T_{D_2})}{d_\irrep} = \frac{\chi^{S_2}(T_{D_2})}{d_{S_2}}}. 
\end{equation}

\end{theorem}
\noindent \textbf{Proof of Theorem \ref{thm: twice refined GCTST sum}} (Constructive)
To prove this, consider the matrix $y(D_1,D_2,C)$ of partition functions
\begin{equation}
	y_{b,b'}(D_1, D_2, \cclass) = Z_{h=1,D_1^b, D_2^{b'},\cclass}^G = \sum_{\irrep \in \irreps{G}} \qty(\frac{\chi^{\irrep}(T_{D_1})}{d_\irrep})^{b} \qty(\frac{\chi^\irrep(T_{D_2})}{d_\irrep})^{b'} \frac{\chi^\irrep(T_\cclass)}{d_\irrep}, \quad b,b' = 1,\dots, l.
\end{equation}
for some non-negative integer $l$. We partition the sum into a sum over level sets
\begin{equation}
	y_{b,b'}(D_1, D_2, \cclass) = \sum_{\irrep'_1} \qty(\frac{\chi^{\irrep'_1}(T_{D_1})}{d_{\irrep'_1}})^{b} \sum_{\irrep \in [(\irrep'_1, D_1)]} \qty(\frac{\chi^\irrep(T_{D_2})}{d_\irrep})^{b'}\frac{\chi^\irrep(T_\cclass)}{d_\irrep}.
\end{equation}
Now we restrict to the partition functions with $b=1,\dots,k$ where $k$ is the number of irreducible representations $\irrep$ with distinct $\frac{\chi^{\irrep}(T_{D_1})}{d_{\irrep}}$. The $k \times k$ matrix
\begin{equation}
	\qty(\frac{\chi^{\irrep'_1}(T_{D_1})}{d_{\irrep'_1}})^{b},
\end{equation}
is a Vandermonde matrix with integer row for $\irrep'_1 = S_1$, by assumption in the theorem. The inverse Vandermonde lemma argument in the proof of Theorem \ref{thm: refined GCTST sum} implies that
\begin{equation}
	\sum_{\irrep \in [(S_1, D_1)]} \qty(\frac{\chi^\irrep(T_{D_2})}{d_\irrep})^{b'} \frac{\chi^\irrep(T_\cclass)}{d_\irrep} \in \mathbb{Z}.
\end{equation}
We further partition this sum into level sets as
\begin{equation}
	\sum_{\irrep \in [(S_1, D_1)]} \qty(\frac{\chi^\irrep(T_{D_2})}{d_\irrep})^{b'} \frac{\chi^\irrep(T_\cclass)}{d_\irrep} = \sum_{\irrep'_2} \qty(\frac{\chi^{\irrep_2'}(T_{D_2})}{d_{\irrep_2'}})^{b'} \sum_{\irrep \in [(S_1, D_1)] \cap [(R_2', D_2)]}\frac{\chi^\irrep(T_\cclass)}{d_\irrep},
\end{equation}
where the sum over $\irrep'_2$ is a set of irreducible representations such that
\begin{equation}
	\frac{\chi^{\irrep_2'}(T_{D_2})}{d_{\irrep_2'}}
\end{equation}
are all distinct, and the union of $[(\irrep'_2, D_2)]$ is all of $[(S_1, D_1)]$. Now we restrict to the partition functions with $b'=1,\dots,k'$ where $k'$ is the number of irreducible representations $\irrep$ with distinct $\frac{\chi^{\irrep}(T_{D_2})}{d_{\irrep}}$.
Again, it follows by the Vandermonde argument that
\begin{equation}
	\sum_{\irrep \in [(S_1, D_1)] \cap [(S_2, D_2)]} \frac{\chi^\irrep(T_\cclass)}{d_\irrep} \in \mathbb{Z}, \quad \forall \cclass \in \cclasses{G}.
\end{equation}
It is straight-forward to generalize the argument to arbitrary numbers of pairs $(S,D)$ with integer normalized characters. This is proven using Galois theory in \eqref{eq: galois r-refined GCTST sum}.

Another generalization involves the insertion of handle creation operators to fix $d_\irrep$ in the sums.
This can be used to engineer integral sums over set
\begin{align}
	&\{\irrep \in \irreps{G} \qq{s.t.} \frac{\chi^\irrep(T_D)}{d_\irrep} =  \frac{\chi^S(T_D)}{d_S} \qq{and} d_\irrep = d_S\} =\\
	&\{\irrep \in \irreps{G} \qq{s.t.} \abs{D} \frac{\chi_D^\irrep}{d_\irrep} = \abs{D} \frac{\chi_D^S}{d_S} \qq{and} d_\irrep = d_S\} =\\
	&\{\irrep \in \irreps{G} \qq{ s.t.}  {\chi_D^\irrep} = {\chi_D^S},  \qq{and} d_\irrep = d_S\}.
\end{align}
giving the following proposition.
\begin{proposition}
	The following sum over the subset of irreducible representations defined above
	\begin{equation}
		\sum_{\irrep} \chi_\cclass^R
	\end{equation}
	is an integer for all $\cclass \in \cclasses{G}$.
\end{proposition}
As we will see in section \ref{sec: Galois Theory}, this is a special case of the Galois result \eqref{eq: galois multiple row sums} where the ordered list of group elements $H=(h_1,e)$ with $h_1 \in D$ and the ordered list of integers $N=(\chi^S_D, d_S)$.

\subsection{Refined integral row sums from $\fusionalg{G}$-CTST}\label{subsec: integer level set sums RG}
In this section we will use $\fusionalg{G}$-CTST to derive results dual to those above. We will prove the following integrality of row sums. 

\begin{theorem}\label{thm: refined RGCTST sums}
	
Let $(S,D) \in \irreps{G} \times \cclasses{G}$ such that
\begin{equation}
	\chi_D^S \in \mathbb{Z}.
\end{equation}
Define the following subset of $\cclasses{G}$
\begin{equation}
	\langle (S,D) \rangle = \qty{\cclass \in \cclasses{G} \qq{s.t. } \chi^S_\cclass = \chi^S_D}.
\end{equation}
Then the sums
\begin{equation}
	\sum_{\cclass \in \langle (S,D) \rangle} \chi_\cclass^\irrep \in \mathbb{Z}, \quad \forall \irrep \in \irreps{G}. \label{eq: integral column sums}
\end{equation}
are integers.
\end{theorem}

\noindent \textbf{Proof of Theorem \ref{thm: refined RGCTST sums}} (Constructive)
To prove this, we follow steps similar to those in section \ref{subsec: integer level set sums}. Consider the $K$-dimensional integer vector
\begin{equation}
	y_b(S,\irrep) = Z^{\fusionalg{G}}_{h=1,S^b,\irrep} = \sum_{\cclass} (\chi_\cclass^S)^b \chi_\cclass^\irrep , \quad b=1,\dots,l.
\end{equation}
where $l$ is a non-negative integer. We partition the sum into level sets
\begin{equation}
	y_b(S,\irrep) = \sum_{\cclass'} (\chi_{\cclass'}^S)^b \sum_{\cclass \in \langle (S,\cclass')\rangle}  \chi_\cclass^\irrep,
\end{equation}
where the first sum is over a set of conjugacy classes $\cclass'_1, \dots, \cclass'_k $ such that
\begin{equation}
	\chi_{\cclass'_1}^S, \dots, \chi_{\cclass'_k}^S,
\end{equation}
are all distinct and
\begin{equation}
	\langle (S, \cclass'_1) \rangle \cup \langle (S, \cclass'_2) \rangle \cup \dots \cup \langle (S, \cclass'_k) \rangle = \cclasses{G}.
\end{equation}
Now we restrict to the partition functions with $b=1,\dots,k$. Define the vector
\begin{equation}
	x_{\cclass'}(S,R) = \sum_{\cclass \in \langle (S,\cclass')\rangle}  \chi_\cclass^\irrep,
\end{equation}
and Vandermonde matrix
\begin{equation}
	V_{b,\cclass'}(S) =  (\chi_{\cclass'}^S)^b. 
\end{equation}
Then
\begin{equation}
	y(S,\irrep) = V(S)x(S,\irrep),
\end{equation}
and since $V(S)$ is a Vandermonde matrix, it is invertible and
\begin{equation}
	x(S,\irrep) = V^{-1}(S)y(S,\irrep).
\end{equation}
By the Vandermonde lemma $(V^{-1}(S))_{D b}$ is rational for all $b=1,\dots,K$. The l.h.s of equation \eqref{eq: integral column sums} reads
\begin{equation}
	x_D(S,\irrep) = \sum_{b=1}^K (V^{-1}(S))_{D b}y_b(S,\irrep). \label{eq: vandermonde used RGCTST}
\end{equation}
The l.h.s. is a sum of algebraic integers while the r.h.s. is a sum of rationals, and therefore they are integers. 

The sum can be further refined by fixing several pairs $(S_1, D_1), (S_2, D_2), \dots$. We omit the constructive proof of this because it mirrors the analogous proof given in the previous subsection. The Galois theoretic proof of this generalization, and the above special case is found in section \ref{sec: Galois Theory} (see \textbf{Proof of Theorem \ref{thm: refined RGCTST sums}$^*$} around equation \eqref{eq: galois proof RGCTST generalization}).

\subsection{Generalized partitions of the integer row sums} 
Consider the character table $\chi^\irrep_\cclass$ of a finite group $G$ and fix a row $S$. The sum
\begin{equation}
	\sum_{\cclass} \chi^S_\cclass = n_S,
\end{equation}
is an integer, as proved in equation \eqref{eq: proof row sum is integer}. We will now see that this integer has further combinatorial structure. We call a triplet $[p,q,r]$ where $p, q$ are integer partitions, and $r$ is a list of zeroes satisfying
\begin{equation}
	\sum_{i} p_i - \sum_{j} q_j + \sum_k r_k = n_S.
\end{equation}
a generalized partition of $n_S$ and write
\begin{equation}
	[p, q, r] \vDash n_S.
\end{equation}
The previous results on integrality of character sums can be used to give a generalized integer partition of $n_S$.

To see this, construct the table $\chi_\cclass^{\irrep\neq S}$, equal to the original character table but with row $S$ removed. From this table, construct $\chi_\cclass^{\irrep\neq S, \mathbb{Z}}$ by keeping only the rows that are entirely integer. The table $\chi_\cclass^{\irrep\neq S, \mathbb{Z}}$ defines an equivalence relation on conjugacy classes by
\begin{equation}
	\cclass \sim_S \cclass' \qq{if $\chi_\cclass^{\irrep\neq S, \mathbb{Z}} = \chi_{\cclass'}^{\irrep\neq S, \mathbb{Z}}$ for all rows.}
\end{equation}
The equivalence relation partitions the set of conjugacy classes into equivalence classes $\kappa_1, \dots, \kappa_l$. The equivalence classes $\kappa_i$ are level sets of the vector-valued function
\begin{equation}
	f^R(C) = \chi_{\cclass}^{\irrep\neq S, \mathbb{Z}},
\end{equation}
on conjugacy classes, defined by the integer rows of the character table with row $S$ removed. They define a set partition of the set of conjugacy classes such that
\begin{equation}
	n_S = \sum_{\cclass \in \cclasses{G}}  \chi^S_\cclass = \sum_{i=1}^l \sum_{\cclass \in \kappa_i} \chi^S_\cclass.
\end{equation}

Let ${\cclass}_1 \in \kappa_1, \dots, \cclass_l \in \kappa_l$ be a set of representatives. The equivalence classes $\kappa_i$ are equal to the sets
\begin{align}
	\kappa_i = \bigcap_{\substack{R \neq S \\ \text{R integer row}}} \langle (R, \cclass_i) {\rangle}.
\end{align}
Consequently, the sums
\begin{equation}
	\sum_{\cclass \in \kappa_i} \chi^S_\cclass,
\end{equation}
are integers for all $S \in \irreps{G}$.

We illustrate the procedure on the character table of $\mathbb{Z}_6$,
\be
\begin{array}{c|cccccc}
         &\cclass_1 & \cclass_2 & \cclass_3 & \cclass_4 & \cclass_5 & \cclass_6\\
\hline
	\irrep_1 &1 & 1 & 1 & 1 & 1 & 1 \\
		\irrep_2 & 1 & \zeta_3 & -\zeta_3 -1 & \zeta_3 +1 & -1 & -\zeta_3 \\
		\irrep_3 & 1 & -\zeta_3 -1 & \zeta_3 & \zeta_3 & 1 & -\zeta_3 -1 \\
		\irrep_4 & 1 & 1 & 1 & -1 & -1 & -1 \\
		\irrep_5 &1 & \zeta_3 & -\zeta_3 -1 & -\zeta_3 -1 & 1 & \zeta_3 \\
		\irrep_6 &1 &{-\zeta_3} -1 & \zeta_3 & -\zeta_3 & -1 & \zeta_3 +1.
\end{array}\ee
With $S=\irrep_1$ we have
\be
 \chi^{\irrep\neq \irrep_1}_\cclass = 
\begin{array}{c|cccccc}
&\cclass_1 & \cclass_2 & \cclass_3 & \cclass_4 & \cclass_5 & \cclass_6 \\
\hline 
		R_2 & 1 & \zeta_3 & -\zeta_3 -1 & \zeta_3 +1 & -1 & -\zeta_3 \\
		R_3 & 1 & -\zeta_3 -1 & \zeta_3 & \zeta_3 & 1 & -\zeta_3 -1 \\
		R_4 & 1 & 1 & 1 & -1 & -1 & -1 \\
		R_5 &1 & \zeta_3 & -\zeta_3 -1 & -\zeta_3 -1 & 1 & \zeta_3 \\
		R_6 &1 &{-\zeta_3} -1 & \zeta_3 & -\zeta_3 & -1 & \zeta_3 +1~
\end{array}
\ee
and
\be
\chi^{R\neq R_1, \mathbb{Z}}_\cclass = 
\begin{array}{c|cccccc}
&\cclass_1 & \cclass_2 & \cclass_3 & \cclass_4 & \cclass_5 & \cclass_6 \\ 
\hline
		R_4 & 1 & 1 & 1 & -1 & -1 & -1.
\end{array}
\ee
This leads to the following partition of the conjugacy classes
\begin{equation}
	\{\cclass_1, \cclass_2, \cclass_3, \cclass_4, \cclass_5, \cclass_6\} = \{\cclass_1, \cclass_2, \cclass_3\} \cup \{\cclass_4, \cclass_5, \cclass_6\} = \kappa_1 \cup \kappa_2.
\end{equation}
In particular, we have
\begin{equation}
	\sum_{\cclass} \chi^{R_1}_\cclass = 6 = \sum_{\cclass=\cclass_1, \cclass_2, \cclass_3} \chi^{R_1}_\cclass + \sum_{\cclass=\cclass_4, \cclass_5,\cclass_6} \chi^{R_1}_\cclass = 3+3.
\end{equation}
For $S=R_2$ we have
\be
\chi^{R\neq R_2}_\cclass = 
\begin{array}{c|cccccc}
& \cclass_1 & \cclass_2 & \cclass_3 & \cclass_4 & \cclass_5 & \cclass_6 \\
\hline
		R_1 &1 & 1 & 1 & 1 & 1 & 1 \\
		R_3 & 1 & -\zeta_3 -1 & \zeta_3 & \zeta_3 & 1 & -\zeta_3 -1 \\
		R_4 & 1 & 1 & 1 & -1 & -1 & -1 \\
		R_5 &1 & \zeta_3 & -\zeta_3 -1 & -\zeta_3 -1 & 1 & \zeta_3 \\
		R_6 &1 &{-\zeta_3} -1 & \zeta_3 & -\zeta_3 & -1 & \zeta_3 +1~
\end{array}
\ee
and
\be 
\chi^{R\neq R_2, \mathbb{Z}}_\cclass = 
\begin{array}{c|cccccc}
&\cclass_1 & \cclass_2 & \cclass_3 & \cclass_4 & \cclass_5 & \cclass_6 \\ \hline
		R_1 &1 & 1 & 1 & 1 & 1 & 1 \\
		R_4 & 1 & 1 & 1 & -1 & -1 & -1~.
\end{array}
\ee
This gives the same partition of the conjugacy classes
\begin{equation}
	\{\cclass_1, \cclass_2, \cclass_3, \cclass_4, \cclass_5, \cclass_6\} = \{\cclass_1, \cclass_2, \cclass_3\} \cup \{\cclass_4, \cclass_5, \cclass_6\} = \kappa_1 \cup \kappa_2
\end{equation}
and therefore
\begin{equation}
	\sum_{\cclass} \chi^{R_2}_\cclass = 0 = \sum_{\cclass=\cclass_1, \cclass_2, \cclass_3} \chi^{R_2}_\cclass + \sum_{\cclass=\cclass_4, \cclass_5, \cclass_6} \chi^{R_2}_\cclass = 0 + 0.
\end{equation}
Notably, the partition depends on the choice of $S$. For example, let $S=R_4$, then
\be
\chi^{R\neq R_4}_\cclass =
\begin{array}{c|cccccc}
 &\cclass_1 & \cclass_2 & \cclass_3 & \cclass_4 & \cclass_5 & \cclass_6 \\ \hline
		R_1 &1 & 1 & 1 & 1 & 1 & 1 \\
		R_2 & 1 & \zeta_3 & -\zeta_3 -1 & \zeta_3 +1 & -1 & -\zeta_3 \\
		R_3 & 1 & -\zeta_3 -1 & \zeta_3 & \zeta_3 & 1 & -\zeta_3 -1 \\
		R_5 &1 & \zeta_3 & -\zeta_3 -1 & -\zeta_3 -1 & 1 & \zeta_3 \\
		R_6 &1 &{-\zeta_3} -1 & \zeta_3 & -\zeta_3 & -1 & \zeta_3 +1
\end{array}
\ee
and
\be 
\chi^{R\neq R_4, \mathbb{Z}}_\cclass = 
\begin{array}{c|cccccc}
  &\cclass_1 & \cclass_2 & \cclass_3 & \cclass_4 & \cclass_5 & \cclass_6 \\ \hline
		R_1 &1 & 1 & 1 & 1 & 1 & 1
\end{array}
\ee
such that the induced partition is trivial
\begin{equation}
	\{\cclass_1, \cclass_2, \cclass_3, \cclass_4, \cclass_5, \cclass_6\} = \{\cclass_1, \cclass_2, \cclass_3, \cclass_4, \cclass_5, \cclass_6\} = \kappa_1.
\end{equation}
Therefore
\begin{equation}
	\sum_{\cclass} \chi^{R_4}_\cclass = 0 = \sum_{\cclass=\cclass_1, \cclass_2, \cclass_3, \cclass_4, \cclass_5,\cclass_6} \chi^{R_4}_\cclass = 0.
\end{equation}
In the language of generalized partitions, we have found that
\begin{equation}
	[[3,3],[],[]] \vDash n_{R_1}, \quad [[],[],[0,0]] \vDash n_{R_2}, \quad [[],[],[0]] \vDash n_{R_4}.
\end{equation}


We now give some examples of character table partitions for a selection of groups. For each row we give: the value of the characters on each subset $\kappa_i$ that partitions the columns, the corresponding generalized partition $[p,q,r]$ and lastly the integer row sum $n_S$. In the familiar example of $\mathbb{Z}_6$ we find
\begin{equation}
	\begin{array}{l|llllll}
		S&\kappa_1 & \kappa_2 & p & q & r &n_S \\ \hline
		R_1& \left[1, 1, 1\right]  &  \left[1, 1, 1\right]  &  \left[3, 3\right]  &  \left[\right]  &  \left[\right]  &  6  \\
		R_2& \left[1, \zeta_{3}, -\zeta_{3} - 1\right]  &  \left[\zeta_{3} + 1, -1, -\zeta_{3}\right]  &  \left[\right]  &  \left[\right]  &  \left[0, 0\right]  &  0  \\
		R_3& \left[1, -\zeta_{3} - 1, \zeta_{3}\right]  &  \left[\zeta_{3}, 1, -\zeta_{3} - 1\right]  &  \left[\right]  &  \left[\right]  &  \left[0, 0\right]  &  0  \\
		R_4& \left[1, -1, 1, -1, 1, -1\right]  & [] &  \left[\right]  &  \left[\right]  &  \left[0\right]  &  0  \\
		R_5& \left[1, \zeta_{3}, -\zeta_{3} - 1\right]  &  \left[-\zeta_{3} - 1, 1, \zeta_{3}\right]  &  \left[\right]  &  \left[\right]  &  \left[0, 0\right]  &  0  \\
		R_6& \left[1, -\zeta_{3} - 1, \zeta_{3}\right]  &  \left[-\zeta_{3}, -1, \zeta_{3} + 1\right]  &  \left[\right]  &  \left[\right]  &  \left[0, 0\right]  &  0 
	\end{array}
\end{equation}
For the Mathieu group M11 we find
\begin{equation}
	\resizebox{.9\linewidth}{!}{%
		$\begin{array}{l|llllllllllll}
			S& \kappa_1 & \kappa_2 & \kappa_3 & \kappa_4 & \kappa_5 & \kappa_6 & \kappa_7 & \kappa_8 & p & q & r & n_S \\ \hline
			R_1 &  \left[1\right]  &  \left[1\right] &  \left[1\right]  &  \left[1\right]  &  \left[1\right]  &  \left[1\right]  &  \left[1, 1\right]  &  \left[1, 1\right]  &  \left[1, 1, 1, 1, 1, 1, 2, 2\right]  &  \left[\right]  &  \left[\right]  &  10  \\
			R_2 &  \left[10\right]  &  \left[2\right]  &  \left[1\right]  &  \left[2\right]  &  \left[0\right]  &  \left[-1\right]  &  \left[0, 0\right]  &  \left[-1, -1\right]  &  \left[10, 2, 1, 2\right]  &  \left[-1, -2\right]  &  \left[0, 0\right]  &  12  \\
			R_3 &  \left[10\right]  &  \left[-2\right]  &  \left[1\right]  &  \left[0\right]  &  \left[0\right]  &  \left[1\right]  &  \left[\zeta_{8}^{3} + \zeta_{8}, -\zeta_{8}^{3} - \zeta_{8}\right]  &  \left[-1, -1\right]  &  \left[10, 1, 1\right]  &  \left[-2, -2\right]  &  \left[0, 0, 0\right]  &  8  \\
			R_4 &  \left[10\right]  &  \left[-2\right]  &  \left[1\right]  &  \left[0\right]  &  \left[0\right]  &  \left[1\right]  &  \left[-\zeta_{8}^{3} - \zeta_{8}, \zeta_{8}^{3} + \zeta_{8}\right]  &  \left[-1, -1\right]  &  \left[10, 1, 1\right]  &  \left[-2, -2\right]  &  \left[0, 0, 0\right]  &  8  \\
			R_5 &  \left[11\right]  &  \left[3\right]  &  \left[2\right]  &  \left[-1\right]  &  \left[1\right]  &  \left[0\right]  &  \left[-1, -1\right]  &  \left[0, 0\right]  &  \left[11, 3, 2, 1\right]  &  \left[-1, -2\right]  &  \left[0, 0\right]  &  14  \\
			R_6 &  \left[16\right]  &  \left[0\right]  &  \left[-2\right]  &  \left[0\right]  &  \left[1\right]  &  \left[0\right]  &  \left[0, 0\right]  &  \left[\zeta_{11}^{9} + \zeta_{11}^{5} + \zeta_{11}^{4} + \zeta_{11}^{3} + \zeta_{11}, -\zeta_{11}^{9} - \zeta_{11}^{5} - \zeta_{11}^{4} - \zeta_{11}^{3} - \zeta_{11} - 1\right]  &  \left[16, 1\right]  &  \left[-2, -1\right]  &  \left[0, 0, 0, 0\right]  &  14  \\
			R_7 &  \left[16\right]  &  \left[0\right]  &  \left[-2\right]  &  \left[0\right]  &  \left[1\right]  &  \left[0\right]  &  \left[0, 0\right]  &  \left[-\zeta_{11}^{9} - \zeta_{11}^{5} - \zeta_{11}^{4} - \zeta_{11}^{3} - \zeta_{11} - 1, \zeta_{11}^{9} + \zeta_{11}^{5} + \zeta_{11}^{4} + \zeta_{11}^{3} + \zeta_{11}\right]  &  \left[16, 1\right]  &  \left[-2, -1\right]  &  \left[0, 0, 0, 0\right]  &  14  \\
			R_8 &  \left[44\right]  &  \left[4\right]  &  \left[-1\right]  &  \left[0\right]  &  \left[-1\right]  &  \left[1\right]  &  \left[0, 0\right]  &  \left[0, 0\right]  &  \left[44, 4, 1\right]  &  \left[-1, -1\right]  &  \left[0, 0, 0\right]  &  47  \\
			R_9 &  \left[45\right]  &  \left[-3\right]  &  \left[0\right]  &  \left[1\right]  &  \left[0\right]  &  \left[0\right]  &  \left[-1, -1\right]  &  \left[1, 1\right]  &  \left[45, 1, 2\right]  &  \left[-3, -2\right]  &  \left[0, 0, 0\right]  &  43  \\
			R_{10} &  \left[55\right]  &  \left[-1\right]  &  \left[1\right]  &  \left[-1\right]  &  \left[0\right]  &  \left[-1\right]  &  \left[1, 1\right]  &  \left[0, 0\right]  &  \left[55, 1, 2\right]  &  \left[-1, -1, -1\right]  &  \left[0, 0\right]  &  55  \\
		\end{array}$}
\end{equation}
and for $\mathbb{Z}_7 \rtimes \mathbb{Z}_3$
\begin{equation}
	\resizebox{.9\linewidth}{!}{%
		$\begin{array}{l|llllllllllll}
			S&\kappa_1 &\kappa_2 &\kappa_3 &\kappa_4 &\kappa_5 &\kappa_6 &\kappa_7 &\kappa_8 &p & q & r &n_S\\ \hline
			R_{1} &  \left[1, 1\right]  &  \left[1\right]  &  \left[1, 1\right]  &  \left[1\right]  &  \left[1, 1\right]  &  \left[1\right]  &  \left[1, 1, 1, 1\right]  &  \left[1, 1\right]  &  \left[2, 1, 2, 1, 2, 1, 4, 2\right]  &  \left[\right]  &  \left[\right]  &  15  \\
			R_{2} &  \left[1, 1\right]  &  \left[-1\right]  &  \left[-1, -1\right]  &  \left[1\right]  &  \left[1, 1\right]  &  \left[1\right]  &  \left[-1, -1, -1, -1\right]  &  \left[1, 1\right]  &  \left[2, 1, 2, 1, 2\right]  &  \left[-1, -2, -4\right]  &  \left[\right]  &  1  \\
			R_{3} &  \left[1, 1\right]  &  \left[-1\right]  &  \left[1, 1\right]  &  \left[1\right]  &  \left[1, 1\right]  &  \left[-1\right]  &  \left[1, 1, 1, 1\right]  &  \left[1, 1\right]  &  \left[2, 2, 1, 2, 4, 2\right]  &  \left[-1, -1\right]  &  \left[\right]  &  11  \\
			R_{4} &  \left[1, 1\right]  &  \left[1\right]  &  \left[-1, -1\right]  &  \left[1\right]  &  \left[1, 1\right]  &  \left[-1\right]  &  \left[-1, -1, -1, -1\right]  &  \left[1, 1\right]  &  \left[2, 1, 1, 2, 2\right]  &  \left[-2, -1, -4\right]  &  \left[\right]  &  1  \\
			R_{5} &  \left[2, -2, 2\right]  &  \left[0\right]  &  \left[0, 0\right]  &  \left[2, -2, 2, -2\right]  &  \left[0\right]  &  \left[0, 0, 0, 0\right]  & [] & [] & \left[2\right]  &  \left[\right]  &  \left[0, 0, 0, 0, 0\right]  &  2  \\
			R_{6} &  \left[2, 2\right]  &  \left[0\right]  &  \left[-2, -2\right]  &  \left[2\right]  &  \left[-1, -1\right]  &  \left[0\right]  &  \left[1, 1, 1, 1\right]  &  \left[-1, -1\right]  &  \left[4, 2, 4\right]  &  \left[-4, -2, -2\right]  &  \left[0, 0\right]  &  2  \\
			R_{7} &  \left[2, 2\right]  &  \left[0\right]  &  \left[2, 2\right]  &  \left[2\right]  &  \left[-1, -1\right]  &  \left[0\right]  &  \left[-1, -1, -1, -1\right]  &  \left[-1, -1\right]  &  \left[4, 4, 2\right]  &  \left[-2, -4, -2\right]  &  \left[0, 0\right]  &  2  \\
			R_{8} &  \left[2, -2\right]  &  \left[0\right]  &  \left[-\zeta_{8}^{3} - \zeta_{8}, \zeta_{8}^{3} + \zeta_{8}\right]  &  \left[0\right]  &  \left[2, -2\right]  &  \left[0\right]  &  \left[-\zeta_{8}^{3} - \zeta_{8}, -\zeta_{8}^{3} - \zeta_{8}, \zeta_{8}^{3} + \zeta_{8}, \zeta_{8}^{3} + \zeta_{8}\right]  &  \left[0, 0\right]  &  \left[\right]  &  \left[\right]  &  \left[0, 0, 0, 0, 0, 0, 0, 0\right]  &  0  \\
			R_{9} &  \left[2, -2\right]  &  \left[0\right]  &  \left[\zeta_{8}^{3} + \zeta_{8}, -\zeta_{8}^{3} - \zeta_{8}\right]  &  \left[0\right]  &  \left[2, -2\right]  &  \left[0\right]  &  \left[\zeta_{8}^{3} + \zeta_{8}, \zeta_{8}^{3} + \zeta_{8}, -\zeta_{8}^{3} - \zeta_{8}, -\zeta_{8}^{3} - \zeta_{8}\right]  &  \left[0, 0\right]  &  \left[\right]  &  \left[\right]  &  \left[0, 0, 0, 0, 0, 0, 0, 0\right]  &  0  \\
			R_{10} &  \left[2, 2\right]  &  \left[0\right]  &  \left[0, 0\right]  &  \left[-2\right]  &  \left[-1, -1\right]  &  \left[0\right]  &  \left[-\zeta_{12}^{3} + 2 \zeta_{12}, \zeta_{12}^{3} - 2 \zeta_{12}, -\zeta_{12}^{3} + 2 \zeta_{12}, \zeta_{12}^{3} - 2 \zeta_{12}\right]  &  \left[1, 1\right]  &  \left[4, 2\right]  &  \left[-2, -2\right]  &  \left[0, 0, 0, 0\right]  &  2  \\
			R_{11} &  \left[2, 2\right]  &  \left[0\right]  &  \left[0, 0\right]  &  \left[-2\right]  &  \left[-1, -1\right]  &  \left[0\right]  &  \left[\zeta_{12}^{3} - 2 \zeta_{12}, -\zeta_{12}^{3} + 2 \zeta_{12}, \zeta_{12}^{3} - 2 \zeta_{12}, -\zeta_{12}^{3} + 2 \zeta_{12}\right]  &  \left[1, 1\right]  &  \left[4, 2\right]  &  \left[-2, -2\right]  &  \left[0, 0, 0, 0\right]  &  2  \\
			R_{12} &  \left[2, -2\right]  &  \left[0\right]  &  \left[-\zeta_{8}^{3} - \zeta_{8}, \zeta_{8}^{3} + \zeta_{8}\right]  &  \left[0\right]  &  \left[-1, 1\right]  &  \left[0\right]  &  \left[-\zeta_{24}^{7} + \zeta_{24}^{3} - \zeta_{24}, \zeta_{24}^{7} + \zeta_{24}^{5}, \zeta_{24}^{7} - \zeta_{24}^{3} + \zeta_{24}, -\zeta_{24}^{7} - \zeta_{24}^{5}\right]  &  \left[\zeta_{12}^{3} - 2 \zeta_{12}, -\zeta_{12}^{3} + 2 \zeta_{12}\right]  &  \left[\right]  &  \left[\right]  &  \left[0, 0, 0, 0, 0, 0, 0, 0\right]  &  0  \\
			R_{13} &  \left[2, -2\right]  &  \left[0\right]  &  \left[-\zeta_{8}^{3} - \zeta_{8}, \zeta_{8}^{3} + \zeta_{8}\right]  &  \left[0\right]  &  \left[-1, 1\right]  &  \left[0\right]  &  \left[\zeta_{24}^{7} + \zeta_{24}^{5}, -\zeta_{24}^{7} + \zeta_{24}^{3} - \zeta_{24}, -\zeta_{24}^{7} - \zeta_{24}^{5}, \zeta_{24}^{7} - \zeta_{24}^{3} + \zeta_{24}\right]  &  \left[-\zeta_{12}^{3} + 2 \zeta_{12}, \zeta_{12}^{3} - 2 \zeta_{12}\right]  &  \left[\right]  &  \left[\right]  &  \left[0, 0, 0, 0, 0, 0, 0, 0\right]  &  0  \\
			R_{14} &  \left[2, -2\right]  &  \left[0\right]  &  \left[\zeta_{8}^{3} + \zeta_{8}, -\zeta_{8}^{3} - \zeta_{8}\right]  &  \left[0\right]  &  \left[-1, 1\right]  &  \left[0\right]  &  \left[-\zeta_{24}^{7} - \zeta_{24}^{5}, \zeta_{24}^{7} - \zeta_{24}^{3} + \zeta_{24}, \zeta_{24}^{7} + \zeta_{24}^{5}, -\zeta_{24}^{7} + \zeta_{24}^{3} - \zeta_{24}\right]  &  \left[-\zeta_{12}^{3} + 2 \zeta_{12}, \zeta_{12}^{3} - 2 \zeta_{12}\right]  &  \left[\right]  &  \left[\right]  &  \left[0, 0, 0, 0, 0, 0, 0, 0\right]  &  0  \\
			R_{15} &  \left[2, -2\right]  &  \left[0\right]  &  \left[\zeta_{8}^{3} + \zeta_{8}, -\zeta_{8}^{3} - \zeta_{8}\right]  &  \left[0\right]  &  \left[-1, 1\right]  &  \left[0\right]  &  \left[\zeta_{24}^{7} - \zeta_{24}^{3} + \zeta_{24}, -\zeta_{24}^{7} - \zeta_{24}^{5}, -\zeta_{24}^{7} + \zeta_{24}^{3} - \zeta_{24}, \zeta_{24}^{7} + \zeta_{24}^{5}\right]  &  \left[\zeta_{12}^{3} - 2 \zeta_{12}, -\zeta_{12}^{3} + 2 \zeta_{12}\right]  &  \left[\right]  &  \left[\right]  &  \left[0, 0, 0, 0, 0, 0, 0, 0\right]  &  0  \\
		\end{array}$}
\end{equation}

\subsection{Generalized partitions of the integer column sums} 
Similar considerations are relevant for column sums. For a fixed conjugacy class $D$, the sum
\begin{equation}
	\sum_{\irrep \in \irreps{G}} \chi^{{\irrep}}_D = n_{D}
\end{equation}
is an integer, as we proved in Proposition \ref{prop: sum over all irreps is integer}. We use a completely analogous set of rules to partition the columns. First construct the table $\chi^{\irrep}_{\cclass \neq D}$, corresponding to the character table without the column $D$. Further, restrict to only integer columns, yielding the table $\chi^{\irrep}_{\cclass \neq D, \mathbb{Z}}$. The table defines an equivalence relation on rows
\begin{equation}
	\irrep \sim_{D} \irrep' \qq{if $\chi^{\irrep}_{\cclass \neq D, \mathbb{Z}} = \chi^{\irrep'}_{\cclass \neq D, \mathbb{Z}}$ for all columns,}
\end{equation}
that partitions column $D$ into subsets $\rho_1, \dots, \rho_l$ such that
\begin{equation}
	n_D = \sum_{\irrep \in \irreps{G}} \chi^{\irrep}_D = \sum_{i=1}^l \sum_{\irrep \in \rho_i} \chi_D^\irrep.
\end{equation}
For $D$ equal to the conjugacy class of the identity, the sums
\begin{equation}
	\sum_{\irrep \in \rho_i} \chi^\irrep_D = \sum_{\irrep \in \rho_i} d_\irrep,
\end{equation}
are manifestly integers. For $D$ different from the identity element, we can prove that they are integers as follows.
Let $\irrep_1 \in \rho_1, \dots, \irrep_l \in \rho_l$ be a set of representatives. The equivalence classes are sets
\begin{equation}
	\rho_i = \{\irrep \in \irreps{G} \qq{s.t. $\chi^{R}_{\cclass} = \chi^{R_i}_{\cclass}$ for all integer integer columns $\cclass \neq D$}\}.
\end{equation}
In particular, the column of the identity conjugacy class is always integer and therefore included in the above set. This forces $d_\irrep = d_{\irrep_i}$. Consequently, the equivalence classes can be understood as intersections of subsets
\begin{equation}
	\rho_i = \qty[\bigcap_{\substack{\cclass \neq D \\ \text{$\cclass$ integer column}}} [(\irrep_i, \cclass)]] \cap [d_{\irrep_i}].
\end{equation}
The intersection with $[d_{\irrep_i}]$ is necessary to force $d_{\irrep} = d_{\irrep_i}$.
It follows that
\begin{equation}
	\sum_{\irrep \in \rho_i} \chi_D^\irrep \in \mathbb{Z}.
\end{equation}

We give examples of generalized partitions of columns for the same groups as above. Note that the rows now label columns in the character tables.
For $\mathbb{Z}_6$ we have
\begin{equation}	
	\begin{array}{l|llllll}
		D& \rho_1 & \rho_2 & p & q & r & n_D \\ \hline
		C_{1} &  \left[1, 1, 1\right]  &  \left[1, 1, 1\right]  &  \left[3, 3\right]  &  \left[\right]  &  \left[\right]  &  6  \\
		C_{2} &  \left[1, \zeta_{3}, -\zeta_{3} - 1\right]  &  \left[\zeta_{3} + 1, -1, -\zeta_{3}\right]  &  \left[\right]  &  \left[\right]  &  \left[0, 0\right]  &  0  \\
		C_{3} &  \left[1, -\zeta_{3} - 1, \zeta_{3}\right]  &  \left[\zeta_{3}, 1, -\zeta_{3} - 1\right]  &  \left[\right]  &  \left[\right]  &  \left[0, 0\right]  &  0  \\
		C_{4} &  \left[1, -1, 1, -1, 1, -1\right]  & [] & \left[\right]  &  \left[\right]  &  \left[0\right]  &  0  \\
		C_{5} &  \left[1, \zeta_{3}, -\zeta_{3} - 1\right]  &  \left[-\zeta_{3} - 1, 1, \zeta_{3}\right]  &  \left[\right]  &  \left[\right]  &  \left[0, 0\right]  &  0  \\
		C_{6} &  \left[1, -\zeta_{3} - 1, \zeta_{3}\right]  &  \left[-\zeta_{3}, -1, \zeta_{3} + 1\right]  &  \left[\right]  &  \left[\right]  &  \left[0, 0\right]  &  0  \\
	\end{array}
\end{equation}
for the Mathieu group M11 we have
\begin{equation}
	\resizebox{.9\linewidth}{!}{%
		$\begin{array}{l|llllllllllll} 
			D& \rho_1 & \rho_2 & \rho_3 & \rho_4 & \rho_5 & \rho_6 & \rho_ 7& \rho_8 & p & q & r & n_D \\ \hline
			C_{1} &  \left[1\right]  &  \left[10\right]  &  \left[10, 10\right]  &  \left[11\right]  &  \left[16, 16\right]  &  \left[44\right]  &  \left[45\right]  &  \left[55\right]  &  \left[1, 10, 20, 11, 32, 44, 45, 55\right]  &  \left[\right]  &  \left[\right]  &  218  \\
			C_{2} &  \left[1\right]  &  \left[2\right]  &  \left[-2, -2\right]  &  \left[3\right]  &  \left[0, 0\right]  &  \left[4\right]  &  \left[-3\right]  &  \left[-1\right]  &  \left[1, 2, 3, 4\right]  &  \left[-4, -3, -1\right]  &  \left[0\right]  &  2  \\
			C_{3} &  \left[1\right]  &  \left[1\right]  &  \left[1, 1\right]  &  \left[2\right]  &  \left[-2, -2\right]  &  \left[-1\right]  &  \left[0\right]  &  \left[1\right]  &  \left[1, 1, 2, 2, 1\right]  &  \left[-4, -1\right]  &  \left[0\right]  &  2  \\
			C_{4} &  \left[1\right]  &  \left[2\right]  &  \left[0, 0\right]  &  \left[-1\right]  &  \left[0, 0\right]  &  \left[0\right]  &  \left[1\right]  &  \left[-1\right]  &  \left[1, 2, 1\right]  &  \left[-1, -1\right]  &  \left[0, 0, 0\right]  &  2  \\
			C_{5} &  \left[1\right]  &  \left[0\right]  &  \left[0, 0\right]  &  \left[1\right]  &  \left[1, 1\right]  &  \left[-1\right]  &  \left[0\right]  &  \left[0\right]  &  \left[1, 1, 2\right]  &  \left[-1\right]  &  \left[0, 0, 0, 0\right]  &  3  \\
			C_{6} &  \left[1\right]  &  \left[-1\right]  &  \left[1, 1\right]  &  \left[0\right]  &  \left[0, 0\right]  &  \left[1\right]  &  \left[0\right]  &  \left[-1\right]  &  \left[1, 2, 1\right]  &  \left[-1, -1\right]  &  \left[0, 0, 0\right]  &  2  \\
			C_{7} &  \left[1\right]  &  \left[0\right]  &  \left[\zeta_{8}^{3} + \zeta_{8}, -\zeta_{8}^{3} - \zeta_{8}\right]  &  \left[-1\right]  &  \left[0, 0\right]  &  \left[0\right]  &  \left[-1\right]  &  \left[1\right]  &  \left[1, 1\right]  &  \left[-1, -1\right]  &  \left[0, 0, 0, 0\right]  &  0  \\
			C_{8} &  \left[1\right]  &  \left[0\right]  &  \left[-\zeta_{8}^{3} - \zeta_{8}, \zeta_{8}^{3} + \zeta_{8}\right]  &  \left[-1\right]  &  \left[0, 0\right]  &  \left[0\right]  &  \left[-1\right]  &  \left[1\right]  &  \left[1, 1\right]  &  \left[-1, -1\right]  &  \left[0, 0, 0, 0\right]  &  0  \\
			C_{9} &  \left[1\right]  &  \left[-1\right]  &  \left[-1, -1\right]  &  \left[0\right]  &  \left[\zeta_{11}^{9} + \zeta_{11}^{5} + \zeta_{11}^{4} + \zeta_{11}^{3} + \zeta_{11}, -\zeta_{11}^{9} - \zeta_{11}^{5} - \zeta_{11}^{4} - \zeta_{11}^{3} - \zeta_{11} - 1\right]  &  \left[0\right]  &  \left[1\right]  &  \left[0\right]  &  \left[1, 1\right]  &  \left[-1, -2, -1\right]  &  \left[0, 0, 0\right]  &  -2  \\
			C_{10} &  \left[1\right]  &  \left[-1\right]  &  \left[-1, -1\right]  &  \left[0\right]  &  \left[-\zeta_{11}^{9} - \zeta_{11}^{5} - \zeta_{11}^{4} - \zeta_{11}^{3} - \zeta_{11} - 1, \zeta_{11}^{9} + \zeta_{11}^{5} + \zeta_{11}^{4} + \zeta_{11}^{3} + \zeta_{11}\right]  &  \left[0\right]  &  \left[1\right]  &  \left[0\right]  &  \left[1, 1\right]  &  \left[-1, -2, -1\right]  &  \left[0, 0, 0\right]  &  -2  \\
		\end{array}$}
\end{equation}
and for $\mathbb{Z}_7 \rtimes \mathbb{Z}_3$
\begin{equation}
	\resizebox{0.9\linewidth}{!}{%
		$\begin{array}{l|lllllllllllll}
			D& \rho_1 & \rho_2 & \rho_3 & \rho_4 & \rho_5 & \rho_6 & \rho_7 & \rho_8 & \rho_9 & p & q & r & n_D\\ \hline
			C_{1} &  \left[1\right]  &  \left[1\right]  &  \left[1\right]  &  \left[1\right]  &  \left[2\right]  &  \left[2, 2\right]  &  \left[2, 2\right]  &  \left[2, 2\right]  &  \left[2, 2, 2, 2\right]  &  \left[1, 1, 1, 1, 2, 4, 4, 4, 8\right]  &  \left[\right]  &  \left[\right]  &  26  \\
			C_{2} &  \left[1, -1\right]  &  \left[-1, 1\right]  &  \left[0\right]  &  \left[0, 0\right]  &  \left[0, 0\right]  &  \left[0, 0\right]  & []& []& \left[0, 0, 0, 0\right]  &  \left[\right]  &  \left[\right]  &  \left[0, 0, 0, 0, 0, 0, 0\right]  &  0  \\
			C_{3} &  \left[1\right]  &  \left[-1\right]  &  \left[1\right]  &  \left[-1\right]  &  \left[0\right]  &  \left[-2, 2\right]  &  \left[-\zeta_{8}^{3} - \zeta_{8}, \zeta_{8}^{3} + \zeta_{8}\right]  &  \left[0, 0\right]  &  \left[-\zeta_{8}^{3} - \zeta_{8}, -\zeta_{8}^{3} - \zeta_{8}, \zeta_{8}^{3} + \zeta_{8}, \zeta_{8}^{3} + \zeta_{8}\right]  &  \left[1, 1\right]  &  \left[-1, -1\right]  &  \left[0, 0, 0, 0, 0\right]  &  0  \\
			C_{4} &  \left[1\right]  &  \left[1\right]  &  \left[1\right]  &  \left[1\right]  &  \left[-2\right]  &  \left[2, 2, -2, -2\right]  &  \left[0, 0\right]  &  \left[0, 0, 0, 0\right]  & []& \left[1, 1, 1, 1\right]  &  \left[-2\right]  &  \left[0, 0, 0\right]  &  2  \\
			C_{5} &  \left[1\right]  &  \left[1\right]  &  \left[1\right]  &  \left[1\right]  &  \left[2\right]  &  \left[2, 2\right]  &  \left[-2, -2\right]  &  \left[2, 2\right]  &  \left[-2, -2, -2, -2\right]  &  \left[1, 1, 1, 1, 2, 4, 4\right]  &  \left[-4, -8\right]  &  \left[\right]  &  2  \\
			C_{6} &  \left[1\right]  &  \left[1\right]  &  \left[1\right]  &  \left[1\right]  &  \left[2\right]  &  \left[-1, -1\right]  &  \left[2, 2\right]  &  \left[-1, -1\right]  &  \left[-1, -1, -1, -1\right]  &  \left[1, 1, 1, 1, 2, 4\right]  &  \left[-2, -2, -4\right]  &  \left[\right]  &  2  \\
			C_{7} &  \left[1, -1\right]  &  \left[1, -1\right]  &  \left[0\right]  &  \left[0, 0\right]  &  \left[0, 0\right]  &  \left[0, 0\right]  &  \left[0, 0, 0, 0\right]  & []& \left[\right]  &  \left[\right]  &  \left[0, 0, 0, 0, 0, 0, 0\right]  &  0  \\
			C_{8} &  \left[1\right]  &  \left[-1\right]  &  \left[1\right]  &  \left[-1\right]  &  \left[0\right]  &  \left[-2, 2\right]  &  \left[\zeta_{8}^{3} + \zeta_{8}, -\zeta_{8}^{3} - \zeta_{8}\right]  &  \left[0, 0\right]  &  \left[\zeta_{8}^{3} + \zeta_{8}, \zeta_{8}^{3} + \zeta_{8}, -\zeta_{8}^{3} - \zeta_{8}, -\zeta_{8}^{3} - \zeta_{8}\right]  &  \left[1, 1\right]  &  \left[-1, -1\right]  &  \left[0, 0, 0, 0, 0\right]  &  0  \\
			C_{9} &  \left[1\right]  &  \left[-1\right]  &  \left[1\right]  &  \left[-1\right]  &  \left[0\right]  &  \left[1, -1\right]  &  \left[-\zeta_{8}^{3} - \zeta_{8}, \zeta_{8}^{3} + \zeta_{8}\right]  &  \left[-\zeta_{12}^{3} + 2 \zeta_{12}, \zeta_{12}^{3} - 2 \zeta_{12}\right]  &  \left[-\zeta_{24}^{7} + \zeta_{24}^{3} - \zeta_{24}, \zeta_{24}^{7} + \zeta_{24}^{5}, -\zeta_{24}^{7} - \zeta_{24}^{5}, \zeta_{24}^{7} - \zeta_{24}^{3} + \zeta_{24}\right]  &  \left[1, 1\right]  &  \left[-1, -1\right]  &  \left[0, 0, 0, 0, 0\right]  &  0  \\
			C_{10} &  \left[1\right]  &  \left[1\right]  &  \left[1\right]  &  \left[1\right]  &  \left[-2\right]  &  \left[-1, -1\right]  &  \left[0, 0\right]  &  \left[1, 1\right]  &  \left[\zeta_{12}^{3} - 2 \zeta_{12}, -\zeta_{12}^{3} + 2 \zeta_{12}, -\zeta_{12}^{3} + 2 \zeta_{12}, \zeta_{12}^{3} - 2 \zeta_{12}\right]  &  \left[1, 1, 1, 1, 2\right]  &  \left[-2, -2\right]  &  \left[0, 0\right]  &  2  \\
			C_{11} &  \left[1\right]  &  \left[1\right]  &  \left[1\right]  &  \left[1\right]  &  \left[2\right]  &  \left[-1, -1\right]  &  \left[-2, -2\right]  &  \left[-1, -1\right]  &  \left[1, 1, 1, 1\right]  &  \left[1, 1, 1, 1, 2, 4\right]  &  \left[-2, -4, -2\right]  &  \left[\right]  &  2  \\
			C_{12} &  \left[1\right]  &  \left[-1\right]  &  \left[1\right]  &  \left[-1\right]  &  \left[0\right]  &  \left[1, -1\right]  &  \left[-\zeta_{8}^{3} - \zeta_{8}, \zeta_{8}^{3} + \zeta_{8}\right]  &  \left[\zeta_{12}^{3} - 2 \zeta_{12}, -\zeta_{12}^{3} + 2 \zeta_{12}\right]  &  \left[\zeta_{24}^{7} + \zeta_{24}^{5}, -\zeta_{24}^{7} + \zeta_{24}^{3} - \zeta_{24}, \zeta_{24}^{7} - \zeta_{24}^{3} + \zeta_{24}, -\zeta_{24}^{7} - \zeta_{24}^{5}\right]  &  \left[1, 1\right]  &  \left[-1, -1\right]  &  \left[0, 0, 0, 0, 0\right]  &  0  \\
			C_{13} &  \left[1\right]  &  \left[-1\right]  &  \left[1\right]  &  \left[-1\right]  &  \left[0\right]  &  \left[1, -1\right]  &  \left[\zeta_{8}^{3} + \zeta_{8}, -\zeta_{8}^{3} - \zeta_{8}\right]  &  \left[-\zeta_{12}^{3} + 2 \zeta_{12}, \zeta_{12}^{3} - 2 \zeta_{12}\right]  &  \left[\zeta_{24}^{7} - \zeta_{24}^{3} + \zeta_{24}, -\zeta_{24}^{7} - \zeta_{24}^{5}, \zeta_{24}^{7} + \zeta_{24}^{5}, -\zeta_{24}^{7} + \zeta_{24}^{3} - \zeta_{24}\right]  &  \left[1, 1\right]  &  \left[-1, -1\right]  &  \left[0, 0, 0, 0, 0\right]  &  0  \\
			C_{14} &  \left[1\right]  &  \left[1\right]  &  \left[1\right]  &  \left[1\right]  &  \left[-2\right]  &  \left[-1, -1\right]  &  \left[0, 0\right]  &  \left[1, 1\right]  &  \left[-\zeta_{12}^{3} + 2 \zeta_{12}, \zeta_{12}^{3} - 2 \zeta_{12}, \zeta_{12}^{3} - 2 \zeta_{12}, -\zeta_{12}^{3} + 2 \zeta_{12}\right]  &  \left[1, 1, 1, 1, 2\right]  &  \left[-2, -2\right]  &  \left[0, 0\right]  &  2  \\
			C_{15} &  \left[1\right]  &  \left[-1\right]  &  \left[1\right]  &  \left[-1\right]  &  \left[0\right]  &  \left[1, -1\right]  &  \left[\zeta_{8}^{3} + \zeta_{8}, -\zeta_{8}^{3} - \zeta_{8}\right]  &  \left[\zeta_{12}^{3} - 2 \zeta_{12}, -\zeta_{12}^{3} + 2 \zeta_{12}\right]  &  \left[-\zeta_{24}^{7} - \zeta_{24}^{5}, \zeta_{24}^{7} - \zeta_{24}^{3} + \zeta_{24}, -\zeta_{24}^{7} + \zeta_{24}^{3} - \zeta_{24}, \zeta_{24}^{7} + \zeta_{24}^{5}\right]  &  \left[1, 1\right]  &  \left[-1, -1\right]  &  \left[0, 0, 0, 0, 0\right]  &  0 \\
		\end{array}$}
\end{equation}

\section{Galois theory for character tables}\label{sec: Galois Theory}
In this section we will review Galois theory of field extensions and its particular application to character theory. By focusing on cyclotomic fields, which are the field extensions of $\mathbb{Q}$ relevant to character theory, we are able to give a concrete exposition of their Galois theory. In particular, this includes explicit descriptions of their Galois groups, associated actions as automorphisms of the fields, and the fundamental theorem of Galois Theory for cyclotomic fields. The Galois action on cyclotomic fields gives rise to Galois actions on characters of $G$ and we use this to study integrality properties of characters.

Consider a finite group $G$ and a representation $R$ of $G$. From Lemma 2.15 in \cite{isaacs1994character}, we know that the matrix $R(g)$ for some element $g \in G$ is similar to a diagonal matrix with eigenvalues $\epsilon_i$. If $N_g$ is the order of $g$ such that $g^{N_g} = 1$ it follows that
\begin{equation}
	R(g^{N_g}) = R(g)^{N_g} = 1,
\end{equation}
and the eigenvalues satisfy $\epsilon_i^{N_g}=1$. Therefore, $\epsilon_i$ are $N_{g}^{\text{th}}$ roots of unity. Because the character of $R(g)$ is given by 
\be
\chi^\irrep(g)= \sum_i \epsilon_i,
\ee
it lies in the cyclotomic field $\mathds{Q}(e^{\frac{2 \pi i}{N_g}})$. This is a number field containing elements of the form
\be 
a + b ~ e^{\frac{2 \pi i}{N_g}}
\ee
where $a,b \in \mathds{Q}$, and their combinations using the field operations $+,\times$. Since roots of unity are algebraic integers, the character of a representation evaluated on $g \in G$ is an algebraic integer in $\mathds{Q}(e^{\frac{2 \pi i}{N_g}})$.

In general, let $E$ be the exponent of the group $G$, that is, the smallest integer $E$ such that $g^E=1$ for all $g \in G$. Then the above argument shows that the characters of all representations of $G$ lie in the cyclotomic field $\mathds{Q}(e^{\frac{2 \pi i}{E}})$. In fact, from Brauer's theorem (Theorem 10.3 in \cite{isaacs1994character}) it is known that any representation $R$ of a group $G$ can be realized in the field  $\mathds{Q}(e^{\frac{2 \pi i}{E}})$. In other words, it is always possible to choose a basis such that the matrix elements of $R(g)$ for any $g \in G$ is a linear combination of $E^{\text{th}}$ roots of unity with rational coefficients.

In the next subsection, we will go through properties of cyclotomic fields and associated Galois actions. Then we will show that elements invariant under the Galois group are necessarily rational. Following this discussion, we will study Galois action on group representations and characters. This will lead to interesting integrality properties of characters. 

\subsection{Cyclotomic fields and Galois actions}

We found that in the study of group representations and their characters, the cyclotomic field $\mathds{Q}(e^{\frac{2\pi i}{E}})$ plays a crucial role\footnote{For a discussion on general field extensions and Galois actions, see Chapters 13, 14 of \cite{dummit1991abstract}.}. As we will now explain, $\mathds{Q}(e^{\frac{2\pi i}{E}})$ can be understood as a vector space. An element $\xi \in \mathds{Q}(e^{\frac{2\pi i}{E}})$ is called a primitive $E^{\text{th}}$ root of unity if
\begin{equation}
	\xi^E = 1 \qq{and } \xi^k \neq 1 \qq{for } k=1,\dots,E-1. 
\end{equation}
$\mathds{Q}(e^{\frac{2\pi i}{E}})$ can be understood as a vector space over $\mathds{Q}$ of dimension $\phi(E)$, where $\phi(E)$ is the Euler's totient function (see Section 13.6 Corollary 42 in \cite{dummit1991abstract}). In general, the set of elements $\{1,e^{\frac{2\pi i}{E}},e^{\frac{2\pi i2}{E}},\dots, e^{\frac{2\pi i (\phi(E)-1)}{E}}\}$ form a basis of $\mathds{Q}(e^{\frac{2\pi i}{E}})$ (for example, see \cite{bosma1990canonical}). Therefore, a general element of this field is of the form
\be
a=\sum_{k=0}^{\phi(E)-1} a_k \xi_k
\ee
where $a_k$ are rational and $\xi_k=e^{\frac{2\pi ik}{E}}$.

%

We now define the Galois group and its action on the field $\mathds{Q}(e^{\frac{2\pi i}{E}})$. The group $\DZ_{E}^{\times}$ is the subset of integers less than $E$ that are also co-prime to $E$. The product is given by multiplication of integers modulo $E$. The Galois group of the number field $\mathds{Q}(e^{\frac{2\pi i}{E}})$ is the group of automorphisms of this field which fixes the rationals. In other words, it is a bijective map $\sigma: \mathds{Q}(e^{\frac{2\pi i}{E}}) \to \mathds{Q}(e^{\frac{2\pi i}{E}})$ such that for $a,b\in \mathds{Q}(e^{\frac{2\pi i}{E}})$
\be
\label{eq:sigprop}
\sigma(a+b)=\sigma(a)+\sigma(b)~, ~ \sigma(a\times b)= \sigma(a) \times \sigma(b) \text{ and } \sigma(a)=a ~ \forall ~ a \in \mathds{Q} ~.
\ee
The Galois group of $\mathds{Q}(e^{\frac{2\pi i}{E}})$ is isomorphic to $\DZ_{E}^{\times}$ (see Section 14.5 Theorem 26 in \cite{dummit1991abstract}).
Let $\sigma \in \DZ_{E}^{\times}$ and $a \in \mathds{Q}(e^{\frac{2\pi i}{E}})$, the action of $\sigma$ on $a$ is given by
\be
\label{eq:Galactgen}
\sigma(a)= \sigma\bigg (\sum_{k=0}^{\phi(E)-1} a_k \xi_k \bigg )= \sum_{k=0}^{\phi(E)-1} a_k \xi_k^{\sigma}
\ee

\subsection{Elements invariant under Galois group}
\label{sec:Galinv}

In the cyclotomic field $\mathds{Q}(e^{\frac{2\pi i}{E}})$ the elements fixed by the full Galois group $\DZ_{E}^{\times}$ are precisely the rationals. To prove this, we will use the Fundamental Theorem of Galois Theory. Stating this theorem requires the definition of field extensions of Galois type (Galois extensions), which we now give.

The field $F$ is called an extension of $\mathds{Q}$ if $\mathds{Q} \subset F$ and the operations in $F$ (multiplication, addition) correspond to operations in $\mathds{Q}$ when restricted to the subset $\mathds{Q}$. In other words, if $\mathds{Q}$ is a subfield of $F$. For example, the real numbers are an extension of $\mathds{Q}$.

A field extension $F$ of $\mathds{Q}$ is called algebraic if for every $\alpha \in F$,
\begin{equation*}
	\text{there exists a $P_\alpha(x) \in \mathbb{Q}[x]$ such that $P_\alpha(\alpha) = 0$.}
\end{equation*}
In fact, for an algebraic field extension $F$ and $\alpha \in F$ there exists a unique polynomial of lowest degree for which $\alpha$ is a root \cite{dummit1991abstract}. This polynomial is called the minimal polynomial associated with $\alpha$.

An extension $F$ is called separable if for every $\alpha \in F$ and minimal polynomial $P_\alpha(x)$ of degree $d$ 
\begin{equation*}
	\abs{\{\beta \, \vert \, P_\alpha(\beta) = 0\}} = d.
\end{equation*}
That is, all roots are distinct.
$F$ is called a normal extension if for every $\alpha \in F$ and minimal polynomial $P_\alpha(x)$
\begin{equation*}
	\text{$P_\alpha(\beta)=0$ implies that $\beta \in F$.}
\end{equation*}
That is, all roots lie in $F$.
A separable and normal extension is called a Galois extension. This assumption on a field extension allows one to prove the fundamental theorem of Galois theory (Theorem 14 in \cite{dummit1991abstract}).

To apply the fundamental theory of Galois theory to the cyclotomic fields, we first show that it is a Galois extension.
The field $\mathds{Q}(e^{\frac{2\pi i}{E}})$ is the minimal field containing all the roots of the polynomial
\begin{equation}
	p(x)=x^E -1.
\end{equation}
It follows that $\mathds{Q}(e^{\frac{2\pi i}{E}})$ is a Galois extension of $\mathds{Q}$ (see Theorem 13 in \cite{dummit1991abstract}). Therefore, we are allowed to use the fundamental theorem of Galois theory, which says

\noindent {\bf Fundamental Theorem of Galois Theory for cyclotomic fields:} There is a one-to-one correspondence between the subgroups of $\DZ_{E}^{\times}$ and the subfields of $\mathds{Q}(e^{\frac{2\pi i}{E}})$ given by
\bea
K \subset \mathds{Q}(e^{\frac{2\pi i}{E}}) \rightarrow H \subset \mathds{Z}_{E}^{\times} \text{ which fixes all elements in $K$.} \\
H \subset \mathbb{Z}_{E}^{\times} \rightarrow \text{Subfield $K\subset \mathds{Q}(e^{\frac{2\pi i}{E}})$ of fixed points under $H$-action.}
\eea
In particular, this theorem also implies that the trivial subgroup of $\mathbb{Z}_{E}^{\times}$ fixes the full field $\mathds{Q}(e^{\frac{2\pi i}{E}})$ and the subfield fixed by the full Galois group is the field of rationals.

\subsection{Galois action on representations and characters}
\label{Galactrep}

In this section, we will study how the Galois group acts on irreducible representations and conjugacy classes \cite{navarro2018character}\cite{gannon2007galois}. Given a group $G$ and a representation $R$, recall that $R(g)$ can be realized as a matrix with components in the field $\mathds{Q}(e^{\frac{2 \pi i}{E}})$. An element $\sigma$ in the Galois group of this field acts on the matrix $R(g)$ component-wise resulting in a new matrix $\sigma(R(g))$. We know that $R(g)$ satisfies
\be
R(g) R(h)= R(gh) ~ \forall g,h \in G
\ee
This is an algebraic relation
\begin{equation}
	\sum_{j} R_{ij}(g)R_{jk}(h) - R_{ik}(gh) = 0
\end{equation}
between matrix elements. Since $\sigma(0) = 0$ for all $\sigma$ and using \eqref{eq:sigprop} we have
\begin{equation}
	\sigma\qty(\sum_{j} R_{ij}(g)R_{jk}(h) - R_{ik}(gh)) = \sum_{j} \sigma(R_{ij}(g))\sigma(R_{jk}(h)) - \sigma(R_{ik}(gh) )= 0,
\end{equation}
or equivalently
\be
\sigma(R(g)) \sigma(R(h)) = \sigma(R(gh)) ~ \forall g,h \in G
\ee
Therefore, $\sigma(R)$ is also a representation of the group $G$.

If $R$ is an irreducible representation, so is $\sigma(R)$. This is understood as follows. The character of the representation $R$ is an algebraic combination of elements of the matrix $R$. Therefore, under a Galois action on the representation $R$, using \eqref{eq:sigprop} we get
\be
\label{Galactchar1}
\chi^{\sigma(\irrep)}(g)= \sum_i \sigma(R_{ii}(g)) =\sigma\Big (\sum_i R_{ii}(g) \Big )= \sigma(\chi^\irrep(g)).
\ee
By character orthogonality, a representation $R$ is irreducible if and only if
\begin{equation}
	\frac{1}{\abs{G}} \sum_{g \in G} \chi^\irrep(g) \chi^\irrep(g^{-1}) = 1.
\end{equation}
Now assume that $R$ is irreducible.
The character of $\sigma(R(g))$ is simply $\sigma(\chi^\irrep(g))$ and
\begin{equation}
	\frac{1}{\abs{G}} \sum_{g \in G} \sigma(\chi^\irrep(g)) \sigma(\chi^\irrep(g^{-1}) ) = \sigma\qty( \frac{1}{\abs{G}} \sum_{g \in G} \chi^\irrep(g) \chi^\irrep(g^{-1}) ) = \sigma(1) = 1.
\end{equation}
where in the first equality we used \eqref{eq:sigprop}. Very similar manipulations can be used to show that the Galois action preserves orthogonality of irreducible characters. This implies that two non-isomorphic irreducible representations remain non-isomorphic after the action of the Galois group. This discussion implies that the Galois action on the set of irreducible representations is a permutation (bijection) of the irreducible representations.

Recall that the character of a representation evaluated on an element $g \in G$ can be written as
\be
\chi^\irrep(g)= \sum_i \epsilon_i 
\ee
where $\epsilon_i$ are $E^{\text{th}}$ roots of unity. Then, using \eqref{eq:sigprop} the action of $\sigma \in \mathbb{Z}_{E}^\times$ is
\be
\label{Galactchar2}
\sigma(\chi^\irrep(g))= \sum_i \epsilon^{m_{\sigma}}_i= \chi^\irrep(g^{m_{\sigma}}).
\ee
where $m_{\sigma}$ is an integer coprime to the order of $g$ corresponding to the Galois action $\sigma$. For ease of notation, from now on we will write $g^{m_{\sigma}}$ as $g^{\sigma}$. The action of the Galois group $g \rightarrow g^{\sigma}$ takes conjugacy classes to conjugacy classes since
\begin{equation}
	(hgh^{-1})^{\sigma}= h g^{\sigma} h^{-1}
\end{equation}
Therefore, Galois group also acts on conjugacy classes as a permutation. If $g$ is in the conjugacy class $\cclass \in \cclasses{G}$, and $g^\sigma \in \cclass'$ we write $\sigma(C) = C'$ for the action on conjugacy classes.

To summarize, equations \eqref{Galactchar1} and \eqref{Galactchar2} are two equivalent ways in which we can describe the action of the Galois group on characters. The former action is on the irreducible representations by a permutation while the latter action permutes the conjugacy classes. This permutation of the conjugacy classes may not be an automorphism of the group. 
%
%

\subsection{Integrality of characters}
We now study the following question: Given the fundamental theorem of Galois theory for cyclotomic fields, and the action of the Galois group on representations, can we deduce when a character is integer valued? As we will now prove, $\chi^\irrep(g)$ is an integer if  $g^{\sigma}$ belongs to the same conjugacy class as $g$ for all $\sigma$ co-prime to $E$.

The proof goes as follows. From our discussion in the previous section (see equation \eqref{Galactchar2}), we know that the Galois action on $\chi^\irrep(g)$ gives $\chi^\irrep(g^{\sigma})$. Suppose $g^{\sigma}$ belongs to the same conjugacy class as $g$ for all $\sigma$, then
\be
\chi^\irrep(g^{\sigma})=\chi^\irrep(g).
\ee
In other words, $\chi^\irrep(g)$ is invariant under the full Galois group $\DZ_{E}^{\times}$. Then from our discussion in section \ref{sec:Galinv}, we know that $\chi^\irrep(g)$ is a rational number. Since $\chi^\irrep(g)$ is an algebraic integer, $\chi^\irrep(g)$ being rational implies that $\chi^\irrep(g)$ is an integer. 
In fact, it is sufficient to consider a subgroup $\DZ_{N_g}^{\times} \subset \DZ_{E}^{\times}$ of integers co-prime to $N_g$, the order of $g$. We will not prove this here.

A famous example of this is the symmetric group $S_N$. In $S_N$, both $g$ and $g^{\sigma}$ for any $\sigma$ co-prime to the order of $g$ are in the same conjugacy class. Therefore, all characters are valued in the integers (see page 103, corollary 2 and following examples in \cite{serre1977linear}). 


\subsection{Integrality of sums of characters along columns}
As we just learned, the integrality of particular characters requires checking non-trivial properties of powers of group elements. This depends strongly on the group $G$. In this section we will find that there exists combinations of characters that are integral for general $G$.

\begin{proposition}\label{prop: galois sum along cols 1}
For any $n \in \mathbb{N}$, define the following sum of irreducible characters
\begin{equation}
	\label{Lndef}
	L_n(g) = \sum_{\irrep } \chi^\irrep(g),
\end{equation}
where the sum is over the set
\begin{equation}
	\{\irrep \in \irreps{G} \, \vert \, d_\irrep =n \} \label{eq: dim irrep set galois}
\end{equation}
of irreducible representations of dimension equal to $n$.
\end{proposition}
We will now use Galois theory to show that the sum is an integer for any $n,g$.

\noindent\textbf{Proof of Proposition \ref{prop: galois sum along cols 1}}
To show this, consider the Galois action on $L_n(g)$ by an element $\sigma \in \DZ_{E}^{\times}$,
\be
\sigma(L_n(g))= \sum_{\irrep} \sigma(\chi^\irrep(g)).
\ee
Using \eqref{Galactchar1}, we get
\be
\label{GalactLn}
\sigma(L_n(g))= \sum_{\irrep} \chi^{\sigma(\irrep)}(g),
\ee
where $\sigma(\irrep)$ is the permutation action of irreducible representations.
Since $d_\irrep$ is an integer, we have
\be
d_{\sigma(\irrep)} = \sigma(\chi^\irrep(1))= \chi^\irrep(1)= d_\irrep,
\ee
for every $\sigma \in \DZ_{E}^{\times}$.
Therefore, $\sigma(\irrep)$ and $\irrep$ have the same dimension. In particular, the set in equation \eqref{eq: dim irrep set galois} is invariant under the action of $\DZ_{E}^{\times}$. Equivalently, it is a permutation of terms in the sum \eqref{Lndef}. Therefore, $L_n(g)$ is invariant under the action of the Galois group $\DZ_{E}^{\times}$. From our discussion in section \ref{sec:Galinv}, we know that $L_n(g)$ has to be rational. Since $L_n(g)$ is a sum of algebraic integers, it is in fact an integer. We have proved that $L_n(g)$ is an integer for any $n$ and $g \in G$. For $n=d_{\irrep'}$ for some $\irrep' \in \irreps{G}$, this is the Galois theory proof of equation \eqref{eq: TQFT dim sum is integer}.

More generally, we can consider the following sum of characters.
\begin{proposition}\label{prop: galois sum along cols 2}
Fix an element $h \in G$, and an integer $n$. Define the sum of characters
\be
\label{Lnhdef}
L_{n,h}(g) =\sum_{\irrep } \chi^\irrep(g),
\ee
where the sum is over the set
\begin{equation}
	\{\irrep \in \irreps{G} \, \vert \, \chi^\irrep(h) = n\}. \label{eq: chi h irrep set galois}
\end{equation}
If there are no representations satisfying the constraint in the sum above, we define the sum to be zero. We will show that $L_{n,h}(g)$ is an integer for any $n,g,h$.
\end{proposition}
In fact, the proof is the same as above.

\noindent\textbf{Proof of Proposition \ref{prop: galois sum along cols 2}}
Consider the Galois action on $L_{n,h}(g)$ by an element $\sigma \in \DZ_{E}^{\times}$. We have 
\be
\sigma(L_{n,h}(g))= \sum_{\irrep }  \sigma(\chi^\irrep(g)),
\ee
and using \eqref{Galactchar1} we get
\be
\sigma(L_{n,h}(g))= \sum_{\irrep} \chi^{\sigma(\irrep)}(g).
\ee
Since $n$ is an integer we have
\be
\sigma(\chi^\irrep(h))= \chi^\irrep(h) = n,
\ee
or equivalently $\sigma(\irrep)$ satisfies $\chi^{\sigma(\irrep)}(h)=n$. The Galois action leaves the set in \eqref{eq: chi h irrep set galois} invariant. That is, it only permutes the terms in equation \eqref{Lnhdef} for $L_{n,h}(g)$. Therefore, $L_{n,h}(g)$ is invariant under the action of the Galois group $\DZ_{E}^{\times}$. From our discussion in section \ref{sec:Galinv}, we know that $L_{n,h}(g)$ has to be rational. Since $L_{n,h}(g)$ is a sum of algebraic integers, it is in fact an integer. We have proved that $L_{n,h}(g)$ is an integer for any $n$ and $g,h \in G$.

\noindent \textbf{Proof of Theorem \ref{thm: refined GCTST sum}} (Galois)
A small variation of this is to consider a sum over the set
\begin{equation}
	\{\irrep \in \irreps{G}  \, \vert \frac{\chi^\irrep(h)}{d_\irrep} = q\}, \label{eq: galois set variation}
\end{equation}
where $\irrep$ is an irreducible representation, $d_\irrep$ its dimension and $q \in \mathbb{Q}$. 
Since
\begin{equation}
	\sigma\qty(\frac{\chi^\irrep(h)}{d_\irrep}) = \sigma\qty(\frac{\chi^\irrep(h)}{\chi^R(1)}) = \frac{\sigma(\chi^{\irrep}(h))}{\sigma(\chi^{R}(1))} = \frac{\chi^{\sigma(\irrep)}(h)}{\chi^{\sigma(R)}(1)} = \sigma(q) = q,
\end{equation}
the sum
\begin{equation}
	\sum_R \frac{\chi^\irrep(g)}{d_\irrep},
\end{equation}
is invariant under Galois action and therefore an integer.
Picking $h \in D, S \in \irreps{G}$ such that
\begin{equation}
	q = \frac{\chi_D^S}{d_S} \in \mathbb{Z}
\end{equation}
gives Theorem \ref{thm: refined GCTST sum}.
\begin{proposition}\label{prop: galois sum along cols 3}
Even more generally, for a given set of group elements $H=\{h_1,h_2,\dots,h_r\}$ and integers $N=\{n_1,n_2,\dots,n_r\}$ we can define the sum of characters
\be
L_{N,H}(g) = \sum_{\irrep} \chi^{\irrep}(g)
\ee
where the sum is over the set
\begin{equation}
	\{\irrep \in \irreps{G} \, \vert \, \chi^\irrep(h_1) = n_1, \dots, \chi^\irrep(h_r) = n_r\}. \label{eq: galois multiple row sums}
\end{equation}
If there are no representations satisfying the constraint in the sum above, we define the sum to be zero.
\end{proposition}
It is straightforward to generalize the arguments used above to show that $L_{N,H}(g)$ is an integer for any choice of $H,N$ and $g \in G$.

\noindent \textbf{Proof of Theorem \ref{thm: twice refined GCTST sum}$^*$} (Galois)
As before, we can make contact with section \ref{sec:implications} by consider the set
	\begin{equation}
		\{\irrep \in \irreps{G} \, \vert \, \frac{ \chi^\irrep(h_1)}{d_\irrep} = q_1, \dots, \frac{ \chi^\irrep(h_r)}{d_\irrep} = q_r \}. \label{eq: galois r-refined GCTST sum}
	\end{equation}
	The sum
	\begin{equation}
		\sum_R \frac{ \chi^R(g)}{d_R},
	\end{equation}
	is an integer for $q_1, \dots, q_r \in \mathbb{Q}$. Picking $h_i \in D_i$ and $S_i \in \irreps{G}$ such that
	\begin{equation}
		q_i = \frac{\chi^{S_i}_{D_i}}{d_{S_i}} \in \mathbb{Z},
	\end{equation}
	gives the generalization of Theorem \ref{thm: twice refined GCTST sum}. The special case of $r=2$ is Theorem \ref{thm: twice refined GCTST sum}.

\subsection{Integrality of sums of characters along rows}
We can also prove integrality of sums of characters along rows.
\begin{proposition}\label{prop: galois sum along rows 1}
Consider the following sum of characters. Let $R_1$ be a fixed representation of the group and let $n$ be fixed integer. We define the sum of characters
\be
\label{KnRdef}
K_{n,R_1}(R):=\sum_{g, \chi^{R_1}(g)=n} \chi^\irrep(g)
\ee
If there are no conjugacy classes satisfying the constraint in the sum above, we define the sum to be zero. We will show that $K_{n,R_1}(R)$ is an integer for any $n,R_1,R$.
\end{proposition}

\noindent \textbf{Proof of Proposition \ref{prop: galois sum along rows 1}}
Consider the Galois action on $K_{n,R_1}(R)$ by an element $\sigma \in \DZ_{E}^{\times}$. We have 
\be
\sigma(K_{n,R_1}(R)):=\sum_{g} \sigma(\chi^\irrep(g))
\ee
where the sum is over the set
\be
\{g\in G \, | \, \chi^{\irrep_1}(g)=n\}
\ee
Using \eqref{Galactchar2}, we get
\be
\sigma(K_{n,R_1}(R)):=\sum_{g} \chi^\irrep(g^{\sigma})
\ee
Since $n$ is an integer, we have
\be
\sigma(\chi^{R_1}(g))= \chi^\irrep(g^{\sigma})= n
\ee
Therefore, $g^{\sigma}$ satisfies $\chi^{\irrep_1}(g^{\sigma})=n$. The Galois action on $K_{n,R_1}(R)$ is a permutation of terms in the sum \eqref{KnRdef}. Therefore, $K_{n,R_1}(R)$ is invariant under the action of the Galois group $\DZ_{E}^{\times}$. From our discussion in section \ref{sec:Galinv}, we know that $K_{n,R_1}(R)$ has to be rational. Since $K_{n,R_1}(R)$ is a sum of algebraic integers, it is in fact an integer.

We have proved that $K_{n,R_1}(R)$ is an integer for any $n$ and representations $R,R_1$ of $G$.
\begin{proposition}\label{prop: galois sum along rows 2}
More generally, for a given set of group representations $\Pi=\{R_1,R_2,...,R_r\}$ and integers $N=\{n_1,n_2,...n_r\}$ we can define the sum of characters
\be
K_{N,\Pi}(R):=\sum_{g} \chi^\irrep(g) \label{eq: row sum galois proof}
\ee
where the sum is over the set
\be
\{g\in G \, | \, \chi^{R_i}(g)=n_i\}
\ee
If there are no group elements satisfying the constraint in the sum above, we define the sum to be zero. This sum is an integer for any choice of $\Pi, N$ and representation $R$ of $G$.
\end{proposition}
It is straightforward to generalize the arguments used above to show that $K_{N,\Pi}(R)$ is an integer.

\begin{proposition}\label{prop: galois sum along rows 3}
For a fixed integer $n$, another combination of characters that we have considered is
\be
\label{Mndef}
M_{n}(R):=\sum_{C} \chi^\irrep_\cclass
\ee
where the sum is over the set
\be
\{C\in  \cclasses{G} \, | \, |C|=n\}
\ee
If no such conjugacy classes exist, we define the sum to be zero. We will show that $M_{n}(R)$ is an integer for any $n$ and representation $R$.
\end{proposition}

\noindent\textbf{Proof of Proposition \ref{prop: galois sum along rows 3}}
To see this, we again use Galois theory. 
Consider the Galois action on $M_{n}(R))$ by an element $\sigma \in \DZ_{E}^{\times}$. We have 
\be
\sigma(M_{n}(R)):=\sum_{C} \sigma(\chi^\irrep_\cclass)
\ee
Using \eqref{Galactchar2}, we get
\be
\sigma(M_{n}(R)):=\sum_{C} \chi^\irrep_{ \sigma(C)}
\ee
We will show that $|C|=|\sigma(C)|$. Indeed, 
\be
\label{eq:Galconjsize}
hgh^{-1}=l \implies hg^{\sigma}h=l^{\sigma} \text{ and } hg^{\sigma}h=l^{\sigma} \implies hg^{\sigma\tau}h=l^{\sigma\tau} \implies  hgh^{-1}=l
\ee
where $\tau$ is the inverse of $\sigma$ in $\DZ_{E}^{\times}$, and therefore $g^{\sigma'\tau}=g$. This shows that the conjugacy classes containing $|C|$ and $\sigma(C)$ contains the same number of elements. Therefore, the Galois action by $\sigma$ on \eqref{Mndef} is just a permutation of the terms in the sum \eqref{Mndef}. Therefore, $M_{n}(R)$ is invariant under the action of the Galois group $\DZ_{E}^{\times}$. From our discussion in section \ref{sec:Galinv}, we know that $M_{n}(R)$ has to be rational. Since $M_{n}(R)$ is a sum of algebraic integers, it is in fact an integer.

We now prove Theorem \ref{thm: refined RGCTST sums} and its generalizations.

\noindent \textbf{Proof of Theorem \ref{thm: refined RGCTST sums}$^*$} (Galois)
Using the above, we can prove the generalization of Theorem \ref{thm: refined RGCTST sums}. Consider the set
	\begin{equation}
		\{C \in \cclasses{G} \, \vert \, \chi^{S_1}_{C} = \chi^{S_1}_{D_1}, \dots,  \chi^{S_r}_{C} = \chi^{S_r}_{D_r} \}, \label{eq: galois proof RGCTST generalization}
	\end{equation}
	where $S_i, \in \irreps{G}, D_i \in \cclasses{G}$ is a collection of irreducible representation such that $\chi^{S_i}_{D_i} \in \mathbb{Z}$.
	It follows that
	\begin{equation}
		\sum_{\cclass \in \cclasses{G}} \chi^R_\cclass
	\end{equation}
	is an integer for all $\irrep \in \irreps{G}$, where sum is over the set of conjugacy classes in \eqref{eq: galois proof RGCTST generalization}. Theorem \ref{thm: refined RGCTST sums} is the special case of $r=1$.
	
\section{Counting rows and columns in a number field}
\label{sec: int rows and int cols}

In this section, we will find sufficient conditions for the number of rows of the character table belonging to a number field $F$ to be equal to the number of columns of the character table belonging to $F$. We will begin with a review of general group actions on character tables. We will state and prove Brauer's Lemma on Character Tables. We will then use this lemma to show that if the Galois group of a number field containing the character table is cyclic, then the number of integer rows and columns of the character table are equal \cite{navarro2018character}. We will introduce the minimal number field $F_G$ containing the character table of $G$ and derive necessary conditions for its Galois group to be cyclic. We will refine these results by showing that the the Galois group of $F_G$ being cyclic implies that the number of rows in a subfield of $F_G$ is equal to the number of columns in the same subfield. We will introduce two decorated graphs which capture the number-theoretic properties of rows and columns of the character table, respectively. Finally, we will end this section with various explicitly worked out examples which illustrates our results. 

\subsection{Counting integer rows and columns}
\label{sec:introwcol}

For a group with a cyclic Galois group, the character table has equal number of integer rows and columns. The main results that goes into proving this statement are Brauer's lemma on character tables and the Fundamental Theorem of Galois theory. Let us recall Brauer's lemma on character tables and its proof \cite{brauer1941connection}\cite{navarro2018character} (see also Chapter 18 of \cite{huppertcharacter}). Before stating the lemma and proving it, let us fix some notation. For a group $G$, let Cl$(G)$ be the set of conjugacy classes of $G$. Let $g_i \in \text{Cl}(G)$ be some representatives of the conjugacy classes of $G$ where $i \in I$ belongs to some index set. Similarly, let Irr$(G)$ be the set of equivalence classes of irreducible representations of $G$ and let $R_i \in \text{Irr}(G)$ be some representatives where $i$ again belongs to the same index set $I$. Given this choice of representatives, let $X$ be the character matrix with components 
\be
X_{ij}=\chi^{\irrep_i}(g_j)
\ee

\noindent  {\textbf{Brauer's Lemma on Character Tables:}}  Let $A$ and $G$ be groups and suppose $A$ acts on the set Irr$(G)$ and Cl$(G)$.  Suppose this action satisfies
\be
\label{Aactionconst}
\chi^{\irrep_i}(a(g_j))= \chi^{a(\irrep_i)}(g_j) ~ \forall i,j\in I
\ee
where $a \in A$. Then $a$ fixes the same number of irreducible representations as conjugacy classes. 

Note that the statement of this lemma is slightly different from that in \cite{navarro2018character}. Indeed, Brauer's lemma on character table is usually proven assuming the condition 
\be
\label{Autactionconst}
\chi^{a(\irrep_i)}(a(g_j))= \chi^{\irrep_i}(g_j) ~ \forall i,j\in I
\ee
which is different from what we have in \eqref{Aactionconst}. Action of the automorphism group on the character table satisfies \eqref{Autactionconst} instead of \eqref{Aactionconst}. For an automorphism $\alpha \in \text{Aut}(G)$ acting on the group, the action on the irreducible representations is defined as
\be
\chi^{\alpha(\irrep)}(g):= \chi^\irrep(\alpha^{-1}(g)))
\ee
The inverse action of $\alpha$ on $g$ is crucial to get the equality
\be
\chi^{(\alpha\beta)(\irrep)}(g)= \chi^{\irrep}((\alpha \beta)^{-1}(g))=\chi^{\irrep}(\beta^{-1}(\alpha^{-1} (g)))= \chi^{\beta(\irrep)}(\alpha^{-1}(g))=\chi^{\alpha(\beta(\irrep))}(g)
\ee
for $\alpha,\beta \in \text{Aut}(G)$. On the other hand, using \eqref{Galactchar1} and \eqref{Galactchar2}, the action of the Galois group satisfies
\be
\chi^{\sigma(\irrep)}(g)= \chi^\irrep(g^{\sigma})
\ee
Therefore, when $A$ is the Galois group action on the characters it satisfies \eqref{Aactionconst}. To prove Brauer's Lemma on Character Tables, let us introduce the matrices $P(a)$ and $Q(a)$ where $a \in A$. The matrix $P(a)$ is defined as follows
\bea
(P(a))_{ij}=1 \text{ if } \chi^{a(\irrep_i)} = \chi^{\irrep_j} \\
(P(a))_{ij}=0 \text{ otherwise }
\eea
 The matrix $Q(a)$ is defined as follows
\bea
(Q(a))_{ij}=1 \text{ if } K_{g_i} = K_{a(g_j)} \\
(Q(a))_{ij}=0 \text{ otherwise }
\eea
 We have
\be
(P(a)X)_{ij}= \sum_k (P(a))_{ik} X_{kj}=  \chi^{a(\irrep_i)}(g_j) 
\ee
Similarly, we find that 
\be
(XQ(a))_{ij}= \sum_k X_{ik} (Q(a))_{kj}=  \chi^{\irrep_i}(a(g_j)) 
\ee
Therefore, using \eqref{Aactionconst} we find that $P(a)X=XQ(a)$. Therefore, $P(a)=XQ(a)X^{-1}$. Hence Tr$(P(a))=$Tr$(Q(a))$. Trace of $P(a)/Q(a)$ precisely gives the number of irreps/conjugacy classes left invariant by the $a \in A$ action. If $A$ is cyclic, then we can apply this lemma to a generator of this cyclic group showing that $A$ acts trivially on equal number of irreps and conjugacy classes. 

If $A$ is the Galois group of the Cyclotomic field $\mathds{Q}(e^{\frac{2 \pi i}{E}})$, then the action of this Galois group on the irreducible representations and conjugacy classes satisfies \eqref{Aactionconst} (see section \ref{Galactrep}). We are interested in the case where the Galois group $\DZ_{E}^{\times}$ of $\mathds{Q}(e^{\frac{2 \pi i}{E}})$ is cyclic. This is the case if and only if the exponent of the group $E$ is one of the following form: $1,2,4,p^k,2p^k$ where $p$ is an odd prime and $k$ is a positive integer \cite{gauss1966disquisitiones}. In particular, groups of these orders have a cyclic Galois group. 

While the character table is guaranteed to be contained in the cyclotomic field  $\mathds{Q}(e^{\frac{2 \pi i}{E}})$, this field is often much bigger than the minimal field containing the character table. For example, in the case of the group $S_n$, the character table contains only integers while the dimension of the field $\mathds{Q}(e^{\frac{2 \pi i}{E}})$ as a vector space  increases with $n$. Therefore, it is convenient to define the field $F_G$ for a group $G$ which is the minimal field containing the character table of $G$. $F_G$ is a subfield of $\mathds{Q}(e^{\frac{2 \pi i}{E}})$. Moreover, $F_G$ is a Galois extension of the rationals. This follows from the Fundamental Theorem of Galois Theory. 

Consider the tower of field extensions $F/I/\mathds{Q}$, where $F$ is a Galois extension and let $N$ be the subgroup of Gal$(F)$ which fixes the field $I$. 
\be
N=\{\sigma \in \text{Gal}(F)| \sigma(\alpha)=\alpha ~ \forall \alpha \in I \}
\ee
A consequence of the Fundamental Theorem of Galois Theory (see Chapter 14.2 Theorem 14 (4) in \cite{dummit1991abstract}) is that the intermediate field $I$ is a Galois extension if and only if the subgroup $N$ is normal in Gal$(F)$. Then the Galois group of $I$ is
\be
\text{Gal}(I)=\text{Gal}(F)/N
\ee
In our case, $F=\mathds{Q}(e^{\frac{2 \pi i}{E}})$ is a cyclotomic field for which all subgroups of the Galois group are normal. Therefore, all intermediate fields are Galois extensions. In particular, $F_G$ is a Galois extension with Galois group
\be
\text{Gal}(F_G)=\DZ_{E}^{\times}/N
\ee
for some normal subgroup $N$. If $\DZ_{E}^{\times}$ is cyclic, then its quotient by a normal subgroup is also cyclic. Therefore, if the exponent of the group $G$ is of the form $1,2,4,p^k,2p^k$, where $p$ is an odd prime and $k$ is a positive integer, then $\text{Gal}(F_G)$ is cyclic. For the rest of this section, we will assume that $\text{Gal}(F_G)$ is cyclic.

From Brauer's lemma on character tables applied to the case of $\text{Gal}(F_G)$ acting on irreducible representations and conjugacy classes, it is clear that the number of conjugacy classes left invariant by the full Galois group is same as the number of irreducible representations left invariant by it. From the Fundamental Theorem of Galois theory (see section \eqref{sec:Galinv} for a discussion), if an element of the field extension $F_G$ is left invariant under the   
full Galois group, then it should be rational. Since characters are algebraic integers, if $\chi_{R_i}(g_j)$ is invariant under the Galois group then it should be an integer. Therefore, we get the following theorem.
\vspace{0.2cm}

\noindent {\bf Equal number of integer rows and columns:} For a finite group $G$, let $F_G$ be the minimal field containing the character table of $G$. If the Galois group of $F_G$ is cyclic, then the character table of $G$ contains equal number of integer rows and integer columns.

\vspace{0.2cm}
A version of this result is stated and proved in \cite[Theorem 3.3]{navarro2018character}. 

\subsection{Counting rows and columns in a number field}
\label{subsec:rowcolnumfield}

Given the results in the previous section, it is natural to wonder whether the number of rows and columns of the character table belonging to a given subfield of $F_G$ is the same. This is indeed the case. To understand this, note that Brauer's lemma on character tables is not just about invariance under the full cyclic group $A$, but invariance under the action of any element $a \in A$. In fact, if $\chi_{R_i}(g_j)$ is invariant under the action of $a \in A$, then it is invariant under the full cyclic subgroup generated by $a$. Therefore, it is natural to consider invariants under subgroups of $A$. Then Brauer's lemma tells us that the number of rows and columns of the character table invariant under a given subgroup, say $B$, of the cyclic group $A$ is the same.

When $A=\text{Gal}(F_G)$ is the Galois group of $F_G$, then the Fundamental Theorem of Galois theory states that subgroups of $\text{Gal}(F_G)$ and subfields of $F_G$ are in one-to-one correspondence. For every subgroup of $H \subset \text{Gal}(F_G)$, the corresponding subfield is the set of elements in $F_G$ fixed by the action of $H$. In the opposite direction, given a subfield, the corresponding subgroup of the Galois group is that which fixes the chosen subfield. Therefore, by identifying columns and rows of the character table invariant under a subgroup of the Galois group, we are equivalently looking at rows and columns of the character table belonging to a subfield of $F_G$. Using Brauer's lemma in this case we get the following theorem.
\vspace{0.2cm}

\noindent {\bf Equal number of rows and columns in a subfield:} For a finite group $G$, let $F_G$ be the minimal field containing the character table of $G$. If the Galois group of $F_G$ is cyclic, then the number of rows in the character table with entries in a subfield $I \subset F_G$ is the same as the number of columns with entries in $I$. 

\vspace{0.2cm}

This equality can be depicted using a decorated graph. Given the character table of a group, we define a graph with vertices labelled by the field $F_G$ and its subfields. Two vertices labelled by fields $F,K$ have a directed edge pointing from $K$ to $F$ if $F$ is a subfield of $K$. Once we have this graph, we can further decorate the vertices in the following two ways
\begin{enumerate}
\item $G1$: Label the vertex corresponding to the field $F$ as the number of rows of the character table contained in that field.
\item $G2$: Label the vertex corresponding to the field $F$ as the number of columns of the character table contained in that field.
\end{enumerate}
Then our results above show that the decorated graphs G1 and G2 are the same. Let us study a few examples to see this explicitly.

\subsubsection{$\DZ_6$}

Consider the cyclic group $\DZ_6$ with character table given in table \ref{ctZ6}.
\begin{table}[h!]
\begin{center}
\begin{tabular}{c|cccccc}
& 1 & $g$ & $g^2$ & $g^3$  & $g^4$  & $g^5$  \\ \hline
$R_1$ & 1 & 1  & 1 & 1 & 1 & 1  \\
$R_2$ & 1 & $e^{\frac{2 \pi i}{6}}$  & $e^{\frac{2 \pi i2 }{6}}$ & -1 & $e^{\frac{2 \pi i 4}{6}}$ & $e^{\frac{2 \pi i 5}{6}}$  \\
$R_3$ & 1 & $e^{\frac{2 \pi i}{3}}$  & $e^{\frac{2 \pi i2 }{3}}$ & $1$ & $e^{\frac{2 \pi i }{3}}$ & $e^{\frac{2 \pi i 2}{3}}$  \\
$R_4$ & 1 & -1  & 1 & -1 & 1 & -1  \\
$R_5$ & 1 & $e^{\frac{2 \pi i 2 }{3}}$  & $e^{\frac{2 \pi i }{3}}$ & $1$ & $e^{\frac{2 \pi i 2 }{3}}$ & $e^{\frac{2 \pi i }{3}}$  \\
$R_6$ & 1 & $e^{\frac{2 \pi i 5}{6}}$  & $e^{\frac{2 \pi i 4}{6}}$ & -1 & $e^{\frac{2 \pi i 2}{6}}$ & $e^{\frac{2 \pi i}{6}}$  \\
\end{tabular}
\end{center}
\caption{ 
Character table of $\DZ_6$.
}
\label{ctZ6}
\end{table}
We find that all entries in the character table are contained in the field $\mathds{Q}(e^{\frac{2\pi i}{6}})$. Since the primitive $6^{\text{th}}$ roots of unity are negatives of primitive $3^{\text{rd}}$ roots of unity, $\mathds{Q}(e^{\frac{2\pi i}{6}}) \cong \mathds{Q}(e^{\frac{2\pi i}{3}})$. Therefore, the only non-trivial subfield is $\mathds{Q}$. The Galois group of $\mathds{Q}(e^{\frac{2\pi i}{6}})$ is $\DZ_2$. The character table has some rows and columns belonging to the subfield $\mathds{Q}$. The graph $G1$ defined above constructed for this particular case is given in figure \ref{fig:grZ6}. 
\begin{figure}[h!]
\centering

\tikzset{every picture/.style={line width=0.75pt}} 

\tikzset{every picture/.style={line width=0.75pt}} 

\begin{tikzpicture}[x=0.75pt,y=0.75pt,yscale=-1,xscale=1]

\draw  [fill={rgb, 255:red, 255; green, 255; blue, 255 }  ,fill opacity=1 ] (296,67.6) .. controls (296,61.63) and (300.84,56.79) .. (306.81,56.79) .. controls (312.78,56.79) and (317.63,61.63) .. (317.63,67.6) .. controls (317.63,73.58) and (312.78,78.42) .. (306.81,78.42) .. controls (300.84,78.42) and (296,73.58) .. (296,67.6) -- cycle ;
\draw    (306.42,78.21) -- (306.21,153) ;
\draw  [fill={rgb, 255:red, 255; green, 255; blue, 255 }  ,fill opacity=1 ] (296,159.6) .. controls (296,153.63) and (300.84,148.79) .. (306.81,148.79) .. controls (312.78,148.79) and (317.63,153.63) .. (317.63,159.6) .. controls (317.63,165.58) and (312.78,170.42) .. (306.81,170.42) .. controls (300.84,170.42) and (296,165.58) .. (296,159.6) -- cycle ;
\draw   (311.23,109.23) -- (306.25,119.14) -- (301.09,109.32) ;

\draw (322,40.4) node [anchor=north west][inner sep=0.75pt]    {$\mathbb{Q}\left( e^{\frac{2\pi i}{6}}\right)$};
\draw (323,150.4) node [anchor=north west][inner sep=0.75pt]    {$\mathbb{Q}$};
\draw (302,59.4) node [anchor=north west][inner sep=0.75pt]    {$6$};
\draw (302,151.4) node [anchor=north west][inner sep=0.75pt]    {$2$};

\end{tikzpicture}
\caption{Graph $G1$ for $\DZ_6$.}
\label{fig:grZ6}
\end{figure}
\noindent The number in the vertices of the graph is the number of rows of the character table contained in the field attached to the vertex. It is easily verified that graph $G2$ constructed as defined above is same as this graph. 

\subsubsection{$\DZ_7 \rtimes \DZ_3$}

The group $\DZ_7 \rtimes \DZ_3$ can be represented by the following generators and relations
\be
\DZ_7 \rtimes \DZ_3 = \langle a,b | a^7=b^3=1, bab^{-1}=a^4 \rangle
\ee
It has the character table given in Table \ref{ctZ3Z7}.
\begin{table}[h!]
\begin{center}
\begin{tabular}{c|ccccc}
& 1 & $a$ & $b$ & $b^2$  & $a^3$   \\ \hline
$R_1$ & 1 & 1  & 1 & 1 & 1   \\
$R_2$ & 1 & $e^{\frac{2\pi i 2}{3}}$  & 1 & $e^{\frac{2\pi i}{3}}$ & 1   \\
$R_3$ & 1 & $e^{\frac{2\pi i }{3}}$  & 1 & $e^{\frac{2\pi i 2}{3}}$ & 1   \\
$R_4$ & 3 & 0  & $\frac{-1+i\sqrt{7}}{2}$ & 0 & $\frac{-1-i\sqrt{7}}{2}$  \\
$R_5$ & 3 & 0  & $\frac{-1-i\sqrt{7}}{2}$ & 0 & $\frac{-1+i\sqrt{7}}{2}$     
\end{tabular}
\end{center}
\caption{ 
Character table of $\DZ_7 \rtimes \DZ_3$.
}
\label{ctZ3Z7}
\end{table}
We find that all entries in the character table are contained in the field $\mathds{Q}(e^{\frac{2\pi i}{3}},i \sqrt{7})$. $i \sqrt{7}$ is contained in the cyclotomic field $\mathds{Q}(e^{\frac{2\pi i}{7}})$. This follows from the quadratic Gauss sum \cite{berndt1998gauss}
\be
\sum_{n=0}^{p-1} e^{\frac{2 \pi i n^2}{p}} = \begin{cases}
\sqrt{p} &\text{if } p=1 \text{ mod }4 \\ 
i\sqrt{p} &\text{if } p=3 \text{ mod }4
\end{cases}  
\ee
where $p$ is prime. Therefore, the smallest cyclotomic field containing the entries of the character table is $\mathds{Q}(e^{\frac{2\pi i}{21}})$. The Galois group of this number field is $\DZ_{21}^{\times}\simeq \DZ_{7}^{\times} \times \DZ_{3}^{\times}\simeq \DZ_{6} \times \DZ_{2}$. Let $\sigma_1$ and $\sigma_2$ be generators of $\DZ_6$ and $\DZ_2$ factors of the Galois group, respectively with action on the number field given by
\be 
\sigma_1(e^{\frac{2 \pi i}{21}})= e^{\frac{2 \pi i 10}{21}} , ~ \sigma_2(e^{\frac{2 \pi i}{21}})= e^{\frac{2 \pi i 8}{21}}
\ee
$\sigma_2$ acts trivially on $i \sqrt{7}$ while $\sigma_1$ acts as $\sigma_1(i\sqrt{7})=-i\sqrt{7}$ and $\sigma_1$ acts trivially on $e^{\frac{2 \pi i}{3}}$ while $\sigma_2$ acts as $\sigma_2(e^{\frac{2 \pi i}{3}})=e^{\frac{2 \pi i 2}{3}}$ . Therefore, both $\sigma_1$ and $\sigma_2$ have order two actions on the character table. Let us find the rows fixed by $\sigma_1$ and $\sigma_2$.
\be
\sigma_1 \text{ fixes the rows } R_1, R_2, R_3
\ee
\be
\sigma_2 \text{ fixes the rows } R_1,R_4,R_5
\ee
Similarly, let us find the columns fixed by $\sigma_1$ and $\sigma_2$.
\be
\sigma_1 \text{ fixes the columns } 1, a, b^2  
\ee
\be
\sigma_2 \text{ fixes the columns } 1, b, a^3
\ee
Note that $\sigma_1$ fixes equal number of rows and columns. The same is true for $\sigma_2$. This follows from Brauer's lemma on character tables as $\sigma_1$ and $\sigma_2$ generate cyclic subgroups of the Galois group. The rows and columns fixed by the full Galois group are
\be
\text{\{rows fixed by $\sigma_1$\} } \cap \text{ \{rows fixed by $\sigma_2$\}}= R_1
\ee
\be
\text{\{columns fixed by $\sigma_1$\} } \cap \text{ \{columns fixed by $\sigma_2$\}= 1}
\ee
The graph $G1$ defined above constructed for this particular case is given in figure \ref{fig:grZ7Z3}. 
\begin{figure}[h!]
\centering
\tikzset{every picture/.style={line width=0.75pt}} 

\begin{tikzpicture}[x=0.75pt,y=0.75pt,yscale=-1,xscale=1]

\draw  [fill={rgb, 255:red, 255; green, 255; blue, 255 }  ,fill opacity=1 ] (375,161.6) .. controls (375,155.63) and (379.84,150.79) .. (385.81,150.79) .. controls (391.78,150.79) and (396.63,155.63) .. (396.63,161.6) .. controls (396.63,167.58) and (391.78,172.42) .. (385.81,172.42) .. controls (379.84,172.42) and (375,167.58) .. (375,161.6) -- cycle ;
\draw    (379,171) -- (329,246) ;
\draw  [fill={rgb, 255:red, 255; green, 255; blue, 255 }  ,fill opacity=1 ] (310,252.6) .. controls (310,246.63) and (314.84,241.79) .. (320.81,241.79) .. controls (326.78,241.79) and (331.63,246.63) .. (331.63,252.6) .. controls (331.63,258.58) and (326.78,263.42) .. (320.81,263.42) .. controls (314.84,263.42) and (310,258.58) .. (310,252.6) -- cycle ;
\draw    (252,169) -- (313,245) ;
\draw  [fill={rgb, 255:red, 255; green, 255; blue, 255 }  ,fill opacity=1 ] (236,161.6) .. controls (236,155.63) and (240.84,150.79) .. (246.81,150.79) .. controls (252.78,150.79) and (257.63,155.63) .. (257.63,161.6) .. controls (257.63,167.58) and (252.78,172.42) .. (246.81,172.42) .. controls (240.84,172.42) and (236,167.58) .. (236,161.6) -- cycle ;
\draw   (364.18,202.05) -- (354.39,207.26) -- (355.85,196.27) ;
\draw   (279.86,196.17) -- (282.41,206.96) -- (272.15,202.75) ;
\draw    (251.12,151.52) -- (308.19,72.24) ;
\draw  [fill={rgb, 255:red, 255; green, 255; blue, 255 }  ,fill opacity=1 ] (329.28,67.91) .. controls (329.2,73.88) and (324.29,78.65) .. (318.32,78.56) .. controls (312.34,78.48) and (307.57,73.57) .. (307.66,67.6) .. controls (307.74,61.63) and (312.65,56.86) .. (318.62,56.94) .. controls (324.6,57.03) and (329.37,61.94) .. (329.28,67.91) -- cycle ;
\draw    (380.08,152.33) -- (328.17,72.47) ;
\draw   (284.31,113.25) -- (274.69,118.75) -- (275.81,107.72) ;
\draw   (356.74,107.8) -- (358.17,118.79) -- (348.39,113.56) ;

\draw (404,134.4) node [anchor=north west][inner sep=0.75pt]    {$\mathbb{Q}\left( e^{\frac{2\pi i}{3}}\right)$};
\draw (337,245.4) node [anchor=north west][inner sep=0.75pt]    {$\mathbb{Q}$};
\draw (381,153.4) node [anchor=north west][inner sep=0.75pt]    {$3$};
\draw (316,244.4) node [anchor=north west][inner sep=0.75pt]    {$1$};
\draw (167,143.4) node [anchor=north west][inner sep=0.75pt]    {$\mathbb{Q}\left( i\sqrt{7}\right)$};
\draw (242,152.4) node [anchor=north west][inner sep=0.75pt]    {$3$};
\draw (313,58.4) node [anchor=north west][inner sep=0.75pt]    {$5$};
\draw (336,36.4) node [anchor=north west][inner sep=0.75pt]    {$\mathbb{Q}\left( i\sqrt{7} ,e^{\frac{2\pi i}{3}}\right)$};

\end{tikzpicture}
\caption{Graph $G1$ for $\DZ_7 \rtimes \DZ_3$.}
\label{fig:grZ7Z3}
\end{figure}
\noindent It is easily verified that graph $G2$ constructed as defined above is same as this graph. 

This example shows that even if the Galois group is not cyclic, the number of integer rows and columns in the character table can be equal.

\subsubsection{$\DZ_{24} \rtimes \DZ_{2}$}

Let us consider the group $\DZ_{24} \rtimes \DZ_{2}$ of order $48$ which is represented as a Polycyclic group with $5$ generators $f1,f2,f3,f4,5$ in GAP (SmallGroup(48,6)). The character table is given by Table \ref{ctZ2Z24} where $A = -2i, B=\sqrt{3}, C = -e^{\frac{2 \pi i}{24}}-e^{\frac{2 \pi i 11}{24}}, D = -e^{\frac{2 \pi i 17 }{24}}-e^{\frac{2 \pi i 19}{24}}$. 

$A$ belongs to the field $\mathds{Q}(e^{\frac{2\pi i}{4}})$, $B$ belongs to the field $\mathds{Q}(e^{\frac{2\pi i}{12}})$ and $C,D$ belongs to the field $\mathds{Q}(e^{\frac{2\pi i}{24}})$. Therefore, all entries in the character table are contained in the field $\mathds{Q}(e^{\frac{2\pi i}{24}})$. The character table has some rows and columns belonging to the subfields $\mathds{Q}(e^{\frac{2\pi i}{12}}), \mathds{Q}(e^{\frac{2\pi i}{4}})$ and $\mathds{Q}$. 

\begin{table}[h!]
\begin{center}
\begin{tabular}{c|ccccccccccccccc}
         & 1 & f1 & f2 & f3 &f4 &f5 & f1f2 & f2f4 & f2f5 & f3f5 & f4f5  & f2f3f5, & f2f4f5  & f3f4f5 & f2f3f4f5 \\
\hline
$R_1$      & 1  & 1  & 1  & 1  & 1  & 1  &1  &1   &1   &1  &1   &1 &  1 &  1 &  1 \\ 
$R_2$      &1 &-1 &-1  &1  &1  &1  &1 &-1  &-1   &1  &1  &-1  &-1   &1  &-1\\
$R_3$     &1 &-1  &1  &1  &1  &1 &-1  &1  & 1   &1  &1   &1   &1   &1   &1\\
$R_4$      &1  &1 &-1  &1  &1  &1 &-1 &-1  &-1   &1  &1  &-1  &-1   &1  &-1\\
$R_5$      &2  &0  &0 &-2  &2  &2  &0  &0  & 0  &-2  &2   &0   &0  &-2   &0\\
$R_6$      &2  &0 &-2  &2  &2 &-1  &0 &-2   &1  &-1 &-1   &1   &1  &-1   &1\\
$R_7$      &2  &0 & 2  &2  &2 &-1  &0  &2  &-1  &-1 &-1  &-1  &-1  &-1  &-1\\
$R_8$     &2  &0  &A  &0 &-2  &2  &0 &-A   &A   &0 &-2   &A  &-A   &0  &-A\\
$R_9$      &2  &0 &-A  &0 &-2  &2  &0  &A  &-A   &0 &-2  &-A   &A   &0   &A\\
$R_{10}$     &2  &0  &0 &-2  &2 &-1  &0  &0   &B   &1 &-1  &-B   &B   &1  &-B\\
$R_{11}$     &2  &0  &0 &-2  &2 &-1  &0  &0  &-B   &1 &-1   &B  &-B   &1   &B\\
$R_{12}$     &2  &0  &A  &0 &-2 &-1  &0 &-A   &C  &-B  &1   &D  &-C   &B  &-D\\
$R_{13}$     &2  &0  &A  &0 &-2 &-1  &0 &-A   &D   &B  &1   &C  &-D  &-B  &-C\\
$R_{14}$     &2  &0 &-A  &0 &-2 &-1  &0  &A  &-D   &B  &1  &-C   &D  &-B   &C\\
$R_{15}$     &2  &0 &-A  &0 &-2 &-1  &0  &A  &-C  &-B  &1  &-D   &C   &B   &D  
\end{tabular}
\end{center}
\caption{ 
Character table of $\DZ_{24} \rtimes \DZ_{2}$.
}
\label{ctZ2Z24}
\end{table} 
The graph $G1$ defined above constructed for this particular case is given in figure \ref{fig:grZ24Z2}.
\begin{figure}[h!]
\centering

\tikzset{every picture/.style={line width=0.75pt}} 

\begin{tikzpicture}[x=0.75pt,y=0.75pt,yscale=-1,xscale=1]

\draw  [fill={rgb, 255:red, 255; green, 255; blue, 255 }  ,fill opacity=1 ] (296,176.6) .. controls (296,170.63) and (300.84,165.79) .. (306.81,165.79) .. controls (312.78,165.79) and (317.63,170.63) .. (317.63,176.6) .. controls (317.63,182.58) and (312.78,187.42) .. (306.81,187.42) .. controls (300.84,187.42) and (296,182.58) .. (296,176.6) -- cycle ;
\draw    (307.02,188) -- (306.81,230.79) ;
\draw  [fill={rgb, 255:red, 255; green, 255; blue, 255 }  ,fill opacity=1 ] (296,241.6) .. controls (296,235.63) and (300.84,230.79) .. (306.81,230.79) .. controls (312.78,230.79) and (317.63,235.63) .. (317.63,241.6) .. controls (317.63,247.58) and (312.78,252.42) .. (306.81,252.42) .. controls (300.84,252.42) and (296,247.58) .. (296,241.6) -- cycle ;
\draw  [fill={rgb, 255:red, 255; green, 255; blue, 255 }  ,fill opacity=1 ] (297,111.6) .. controls (297,105.63) and (301.84,100.79) .. (307.81,100.79) .. controls (313.78,100.79) and (318.63,105.63) .. (318.63,111.6) .. controls (318.63,117.58) and (313.78,122.42) .. (307.81,122.42) .. controls (301.84,122.42) and (297,117.58) .. (297,111.6) -- cycle ;
\draw    (307.02,123) -- (306.81,165.79) ;
\draw  [fill={rgb, 255:red, 255; green, 255; blue, 255 }  ,fill opacity=1 ] (297,47.6) .. controls (297,41.63) and (301.84,36.79) .. (307.81,36.79) .. controls (313.78,36.79) and (318.63,41.63) .. (318.63,47.6) .. controls (318.63,53.58) and (313.78,58.42) .. (307.81,58.42) .. controls (301.84,58.42) and (297,53.58) .. (297,47.6) -- cycle ;
\draw    (307.02,59) -- (306.81,101.79) ;
\draw   (312.28,71.29) -- (307.2,81.14) -- (302.14,71.27) ;
\draw   (312.28,138.29) -- (307.2,148.14) -- (302.14,138.27) ;
\draw   (312.28,201.29) -- (307.2,211.14) -- (302.14,201.27) ;

\draw (325,151.4) node [anchor=north west][inner sep=0.75pt]    {$\mathbb{Q}\left( e^{\frac{2\pi i}{4}}\right)$};
\draw (323,232.4) node [anchor=north west][inner sep=0.75pt]    {$\mathbb{Q}$};
\draw (302,168.4) node [anchor=north west][inner sep=0.75pt]    {$9$};
\draw (302,233.4) node [anchor=north west][inner sep=0.75pt]    {$7$};
\draw (324,86.4) node [anchor=north west][inner sep=0.75pt]    {$\mathbb{Q}\left( e^{\frac{2\pi i}{12}}\right)$};
\draw (303,103.4) node [anchor=north west][inner sep=0.75pt]    {$9$};
\draw (324,20.4) node [anchor=north west][inner sep=0.75pt]    {$\mathbb{Q}\left( e^{\frac{2\pi i}{24}}\right)$};
\draw (299,39.4) node [anchor=north west][inner sep=0.75pt]    {$15$};

\end{tikzpicture}
\caption{Graph $G1$ for $\DZ_{24} \rtimes \DZ_2$.}
\label{fig:grZ24Z2}
\end{figure}
It is easily verified that graph $G2$ constructed as defined above is same as this graph. Note that in this case the Galois group of the cyclotomic field $\mathds{Q}(e^{\frac{2\pi i}{24}})$ is $\DZ_{24}^{\times}\simeq \DZ_{8}^{\times} \times \DZ_3^{\times} \simeq \DZ_2 \times \DZ_2 \times \DZ_2$ which is not cyclic. Let $\sigma_1,\sigma_2,\sigma_3$ be the three generators of the cyclic factors which act as 
\be
\sigma_1(e^{\frac{2 \pi i}{24}})=e^{\frac{2 \pi i 13 }{24}},~ \sigma_2(e^{\frac{2 \pi i}{24}})=e^{\frac{2 \pi i 11 }{24}},~ \sigma_3(e^{\frac{2 \pi i 5}{24}})=e^{\frac{2 \pi i 13 }{24}}
\ee
Similar to the previous example, we can verify that equal number of rows and columns are left invariant by the full Galois group even though the Galois group is not cyclic. However, as the next example shows, there are groups for which the number of integer rows and the number of integer columns of the character table are different.

\subsubsection{$Q_8 \rtimes \DZ_4$}

The character table for a semi-direct product of $Q_8$ and $\DZ_4$ denoted in the GAP library as SmallGroup(32,10) is given by table \ref{ctQ8Z4},
\begin{table}[h!]
\begin{center}
\begin{tabular}{c|cccccccccccccc}
         & 1 & f1 & f2 &f3 &f4 &f5 &f1f2 & f1f4 
  &f2f4 &f3f4 &f4f5 &f1f2f4 &f1f2f5 &f1f2f4f5\\
\hline
$R_1$      &1 & 1  &1  &1 & 1  &1  &1 & 1 & 1 & 1 & 1  &1 & 1 & 1\\
$R_2$      &1 &-1  &1  &1 & 1  &1 &-1 &-1  &1  &1  &1 &-1 &-1 &-1 \\
$R_3$      &1 & 1 &-1  &1 & 1  &1 &-1 & 1 &-1 & 1 & 1 &-1 &-1 &-1 \\
$R_4$     &1 &-1 &-1  &1 & 1  &1  &1 &-1 &-1  &1  &1  &1  &1 & 1\\
$R_5$     &1 & A  &1  &1 &-1  &1  &A &-A &-1 &-1 &-1 &-A & A &-A\\
$R_6$     &  1 &-A  &1  &1 &-1  &1 &-A  &A &-1 &-1 &-1  &A &-A  &A\\
$R_7$     &  1 & A &-1  &1 &-1  &1 &-A &-A & 1 &-1 &-1 & A &-A & A\\
$R_8$     &  1 &-A &-1  &1 &-1  &1 & A & A & 1 &-1 &-1 &-A & A &-A \\
$R_9$     &  2 & 0  &0 &-2 & 2  &2 & 0 & 0  &0 &-2 & 2 & 0 & 0&  0\\
$R_{10}$    &  2 & 0  &0 &-2 &-2 & 2 & 0 & 0 & 0 & 2 &-2 & 0 & 0 & 0\\
$R_{11}$    &  2 & 0  &0  &0 & 2 &-2 & B & 0 & 0 & 0 &-2 & B &-B &-B\\
$R_{12}$    &  2 & 0  &0  &0 & 2 &-2 &-B & 0 & 0 & 0 &-2 &-B & B & B\\
$R_{13}$    & 2 & 0  &0  &0 &-2 &-2  &C & 0 & 0 & 0 & 2 &-C &-C & C\\
$R_{14}$    &  2 & 0  &0  &0 &-2 &-2 &-C & 0 & 0 & 0 & 2&  C&  C &-C
\end{tabular}
\end{center}
\caption{ 
Character table of $Q_8 \rtimes \DZ_4$.
}
\label{ctQ8Z4}
\end{table} 
where $A=i, B=e^{\frac{2\pi i}{8}}+e^{\frac{2 \pi i 3}{8}}=i\sqrt{2}$ and $C=\sqrt{2}$. We see that the minimal field containing the character table is $F_G=\mathds{Q}(e^{\frac{2 \pi i}{8}})=\mathds{Q}(i,\sqrt{2})$. The Galois group for this field is $\DZ_2 \times \DZ_2$. We will denote the elements of this group as $\sigma_1,\sigma_3,\sigma_5,\sigma_7$. These elements act as follows
\be
\sigma_a(e^{\frac{2\pi i}{8}})=e^{\frac{2\pi i a}{8}}, ~ a=1,3,5,7
\ee
$\sigma_3, \sigma_5$ and $\sigma_7$ generate three $\DZ_2$ subgroups of the full Galois group. These subgroups fixes the following subfields of $F_G=\mathds{Q}(i,\sqrt{2})$. $\sigma_3$ fixes the field $\mathds{Q}(i\sqrt{2})$, $\sigma_5$ fixes the field $\mathds{Q}(i)$ and $\sigma_7$ fixes the field $\mathds{Q}(\sqrt{2})$. From this information, we can look at the rows and columns of the character table fixed under the action of the generators of $\DZ_2 \times \DZ_2$, say $\sigma_5$ and $\sigma_7$.
\be
\sigma_5 \text{ fixes the rows } R_1, R_2, R_3, R_4, R_5, R_6, R_7, R_8, R_9, R_{10}  
\ee
\be
\sigma_7 \text{ fixes the rows } R_1,R_2,R_3,R_4,R_9,R_{10},R_{13},R_{14}
\ee
Similarly, let us find the columns fixed by $\sigma_5$ and $\sigma_7$.
\be
\sigma_5 \text{ fixes the columns 1, f1, f2, f3, f4, f5, f1f4, f2f4, f3f4, f4f5}  
\ee
\be
\sigma_7 \text{ fixes the columns  1, f2, f3, f4, f5, f2f4, f3f4, f4f5}
\ee
We find that $\sigma_5$ fixes the same number of rows and columns implying that the number of rows and columns contained in the field $\mathds{Q}(i)$ are equal. Similarly, we find that $\sigma_7$ fixes the same number of rows and columns implying that the number of rows and columns contained in the field $\mathds{Q}(\sqrt{2})$ are equal. This is consistent with Brauer's lemma on character tables. Indeed, the subgroup of the full Galois group generated by $\sigma_5$ is isomorphic to the cyclic group $\DZ_2$. Therefore, Brauer's lemma applied to this group tells us that the number of rows and number of columns columns fixed by $\sigma_5$ are equal. This implies that the number of rows and number of columns columns contained in $\mathds{Q}(i)$ are equal. Similar argument holds for $\sigma_7$ and $\sigma_3$.

Since $\sigma_5$ and $\sigma_7$ generate the full Galois group, the rows fixed under the action of the full Galois group are given by 
\be
\text{\{rows fixed by $\sigma_5$\} } \cap \text{ \{rows fixed by $\sigma_7$\}}= R_1,R_2,R_3,R_4,R_9,R_{10}
\ee
which are precisely the $6$ integer rows of the character table. Similarly, the columns fixed under under the action of the full Galois group are given by 
\be
\text{\{columns fixed by $\sigma_5$\} } \cap \text{ \{columns fixed by $\sigma_7$\}= 1,f2,f3,f4,f5,f2f4,f3f4,f4f5}
\ee
which are precisely the $8$ integer columns of the character table. Therefore, this example explicitly shows that for non-cyclic Galois groups, the number of rows and number of columns fixed by the Galois group action may not be equal. The graphs $G1$ and $G2$ below illustrates the result above. 

\begin{figure}[h!]
\centering

\tikzset{every picture/.style={line width=0.75pt}} 

\begin{tikzpicture}[x=0.75pt,y=0.75pt,yscale=-0.8,xscale=0.8]

\draw  [fill={rgb, 255:red, 255; green, 255; blue, 255 }  ,fill opacity=1 ] (422.42,230.48) .. controls (422.42,220.89) and (430.14,213.12) .. (439.67,213.12) .. controls (449.2,213.12) and (456.92,220.89) .. (456.92,230.48) .. controls (456.92,240.07) and (449.2,247.85) .. (439.67,247.85) .. controls (430.14,247.85) and (422.42,240.07) .. (422.42,230.48) -- cycle ;
\draw    (428.8,245.57) -- (349.04,366.03) ;
\draw  [fill={rgb, 255:red, 255; green, 255; blue, 255 }  ,fill opacity=1 ] (318.73,376.63) .. controls (318.73,367.04) and (326.45,359.27) .. (335.98,359.27) .. controls (345.5,359.27) and (353.22,367.04) .. (353.22,376.63) .. controls (353.22,386.23) and (345.5,394) .. (335.98,394) .. controls (326.45,394) and (318.73,386.23) .. (318.73,376.63) -- cycle ;
\draw    (226.2,242.36) -- (323.51,364.42) ;
\draw  [fill={rgb, 255:red, 255; green, 255; blue, 255 }  ,fill opacity=1 ] (200.67,230.48) .. controls (200.67,220.89) and (208.4,213.12) .. (217.92,213.12) .. controls (227.45,213.12) and (235.17,220.89) .. (235.17,230.48) .. controls (235.17,240.07) and (227.45,247.85) .. (217.92,247.85) .. controls (208.4,247.85) and (200.67,240.07) .. (200.67,230.48) -- cycle ;
\draw   (405.16,295.44) -- (389.55,303.81) -- (391.87,286.16) ;
\draw   (270.64,286) -- (274.71,303.33) -- (258.34,296.57) ;
\draw    (224.79,214.28) -- (315.84,86.96) ;
\draw  [fill={rgb, 255:red, 255; green, 255; blue, 255 }  ,fill opacity=1 ] (349.49,80) .. controls (349.35,89.59) and (341.52,97.25) .. (331.99,97.12) .. controls (322.46,96.98) and (314.85,89.1) .. (314.99,79.51) .. controls (315.13,69.92) and (322.96,62.25) .. (332.49,62.39) .. controls (342.01,62.52) and (349.62,70.41) .. (349.49,80) -- cycle ;
\draw    (430.52,215.59) -- (347.72,87.33) ;
\draw   (277.75,152.82) -- (262.39,161.66) -- (264.18,143.95) ;
\draw   (393.28,144.07) -- (395.57,161.73) -- (379.97,153.32) ;
\draw    (331.99,97.12) -- (333.08,213.45) ;
\draw  [fill={rgb, 255:red, 255; green, 255; blue, 255 }  ,fill opacity=1 ] (316.13,232.09) .. controls (316.13,222.5) and (323.85,214.72) .. (333.38,214.72) .. controls (342.91,214.72) and (350.63,222.5) .. (350.63,232.09) .. controls (350.63,241.68) and (342.91,249.45) .. (333.38,249.45) .. controls (323.85,249.45) and (316.13,241.68) .. (316.13,232.09) -- cycle ;
\draw    (334.38,249.06) -- (334.88,359.15) ;

\draw (463.55,213.76) node [anchor=north west][inner sep=0.75pt]    {$\mathbb{Q}\left( i\sqrt{2}\right)$};
\draw (366.26,369.67) node [anchor=north west][inner sep=0.75pt]    {$\mathbb{Q}$};
\draw (434.57,221.91) node [anchor=north west][inner sep=0.75pt]    {$8$};
\draw (330.87,368.06) node [anchor=north west][inner sep=0.75pt]    {$6$};
\draw (274.05,221.19) node [anchor=north west][inner sep=0.75pt]    {$\mathbb{Q}( i)$};
\draw (213.41,222.31) node [anchor=north west][inner sep=0.75pt]    {$8$};
\draw (323.08,70.34) node [anchor=north west][inner sep=0.75pt]    {$14$};
\draw (362.08,52.37) node [anchor=north west][inner sep=0.75pt]    {$\mathbb{Q}\left( i,\sqrt{2}\right) =\mathbb{Q}\left( e^{\frac{2\pi i}{8}}\right)$};
\draw (324.27,223.52) node [anchor=north west][inner sep=0.75pt]    {$10$};
\draw (133.0,213.76) node [anchor=north west][inner sep=0.75pt]    {$\mathbb{Q}\left(\sqrt{2}\right)$};
\end{tikzpicture}
\caption{Graph $G1$ for $Q_8 \rtimes \DZ_4$}
\end{figure}

\begin{figure}[h!]
\centering

\tikzset{every picture/.style={line width=0.75pt}} 

\begin{tikzpicture}[x=0.75pt,y=0.75pt,yscale=-0.8,xscale=0.8]

\draw  [fill={rgb, 255:red, 255; green, 255; blue, 255 }  ,fill opacity=1 ] (422.42,230.48) .. controls (422.42,220.89) and (430.14,213.12) .. (439.67,213.12) .. controls (449.2,213.12) and (456.92,220.89) .. (456.92,230.48) .. controls (456.92,240.07) and (449.2,247.85) .. (439.67,247.85) .. controls (430.14,247.85) and (422.42,240.07) .. (422.42,230.48) -- cycle ;
\draw    (428.8,245.57) -- (349.04,366.03) ;
\draw  [fill={rgb, 255:red, 255; green, 255; blue, 255 }  ,fill opacity=1 ] (318.73,376.63) .. controls (318.73,367.04) and (326.45,359.27) .. (335.98,359.27) .. controls (345.5,359.27) and (353.22,367.04) .. (353.22,376.63) .. controls (353.22,386.23) and (345.5,394) .. (335.98,394) .. controls (326.45,394) and (318.73,386.23) .. (318.73,376.63) -- cycle ;
\draw    (226.2,242.36) -- (323.51,364.42) ;
\draw  [fill={rgb, 255:red, 255; green, 255; blue, 255 }  ,fill opacity=1 ] (200.67,230.48) .. controls (200.67,220.89) and (208.4,213.12) .. (217.92,213.12) .. controls (227.45,213.12) and (235.17,220.89) .. (235.17,230.48) .. controls (235.17,240.07) and (227.45,247.85) .. (217.92,247.85) .. controls (208.4,247.85) and (200.67,240.07) .. (200.67,230.48) -- cycle ;
\draw   (405.16,295.44) -- (389.55,303.81) -- (391.87,286.16) ;
\draw   (270.64,286) -- (274.71,303.33) -- (258.34,296.57) ;
\draw    (224.79,214.28) -- (315.84,86.96) ;
\draw  [fill={rgb, 255:red, 255; green, 255; blue, 255 }  ,fill opacity=1 ] (349.49,80) .. controls (349.35,89.59) and (341.52,97.25) .. (331.99,97.12) .. controls (322.46,96.98) and (314.85,89.1) .. (314.99,79.51) .. controls (315.13,69.92) and (322.96,62.25) .. (332.49,62.39) .. controls (342.01,62.52) and (349.62,70.41) .. (349.49,80) -- cycle ;
\draw    (430.52,215.59) -- (347.72,87.33) ;
\draw   (277.75,152.82) -- (262.39,161.66) -- (264.18,143.95) ;
\draw   (393.28,144.07) -- (395.57,161.73) -- (379.97,153.32) ;
\draw    (331.99,97.12) -- (333.08,213.45) ;
\draw  [fill={rgb, 255:red, 255; green, 255; blue, 255 }  ,fill opacity=1 ] (316.13,232.09) .. controls (316.13,222.5) and (323.85,214.72) .. (333.38,214.72) .. controls (342.91,214.72) and (350.63,222.5) .. (350.63,232.09) .. controls (350.63,241.68) and (342.91,249.45) .. (333.38,249.45) .. controls (323.85,249.45) and (316.13,241.68) .. (316.13,232.09) -- cycle ;
\draw    (334.38,249.06) -- (334.88,359.15) ;

\draw (463.55,213.76) node [anchor=north west][inner sep=0.75pt]    {$\mathbb{Q}\left( i\sqrt{2}\right)$};
\draw (366.26,369.67) node [anchor=north west][inner sep=0.75pt]    {$\mathbb{Q}$};
\draw (434.57,221.91) node [anchor=north west][inner sep=0.75pt]    {$8$};
\draw (330.87,368.06) node [anchor=north west][inner sep=0.75pt]    {$8$};
\draw (274.05,221.19) node [anchor=north west][inner sep=0.75pt]    {$\mathbb{Q}( i)$};
\draw (213.41,222.31) node [anchor=north west][inner sep=0.75pt]    {$8$};
\draw (323.08,70.34) node [anchor=north west][inner sep=0.75pt]    {$14$};
\draw (362.08,52.37) node [anchor=north west][inner sep=0.75pt]    {$\mathbb{Q}\left( i,\sqrt{2}\right) =\mathbb{Q}\left( e^{\frac{2\pi i}{8}}\right)$};
\draw (324.27,223.52) node [anchor=north west][inner sep=0.75pt]    {$10$};
\draw (133.0,213.76) node [anchor=north west][inner sep=0.75pt]    {$\mathbb{Q}\left(\sqrt{2}\right)$};

\end{tikzpicture}
\caption{Graph $G2$ for $Q_8 \rtimes \DZ_4$}
\end{figure}

\section{Constructing row-column duality}
\label{ConstRCD}

In this section we will give an algorithm for going from the $G$-CTST to the $\fusionalg{G}$-CTST and vice versa.

In the $G$-CTST we have a basis of elements $T_{\cclass}$, where $\cclass \in \cclasses{G}$ is a conjugacy class of $G$. The size of the conjugacy class is $\abs{\cclass}$.
\bea 
T_{ \cclass } = \sum_{ g \in \cclass } g 
\eea
The Frobenius form 
\begin{equation}
	g( T_{ \cclass_1} T_{ \cclass_2} ) = \frac{\abs{\cclass_1}}{\abs{G}}\delta_{ \cclass_1, S ( \cclass_2) }, 
\end{equation}
where $S( \cclass_2) $ is the conjugacy class which contains the inverses of the elements in $\cclass_1$. For real groups such as $S_n$, $S( \cclass )  = \cclass$.

The amplitude for three circles going to the vacuum is 
\bea 
&& Z^G_{h=0;\cclass_1, \cclass_2 \cclass_3} ={ 1 \over |G| } \delta ( T_{ \cclass_1 } T_{ \cclass_2} T_{ \cclass_3} ) = 
{ 1 \over |G|^2 } \sum_{ R }   {  | \cclass_1 | | \cclass_2 |  | \cclass_3 | \over d_\irrep }     \chi^\irrep_{ \cclass_1} \chi^\irrep_{ \cclass_2}\chi^\irrep_{ \cclass_3}, \cr 
&&
\eea
where $\chi^\irrep_\cclass $ is the character of a fixed element $g \in \cclass$. 

It is convenient to define  $\widehat { T_{ \cclass } } = { T_{ \cclass} \sqrt{ |G| \over  | \cclass | }  }  $. On this, the bilinear form is 
\bea 
g ( \widehat { T_{ \cclass_1} } , \widehat { T_{ \cclass_2} } ) = \delta_{ \cclass_1 ,S(\cclass_2)},
\eea
and the three-to-null amplitude is
\bea 
&& \widehat{Z}^G_{h=0; \cclass_1, \cclass_2 \cclass_3} = { 1 \over |G| } \delta ( \widehat { T_{ \cclass_1 } } \widehat { T_{ \cclass_2}} 
\widehat {  T_{ \cclass_3} }  ) = 
{  1 \over \sqrt { |G| } } \sum_{ R }   { \sqrt{ |{ \cclass_1}|  | { \cclass_2 }|  | { \cclass_3} | }  \over d_R }     \chi^\irrep_{ \cclass_1} \chi^\irrep_{ \cclass_2}\chi^\irrep_{ \cclass_3} .
\eea

We also know that the eigenvectors  of $T_{ \cclass } $ - viewed as an operator by multiplication in the centre of the group algebra,  are labelled by a label $R$  and their eigenvalues are given by 
\bea 
E_{ \irrep } ( T_{\cclass } )  = { |\cclass| \over d_R } \chi^\irrep_\cclass 
\eea
Note that we can take the point of view that we have a bunch of commuting matrices, with non-degenerate pairing.
$R$ is just a label for the common eigenvectors.  
It follows, for the rescaled versions, 
\bea 
E_\irrep( \widehat{ T_{ \cclass } } ) = { \sqrt {  |  G |   |\cclass | }  \over d_R }  \chi^\irrep_{ \cclass } 
\eea

As we will now prove,
\bea 
\mathrm{Max}_{ \irrep } \, \abs{E_\irrep ( \widehat { T_\cclass } ) }= \sqrt { |G| |\cclass|  } \label{eq: max R}
\eea
In particular, we will prove that
\begin{equation}
	\abs{\chi_{\cclass}^\irrep} \leq d_\irrep,
\end{equation}
from which the result follows.
Consider an irreducible representation $D^\irrep(g)$ of an element $g \in G$ with order $k$ ($g^k = 1$).
The representation is similar to a diagonal matrix $\diag(\lambda_1, \dots, \lambda_{d_\irrep})$. Since $g^k = 1$ we have $(D^\lambda(g))^k = 1$ and therefore $\lambda_i$ are $k$th roots of unity. It follows that $\abs{\lambda_i} = 1$. Now consider
\begin{equation}
	\abs{\chi^\irrep(g)} = \abs{\lambda_1 + \dots \lambda_{d_\irrep}} \leq \abs{\lambda_1} + \dots + \abs{\lambda_{d_\irrep}} = d_\irrep,
\end{equation}
where we used the triangle inequality in the second to last step. Equation \eqref{eq: max R} follows.

The handle creation operator in the idempotent basis is 
\bea 
\Pi = \sum_{ R } { |G|^2 \over d_R^2 } P_R 
\eea
Let us call 
\bea 
\Pi_R = { |G|^2  \over d_R^2 } 
\eea
Let us now define a dualisation of the eigenvalues  by using the components of the handle operator and the maximum of the eigenvalues 
\bea\label{dualEvals}   
\widetilde{ E }_{ R } ( \widehat { T }_{ \cclass } ) && = {1 \over \sqrt { \Pi_R  } }  
E_R ( \widehat{ T}_\cclass ){ 1  \over Max_{ R }\abs{ E_R ( \widehat { T }_\cclass )} }  \cr 
&&  = { d_R \over |G| } { \sqrt { |G|  | \cclass| }  \chi^\irrep_{ \cclass } \over d_R } { 1 \over \sqrt { |G| | \cclass |  } }\cr 
&& = { \chi^\irrep_\cclass \over |G|  } 
\eea
Define the constructive dual amplitudes, by using the dualised eigenvalues, fixing a triple of eigenvectors $R_1, R_2 , R_3 $ and summing over the label $\cclass \in \cclasses{G}$: 
\bea\label{dualityeq1}  
\widetilde { Z}^G_{h=0; R_1 , R_2 , R_3} = \sum_{ \cclass } |G| .  \widetilde{ E }_{ R_1 } ( \widehat { T }_{ \cclass } ) 
\widetilde{ E }_{ R_2 } ( \widehat { T }_{ \cclass } )  \widetilde{ E }_{ R_3 } ( \widehat { T }_{ \cclass } )  . ( Max_{ R } ( \abs*{E_R ( \widehat { T}_\cclass  )})^2 
\eea
Now plug in all the ingredients of the duality formula 
\bea 
&& \widetilde { Z}^G_{h=0; R_1 , R_2 , R_3} = \sum_{ R } |G| . { \chi^{R_1}_\cclass  \over |G|  }{ \chi^{R_2}_\cclass  \over |G|  }{ \chi^{R_3}_\cclass  \over |G|  } . |\cclass| |G| \cr 
&& =  \sum_{R }  { |\cclass | \over |G| } \chi^{R_1}_\cclass  \chi^{R_2}_\cclass \chi^{R_3}_\cclass 
\eea
Note that the dualized amplitudes are exactly the amplitudes of the $\fusionalg{G}$-CTST
\bea 
\widetilde { Z}^G_{h=0; R_1 , R_2 , R_3}   = Z^{\fusionalg{G}}_{h=0; R_1 , R_2 , R_3 } 
\eea
The structure constants of the dual ($\fusionalg{G}$)  theory have been constructed using a formula involving an appropriate form of the eigenvalues of the class-algebra theory. We hope that the same construction ( in terms of eigenvalues ) starting from the dual theory should produce back the class algebra theory. 

So let us start now with the $\fusionalg{G}$ theory. 
Here we have 
\bea 
\delta ( a_{R_1} a_{R_2}  )   = \delta_{ R_1 ,  S ( R_2)  } 
\eea
where $S(R_2)$ is the complex conjugate of $R_2$. For real groups such as $S_n$, $S(R_2 ) = R_2$. The three-to-null amplitude is
\bea 
Z^{\fusionalg{G}}_{h=0; R_1 , R_2 , R_3 } = \sum_{ \cclass }  { |\cclass| \over |G| } \chi^{R_1}_{ \cclass } \chi^{R_2}_{ \cclass } \chi^{R_3}_{ \cclass } 
\eea 
Look at the eigenvalues 
\bea 
E_\cclass ( a_R ) = \chi^\irrep_\cclass
\eea
The maximum of the absolute value is
\bea 
Max_{ \cclass } \abs{E_{ \cclass } ( a_R )}  = d_R,
\eea
using the same argument as in the previous case.
The handle creation eigenvalues are 
\bea 
\Pi_\cclass = { |G| \over | \cclass | } 
\eea
Use same formula as above for rescaling the eigenvalue 
\bea 
\widetilde { E }_\cclass ( a_R ) && = { 1 \over \sqrt { \Pi_\cclass } } . E_\cclass ( a_R ) . { 1 \over Max_{ \cclass } (\abs{ E_K ( a_R )} ) } \cr 
&& =    { \sqrt{ | \cclass | } \over \sqrt { |G| } } {  \chi^\irrep_\cclass \over d_R } 
\eea
The dualized amplitude is defined as 
\bea\label{dualityeq2}
&& \widetilde { Z}^{\fusionalg{G}}_{h=0;\cclass_1, \cclass_2, \cclass_3 } = \sum_{ R } |G| . \widetilde { E }_{\cclass_1}  ( a_R )  \widetilde { E }_{\cclass_2}  ( a_R ) \widetilde { E }_{\cclass_3}  ( a_R ) .  ( Max_{ \cclass } \abs{E_{ \cclass } ( a_R )}  )^2 
\eea
Note that this is the same dualization formula as \eqref{dualityeq1} by taking the $\fusionalg{G}$ as the starting point (rather than the conjugacy class algebra as the starting point). Simplify 
\bea
&& \widetilde { Z}^{\fusionalg{G} }_{h=0;\cclass_1, \cclass_2 , \cclass_3 }  = \sum_{ R } |G| .   { \sqrt{ | \cclass  | } \over \sqrt { |G| } } {  \chi^\irrep_{\cclass_1}  \over d_R } { \sqrt{ | \cclass  | } \over \sqrt { |G| } } {  \chi^\irrep_{\cclass_2}  \over d_R } { \sqrt{ | \cclass_3  | } \over \sqrt { |G| } } {  \chi^\irrep_{\cclass_3}  \over d_R } 
. d_R^2 \cr 
&& = \sum_{ R }  { 1 \over \sqrt{ |G| }  }  \sqrt{ | \cclass_1  | }  \sqrt{ | \cclass_2  | }  \sqrt{ | \cclass_3  | } {  \chi^{R}_{\cclass_1}   \chi^\irrep_{\cclass_2}   \chi^\irrep_{\cclass_3}  \over d_R } 
\eea
Note that the amplitude defined by dualising the $\fusionalg{G}$ theory is exactly the one of the $G$-CTST
\bea
\widetilde { Z}^{\fusionalg{G} }_{h=0; \cclass_1, \cclass_2 , \cclass_3 }  = \widehat{Z}^G_{h=0;\cclass_1, \cclass_2, \cclass_3}. 
\eea
So we have applied a duality transformation -  for any system of $k$ commuting $ k \times k $ matrices with non-degenerate bilinear form  - defined by \eqref{dualityeq2}.

We remark that the above dualisation process lands at the amplitude in the normalized basis for the conjugacy class algebra. To get the amplitude in the original, unnormalized, basis we use the combination
\begin{equation}
	\sum_{ R } |G| .\frac{ \widetilde { E }_{\cclass_1}  ( a_R )  \widetilde { E }_{\cclass_2}  ( a_R ) \widetilde { E }_{\cclass_3}  ( a_R ) }{\sqrt{\Pi_{\cclass_1} \Pi_{\cclass_2}  \Pi_{\cclass_3}}}.  ( Max_{ \cclass } \abs{E_{ \cclass } ( a_R ) } )^2 = \sum_{ R }  { 1 \over { |G| }  }  { | \cclass_1  | }  { | \cclass_2  | }  { | \cclass_3  | } {  \chi^{R}_{\cclass_1}   \chi^\irrep_{\cclass_2}   \chi^\irrep_{\cclass_3}  \over d_R }. \label{dualityeq3}
\end{equation}
This is equal to the unnormalized amplitude ${Z}_{h=0;\cclass_1, \cclass_2, \cclass_3}$.

These dualisation procedures have some intriguing properties. Let us call the dualisation procedure \eqref{dualityeq1}
\begin{equation}
	S_1(Z^G) =Z^{\fusionalg{G}},
\end{equation}
and the dualisation \eqref{dualityeq2}
\begin{equation}
	S_2(Z^{\fusionalg{G}}) = \widehat{Z}^G.
\end{equation}
Taking the point of view that we have a set of commuting matrices with non-degenerate bilinear pairing, we observed that $S_1, S_2$ are actually the same procedure. However, since
\begin{equation}
	S_2(S_1(Z^G)) = \widehat{Z}^G \neq Z^G,
\end{equation}
they are not exact inverses. On the other hand, let the dualisation \eqref{dualityeq3}
\begin{equation}
	S_3(Z^{\fusionalg{G}}) = Z^G.
\end{equation}
It has the property that
\begin{equation}
	S_3(S_1(Z^G)) = Z^G,
\end{equation}
but $S_3$ is not the same procedure as $S_1$. It would be interesting to investigate if there is 
  a dualisation procedure that satisfies both properties in the future.

\section{Conclusions}
In this paper we have introduced a new class of combinatorial topological string theories ($\fusionalg{G}$-CTSTs) based on fusion algebras $\fusionalg{G}$ of a finite group $G$. They are dual to $G$-CTSTs based on Dijkgraaf-Witten theories ($G$-TQFTs). $G$-CTSTs have been used to give constructive proofs of integrality properties of characters, otherwise only accessible by Galois theoretic techniques. Integrality properties derived from $G$-CTSTs concern column sums or sums over irreducible normalized characters for a fixed conjugacy class. Integrality properties derived from $\fusionalg{G}$-CTSTs concern row sums, or fixed irreducible characters summed over conjugacy classes. In this sense, $G$-CTSTs  and $\fusionalg{G}$-CTSTs are dual.  
We have made a start towards an explicit construction of the duality map, but there remain issues to resolve in using this construction in getting the map of integrality results from rows to columns.

We reviewed the application of Galois theory to character tables and showed how it can be used to derive refined integrality properties of character sums. Naturally, we asked if such refinements can be constructed using CTST techniques. We answered this question in the affirmative for large classes of refined sums. The derivation of these refinements in the context of CTSTs are novel for $G$-CTSTs as well as $\fusionalg{G}$-CTSTs. The result relies on a lemma on rationality of inverse Vandermonde matrices, which we proved in appendix \ref{apx: inverse vandermonde lemma}.

Following the observation of the refined integer sums, we defined generalized partitions of the integer row and column sums of character tables. The row sums of characters are positive integers while the  column sums are integers, not necessarily positive. The row sums are the subject of a combinatorial construction question highlighted in \cite{StanleyPositivityProblems}. We showed that the row and column sums naturally come equipped with the structure of generalised partitions, determined by integrality properties of the character table. A refinement of the combinatorial question is to find combinatorial constructions  for these generalised partitions. It is also worth observing that the column sums of the normalized characters are positive integers. These would also admit generalised partitions along similar lines. 



Two very natural extensions of our work involve extending the type of TQFTs considered. $G$-CTSTs have been extended to twisted $G$-CTSTs defined by twisted Dijkgraaf-Witten theories (twisted $G$-TQFTs) in \cite{GCTST2}. These are defined using a group $G$ and a 3-cocycle in $H^3(G,U(1))$.  It would be interesting to extend our notion of Row-Column duality of CTSTs to twisted $G$-TQFTs. Such an extension will provide a physical setting to analyse number theoretic properties of characters of projective representations of a group. 

The TQFTs considered in this paper were of the closed type. They compute topological invariants of manifolds with closed boundaries. A TQFT that computes topological invariants of manifolds which can have corners/open boundaries is called an open-closed TQFT. For closed TQFTs determined by semi-simple Frobenius algebras, which includes $G$-TQFTs, there exists a classification of the possible consistent extensions to an open-closed TQFT \cite{MooreSegal}. An extension of the closed $G$-TQFT is described by the full non-commutative group algebra $\mathbb{C}(G)$. It would be interesting to consider the open-closed extensions of $\fusionalg{G}$-TQFTs and their implications for Row-Column duality. A closely related direction is to consider extended operators in 2D TQFTs. It would be interesting to study the number theoretic properties of the group $G$ contained in the correlation functions of line operators in the $G$-TQFT and $R(G)$-TQFT. Properties of the gauge group that can be constructed from topological line and surface operators of higher-dimensional quantum field theories were studied in \cite{Radhakrishnan:2023zcq}.

The Galois group naturally acts on the rows and columns of the character table. Invariants under this action correspond to integer rows and columns, respectively. The action of the Galois groups on irreducible representations and conjugacy classes can be used to define symmetries (in the sense used in \cite{Gukov:2021swm} \cite{Bhardwaj_2022})  of the $G$-TQFT and $\fusionalg{G}$-TQFT  respectively \cite{prrupcoming}. It will be interesting to study states in the Hilbert space of the G-TQFT and $\fusionalg{G}$-TQFT which are invariant under this symmetry and relate them to the various sums of characters studied in this paper.

We explained how Galois theory gives a description of sufficient conditions of the number of integer rows to agree with the number of integer columns of the character table of a group $G$. More generally, we considered the number of rows and columns belonging to particular extensions of the rationals. Formulating a CTST approach to these kinds of integrality questions is an interesting problem.

The $G$-CTST construction in subsection \ref{subsec: integer level set sums} proves that sums of normalized characters over sets of irreducible representations
\begin{equation}
	\{\irrep \in \irreps{G} \, \vert \, \frac{\chi^{\irrep}_{D_i}}{d_\irrep} = \frac{\chi^{S_i}_{D_i}}{d_{S_i}}\}
\end{equation}
are integral. Meanwhile, Galois theory allowed us to prove integrality over sums of irreducible representations
\begin{equation}
	\{\irrep \in \irreps{G} \, \vert \, {\chi^{\irrep}_{D_i}} = n_i\},
\end{equation}
for integers $n_i$. A good future problem is to reconstruct such level sets within the framework of CTST.

Another problem in finite group representation theory which has a direct link with CTST constructions is  Harada's conjecture, which we now describe. The ratio of determinants of the matrix of structure constants in $G$-TQFT and $R(G)$-TQFT is given by the product of conjugacy class sizes divided by the product of dimensions of irreducible representations
\bea 
{ \prod_\cclass  \abs{\cclass} \over \prod_R d_R } 
\eea
This is explained in Appendix \ref{sec:Harada}. The integrality of this ratio is the subject of Harada's conjecture. A better understanding of the construction of row-column duality (developing section \ref{ConstRCD}) may provide a physical perspective on the conjecture. A broader question is to characterize the invariants of the row-column duality between $G$-CTST and $R(G)$-CTST. 

A similar philosophy to the present paper of using stringy constructions to approach problems in combinatorial representation theory was taken in \cite{bipartitegraphs}. Bipartite graphs connected to string theory were used to give a realisation of  Kronecker coefficients, which are tensor product multiplicities for symmetric groups, in terms of the counting of null vectors of combinatorially constructed integer matrices. The construction of such null vectors is amenable to  known combinatorial algorithms for integer matrices. The null vector equations are related to 
eigenvalue problems which can be solved using  efficient quantum algorithms based on quantum phase estimation \cite{Geloun:2023zqa} (see also closely related work \cite{Bravyi:2023veg}). This is along  the lines of the use of quantum algorithms for efficient solution of linear algebra problems \cite{HHL}. The CTST algorithms of the present paper, which are based on matrix equations, would similarly be amenable to quantum algorithms. Finite groups arise as hidden symmetries in  correlator computations for matrix and tensor models (see for example recent work  \cite{Mironov:2022fsr, Ramgoolam:2023vyq}) with continuous group gauge symmetries. This gives an avenue for wider applications  in holography and gauge-string duality for algorithms  related to finite group representation theory. 

\vspace{1cm}
\centerline{\bf{Acknowledgments}}
 We thank Anindya Banerjee, George Barnes, Joseph Ben Geloun, Matthew Buican, Robert de Mello Koch,Gerard Duchamp,  Eric Sharpe  for insightful discussions on the subject of the paper. SR is supported by the STFC consolidated grant ST/P000754/1 ``String Theory, Gauge Theory and Duality''.
S.R.  acknowledges the support of the Institut Henri Poincaré (UAR 839 CNRS-Sorbonne Université), and LabEx CARMIN (ANR-10-LABX-59-01). SR also acknowledges the  support of the Perimeter Institute for Theoretical Physics during a visit in the final stages of completion of the paper.  Research at Perimeter Institute is supported by the Government of Canada through Industry Canada and by the Province of Ontario through the Ministry of Economic Development and Innovation. R.R. thanks ICTP for support. R.R.’s work was partially funded by the Royal Society under the grant “New Aspects of Conformal and Topological Field Theories Across Dimensions”. R.R. is grateful to ICMS, Edinburgh for a stimulating environment during the conference `Gauge Fields in Arithmetic, Topology and Physics' where some results in this paper were announced. 

\vskip2cm 

\appendix

\section{Frobenius structure of $\fusionalg{G}$-TQFT } \label{apx: fusion tqft}
We define a two-dimensional TQFT using the algebra of representations of a finite group $G$, denoted $\fusionalg{G}$. As a vector space, $\fusionalg{G}$ has a basis $a_\irrep$ labelled by the set $\irreps{G}$ of isomorphism classes of irreducible representations of $G$. The product $\mu: \fusionalg{G} \otimes \fusionalg{G} \rightarrow \fusionalg{G}$ is given by
\bea 
	\mu(a_{ R_1 }, a_{ R_2}) = \sum_{ R_3} N_{ R_1 R_2 }^{ R_3 } a_{ R_3 },
\eea
where
\bea
N_{R_1 R_2}^{R_3} = \frac{1}{\abs{G}}\sum_{g \in G} \chi^{R_1}(g)\chi^{R_2}(g)\overline{\chi^{R_3}(g)}. \label{eq: RG struct const}
\eea
We introduce the graphical notation
\begin{equation}
	N_{R_1 R_2}^{R_3} = \vcenter{\hbox{
			\begin{tikzpicture}[tqft/cobordism/.style={draw}, tqft/every boundary component/.style={draw, dashed}]
				\pic[tqft/pair of pants, name=pop
				];
				\node[above] at (pop-incoming boundary 1.north) {$R_3$};
				\node[below] at (pop-outgoing boundary 1.south) {$R_1$};
				\node[below] at (pop-outgoing boundary 2.south) {$R_2$};
	\end{tikzpicture}}}
\end{equation}
The algebra is commutative since $R_1 \otimes R_2 \cong R_2 \otimes R_1$. It is also unital, the trivial representation, denoted $R_0$, plays the role of the identity element. Equivalently, we have a unit $\eta: \mathbb{C} \rightarrow R(G)$ determined by
\bea
\eta(1) = a_{R_0} =  \vcenter{\hbox{
		\begin{tikzpicture}[tqft/cobordism/.style={draw}, tqft/every boundary component/.style={draw, dashed}]
			\pic[tqft/cup, name=c
			];
			\node[above] at (c-incoming boundary 1.north) {$R_0$};
\end{tikzpicture}}}
\eea
and $\eta(c) = c\eta(1)$ for $c \in \mathbb{C}$. This definition of an algebra is the one used in \cite[Definition 2.1.18]{Kock}.

The algebra $R(G)$ is turned into a Frobenius algebra by adding a non-degenerate bilinear form $g: R(G) \otimes R(G) \rightarrow \mathbb{C}$ to its structure (see \cite[Definition 2.2.5]{Kock}). We use the definition
\bea
g_{R_1 R_2} = g(a_{R_1}, a_{R_2}) = \frac{1}{\abs{G}} \sum_{g \in G} \chi^{R_1}(g){\chi^{R_2}(g)} = \delta_{R_1 \overline{R_2}}, \label{eq: RG biform}
\eea
where $\overline{R}$ is the complex conjugate representation of $R$.
Graphically, this corresponds to
\begin{equation}
	g_{R_1 R_2} =  \vcenter{\hbox{
			\begin{tikzpicture}[tqft/cobordism/.style={draw}, tqft/every boundary component/.style={draw, dashed}]
				\pic[tqft, name=g, outgoing boundary components=2, incoming boundary components=0
				];
				\node[below] at (g-outgoing boundary 1.south) {$R_1$};
				\node[below] at (g-outgoing boundary 2.south) {$R_2$};
	\end{tikzpicture}}}
\end{equation}
It has an inverse
\bea
g^{R_1 R_2} = g^{-1}(a_{R_1}, a_{R_2}) = \frac{1}{\abs{G}} \sum_{g \in G} \overline{\chi^{R_1}(g)}\overline{\chi^{R_2}(g)} = \delta^{\overline{R_1} {R_2}} =  \vcenter{\hbox{
		\begin{tikzpicture}[tqft/cobordism/.style={draw}, tqft/every boundary component/.style={draw, dashed}]
			\pic[tqft, name=ginv, outgoing boundary components=0, incoming boundary components=2
			];
			\node[above] at (ginv-incoming boundary 1.north) {$R_1$};
			\node[above] at (ginv-incoming boundary 2.north) {$R_2$};
\end{tikzpicture}}}. \label{eq: RG biform inverse}
\eea
Graphically,
\begin{equation}
	\vcenter{\hbox{
			\begin{tikzpicture}[tqft/cobordism/.style={draw}, tqft/every boundary component/.style={draw, dashed}]
				\pic[tqft, name=g, outgoing boundary components=2, incoming boundary components=0
				];
				\pic[tqft, name=ginv,  anchor=incoming boundary 2, at=(g-outgoing boundary 1), outgoing boundary components=0, incoming boundary components=2
				];
				\node[above] at (ginv-incoming boundary 1.north) {$R_1$};
				\node[below] at (g-outgoing boundary 2.south) {$R_2$};
	\end{tikzpicture}}} =  \vcenter{\hbox{
			\begin{tikzpicture}[tqft/cobordism/.style={draw}, tqft/every boundary component/.style={draw, dashed}]
				\pic[tqft/cylinder, name=id
				];
				\node[above] at (id-incoming boundary 1.north) {$R_1$};
				\node[below] at (id-outgoing boundary 1.south) {$R_2$};
	\end{tikzpicture}}}=\delta_{R_2}^{R_1}.
\end{equation}
Note that the definition \eqref{eq: RG biform} gives a bilinear form satisfying
\bea
g(a_{R_1} a_{R_2}, a_{R_3}) = g(a_{R_1}, a_{R_2}a_{R_3}),
\eea
or
\begin{equation}
	\vcenter{\hbox{
			\begin{tikzpicture}[tqft/cobordism/.style={draw}, tqft/every boundary component/.style={draw, dashed}]
				\pic[tqft/pair of pants, name=pop
				];
				\pic[tqft, name=g, outgoing boundary components=2, incoming boundary components=0, anchor=outgoing boundary 1, at=(pop-incoming boundary)
				];
				\node[below] at (g-outgoing boundary 2.south) {$R_3$};
				\node[below] at (pop-outgoing boundary 1.south) {$R_1$};
				\node[below] at (pop-outgoing boundary 2.south) {$R_2$};
	\end{tikzpicture}}} = \vcenter{\hbox{
			\begin{tikzpicture}[tqft/cobordism/.style={draw}, tqft/every boundary component/.style={draw, dashed}]
				\pic[tqft/pair of pants, name=pop
				];
				\pic[tqft, name=g, outgoing boundary components=2, incoming boundary components=0, anchor=outgoing boundary 2, at=(pop-incoming boundary)
				];
				\node[below] at (g-outgoing boundary 1.south) {$R_1$};
				\node[below] at (pop-outgoing boundary 1.south) {$R_2$};
				\node[below] at (pop-outgoing boundary 2.south) {$R_3$};
	\end{tikzpicture}}}
\end{equation}

The bilinear form defines a counit $\varepsilon: R(G) \rightarrow \mathbb{C}$ by
\bea
\varepsilon(a_R) = g(a_R, a_{R_0}) = \delta_{R \overline{R_0}} = \delta_{R {R_0}} = \vcenter{\hbox{
		\begin{tikzpicture}[tqft/cobordism/.style={draw}, tqft/every boundary component/.style={draw, dashed}]
			\pic[tqft/cap, name=cc
			];
			\node[below] at (cc-outgoing boundary 1.south) {$R$};
\end{tikzpicture}}}, \label{eq: fusion counit}
\eea
where $R_0=\overline{R_0}$ is the trivial representation (see \cite[2.2.3]{Kock}). Graphically we have,
\begin{equation}
	\vcenter{\hbox{
			\begin{tikzpicture}[tqft/cobordism/.style={draw}, tqft/every boundary component/.style={draw, dashed}]
				\pic[tqft/cap, name=cc
				];
				\node[below] at (cc-outgoing boundary 1.south) {$R$};
	\end{tikzpicture}}} = \vcenter{\hbox{
			\begin{tikzpicture}[tqft/cobordism/.style={draw}, tqft/every boundary component/.style={draw, dashed}]
				\pic[tqft, name=g, outgoing boundary components=2, incoming boundary components=0
				];
				\node[below] at (g-outgoing boundary 1.south) {$R$};
				\node[below] at (g-outgoing boundary 2.south) {$R_0$};
	\end{tikzpicture}}}
\end{equation}

A Frobenius algebra has a canonical \cite[Definition 2.3.16]{Kock} coproduct $\Delta: R(G) \rightarrow R(G) \otimes R(G)$ defined as follows,
\bea
\Delta(a_{R_3}) = \sum_{R_1,R_2} N^{R_1 R_2}_{R_3} a_{R_1} \otimes a_{R_2},
\eea
where
\bea
N^{R_1 R_2}_{R_3} =  g^{R_1 R_4} N_{ R_4 R_3 }^{ R_2 } =  N_{ \overline{R_1} R_3 }^{ R_2 }.
\eea
In terms of cobordisms we have
\begin{equation}
	N^{R_1 R_2}_{R_3}=	\vcenter{\hbox{
			\begin{tikzpicture}[tqft/cobordism/.style={draw}, tqft/every boundary component/.style={draw, dashed}]
				\pic[tqft/reverse pair of pants, name=rpop
				];
				\node[above] at (rpop-incoming boundary 1.north) {$R_1$};
				\node[above] at (rpop-incoming boundary 2.north) {$R_2$};
				\node[below] at (rpop-outgoing boundary 1.south) {$R_3$};
	\end{tikzpicture}}} = 
	\vcenter{\hbox{
			\begin{tikzpicture}[tqft/cobordism/.style={draw}, tqft/every boundary component/.style={draw, dashed}]
				\pic[tqft/pair of pants, name=pop
				];
				\pic[tqft, name=ginv, outgoing boundary components=0, incoming boundary components=2, anchor=incoming boundary 2, at=(pop-outgoing boundary 1)];
				\node[above] at (ginv-incoming boundary 1.north) {$R_1$};
				\node[above] at (pop-incoming boundary 1.north) {$R_2$};
				\node[below] at (pop-outgoing boundary 2.south) {$R_3$};
	\end{tikzpicture}}}
\end{equation}

General genus $h$ partition functions come from iterated use of the coproduct $\Delta$ followed by the product $\mu$. To get a closed manifold we cap off both ends with a unit $\eta$ and counit $\varepsilon$,
\bea
Z^{\fusionalg{G}}_{h} = \varepsilon \circ (\mu \circ \Delta)^h \circ \eta(1) = \begin{aligned}
	&\vcenter{\hbox{\begin{tikzpicture}[tqft/cobordism/.style={draw}, tqft/every boundary component/.style={draw, dashed}]
				\pic[tqft/cap, name=cc];
				\pic[tqft/pair of pants, name=pop, anchor=incoming boundary, at=(cc-outgoing boundary)
				];
				\pic[tqft/reverse pair of pants, name=rpop, anchor=incoming boundary 1, at=(pop-outgoing boundary 1)];
	\end{tikzpicture}}} \\
	&\quad \quad \quad \vdots \\
	&\vcenter{\hbox{\begin{tikzpicture}[tqft/cobordism/.style={draw}, tqft/every boundary component/.style={draw, dashed}]
				\pic[tqft/pair of pants, name=pop
				];
				\pic[tqft/reverse pair of pants, name=rpop, anchor=incoming boundary 1, at=(pop-outgoing boundary 1)];
				\pic[tqft/cup, anchor=outgoing boundary, at=(rpop-outgoing boundary 1)];
	\end{tikzpicture}}} 
\end{aligned}
\eea

To evaluate this, note that
\begin{align}
	\mu \circ \Delta(a_{R_1}) &= \sum_{R_2,R_3} N_{R_1}^{R_2 R_3} \mu(a_{R_2} \otimes a_{R_3}) \\
	&= \sum_{R_2, R_3, R_4} N_{R_1}^{R_2 R_3} N_{R_2 R_3}^{R_4} a_{R_4} \\
	&= \sum_{R_2, R_3, R_4} N_{\overline{R_2} R_1}^{R_3} N_{R_2 R_3}^{R_4} a_{R_4}.
\end{align}
We define
\bea
N_{R_1}^{R_4} = \sum_{R_2, R_3} N_{R_1}^{R_2 R_3} N_{R_2 R_3}^{R_4},
\eea
which is equal to
\begin{align}
	N_{R_1}^{R_4}
	&=\frac{1}{\abs{G}^2}\sum_{R_2, R_3, g,h} \chi^{R_1}(g) \overline{\chi^{R_2}(g)} \overline{\chi^{R_3}(g)} \chi^{R_2}(h) \chi^{R_3}(h) \overline{\chi^{R_4}(h)} \\
	&=\sum_{R_2, R_3, \cclass_1, \cclass_2}\frac{\abs{\cclass_1}\abs{\cclass_2} }{\abs{G}^2} \chi^{R_1}_{\cclass_1} \overline{\chi^{R_2}_{\cclass_1}} \overline{\chi^{R_3}_{\cclass_1}} \chi^{R_2}_{\cclass_2} \chi^{R_3}_{\cclass_2} \overline{\chi^{R_4}_{\cclass_2}} \\
	&=\sum_{\cclass_1, \cclass_2} \chi^{R_1}_{\cclass_1} (\delta_{\cclass_1 \cclass_2})^2 \overline{\chi^{R_4}_{\cclass_2}} \\
	&=\sum_\cclass \chi^{R_1}_{\cclass}  \overline{\chi^{R_4}_{\cclass}}.
\end{align}

Using the above, we can write the genus $h$ partition function as
\bea
Z^{\fusionalg{G}}_{h} = \sum_{R_1, \dots, R_{h-1}} N_{R_0}^{R_1} N_{R_1}^{R_2} \dots N_{R_{h-1}}^{R_0},
\eea
which is equal to
\bea
\sum_{\substack{\cclass_1, \dots, \cclass_h \\ R_1, \dots, R_{h-1}}} \chi^{R_0}_{\cclass_1} \overline{\chi^{R_1}_{\cclass_1}}  \chi^{R_1}_{\cclass_2} \overline{\chi^{R_2}_{\cclass_2}} \dots  \chi^{R_{h-2}}_{\cclass_{h-2}} \overline{\chi^{R_{h-1}}_{\cclass_{h-1}}} \chi^{R_{h-1}}_{\cclass_h} \overline{\chi^{R_0}_{\cclass_h}}.
\eea
Summing over the representations and using
\bea
\sum_{R} \chi^{R}_{\cclass_1} \chi^\irrep_{\cclass_2} = \frac{\abs{G}}{\abs{\cclass_1}}  \delta_{\cclass_1 \cclass_2}
\eea
together with $\chi^{R_0}_\cclass = 1$ gives
\bea
Z^{\fusionalg{G}}_h = \sum_{\cclass_1, \dots, \cclass_h} \frac{\delta_{\cclass_1 \cclass_2} \abs{G}}{\abs{\cclass_1}} \dots \frac{\delta_{\cclass_{h-1} \cclass_h} \abs{G}}{\abs{\cclass_h}} = \sum_{\cclass} \qty(\frac{\abs{G}}{\abs{\cclass}})^{h-1} = \sum_\cclass \Sym{\cclass}^{h-1}. \label{eq: fusion genus h}
\eea

The Frobenius structure can also be used to compute partition functions of closed manifolds with boundary.
As a simple example, the genus $h$ partition function with a single in-boundary labeled by $R$ is given by
\begin{align}
	Z^{\fusionalg{G}}_{h; R} &= \varepsilon \circ (\mu \circ \Delta)^h(a_{R}) = \sum_{R_1, \dots, R_{h-1}} N_{R}^{R_1} N_{R_1}^{R_2} \dots N_{R_{h-1}}^{R_0} = \sum_{\cclass} \qty(\frac{\abs{G}}{\abs{\cclass}})^{h-1} \chi^\irrep_\cclass \\
	&=\begin{aligned}
		&\vcenter{\hbox{\begin{tikzpicture}[tqft/cobordism/.style={draw}, tqft/every boundary component/.style={draw, dashed}]
					\pic[tqft/cap, name=cc];
					\pic[tqft/pair of pants, name=pop, anchor=incoming boundary, at=(cc-outgoing boundary)
					];
					\pic[tqft/reverse pair of pants, name=rpop, anchor=incoming boundary 1, at=(pop-outgoing boundary 1)];
		\end{tikzpicture}}} \\
		&\quad \quad \quad \vdots \\
		&\vcenter{\hbox{\begin{tikzpicture}[tqft/cobordism/.style={draw}, tqft/every boundary component/.style={draw, dashed}]
					\pic[tqft/pair of pants, name=pop
					];
					\pic[tqft/reverse pair of pants, name=rpop, anchor=incoming boundary 1, at=(pop-outgoing boundary 1)];
					\node[below] at (rpop-outgoing boundary 1.south) {$R$};
		\end{tikzpicture}}} 
	\end{aligned}
\end{align}
The partition function with $n$ in-boundaries is given by
\bea
Z^{\fusionalg{G}}_{h; R_1, \dots, R_n} &= \varepsilon \circ (\mu \circ \Delta)^h \mu^{n-1}(a_{R_1} \otimes \dots \otimes a_{R_n}). \label{eq: n-boundary Z}
\eea

To simplify this we first need to evaluate
\begin{align}
	\mu^{n-1}(a_{R_1} \otimes \dots \otimes a_{R_n}) = N_{R_1 R_2}^{R_1'} N_{R_1' R_3}^{R_2'} \dots N_{R_{n-2}' R_n}^{R_{n-1}'} a_{R_{n-1}'}.
\end{align}
We define
\bea
N_{R_1 R_2 \dots R_n}^{R} = N_{R_1 R_2}^{R_1'} N_{R_1' R_3}^{R_2'} \dots N_{R_{n-2}' R_n}^{R}.
\eea
and graphically represent it by the cobordism
\begin{equation}
	N_{R_1 R_2 \dots R_n}^{R} = \vcenter{\hbox{\begin{tikzpicture}[tqft/cobordism edge/.style={draw}]
				\pic[tqft, name=nbord, incoming boundary components = 1, offset=-1.5,
				outgoing boundary components = 4, 
				outgoing upper boundary component 1/.style={draw},
				outgoing lower boundary component 1/.style={draw,dashed},
				outgoing upper boundary component 2/.style={draw},
				outgoing lower boundary component 2/.style={draw,dashed},
				outgoing upper boundary component 3/.style={draw, dotted},
				outgoing lower boundary component 3/.style={draw, dotted,thin},		
				outgoing upper boundary component 4/.style={draw},
				outgoing lower boundary component 4/.style={draw,dashed},	
				incoming upper boundary component 1/.style={draw},
				incoming lower boundary component 1/.style={draw, dashed},	
				between outgoing 1 and 2/.style={draw},
				between outgoing 2 and 3/.style={dotted},
				between outgoing 3 and 4/.style={dotted},
				];
				\node[above] at (nbord-incoming boundary 1.north) {$R$};
				\node[below] at (nbord-outgoing boundary 1.south) {$R_1$};
				\node[below] at (nbord-outgoing boundary 2.south) {$R_2$};
				\node[below] at (nbord-outgoing boundary 4.south) {$R_n$};
	\end{tikzpicture}}}
\end{equation}
Note that
\begin{align}
	N_{R_1 R_2 R_3}^{R_2'} &= N_{R_1 R_2}^{R_1'} N_{R_1' R_3}^{R_2'} \\
	&= \frac{1}{\abs{G}^2}\sum_{g_1, g_2} \chi^{R_1}(g_1) \chi^{R_2}(g_1) \overline{\chi^{R_1'}(g_1)} \chi^{R_1'}(g_2) \chi^{R_3}(g_2) \overline{\chi^{R_2'}(g_2)} \\
	&=\sum_{\cclass_1, \cclass_2} \frac{\abs{\cclass_1}\abs{\cclass_2}}{\abs{G}^2}\chi^{R_1}_{\cclass_1} \chi^{R_2}_{\cclass_1} \overline{\chi^{R_1'}_{\cclass_1}} \chi^{R_1'}_{\cclass_2} \chi^{R_3}_{\cclass_2} \overline{\chi^{R_2'}_{\cclass_2}} \\
	&=\frac{1}{\abs{G}}\sum_{g} \chi^{R_1}(g)\chi^{R_2}(g)\chi^{R_3}(g)\overline{\chi^{R_2'}(g)}.
\end{align}
Therefore,
\bea
N_{R_1 R_2 \dots R_n}^{R} = \frac{1}{\abs{G}} \sum_g \chi^{R_1}(g) \chi^{R_2}(g) \dots \chi^{R_n}(g) \overline{\chi^{R}(g)}.
\eea
Going back to \eqref{eq: n-boundary Z} we have
\begin{align}
	Z^{\fusionalg{G}}_{h; R_1 \dots R_n} &= \sum_R N_{R_1 R_2 \dots R_n}^{R} \varepsilon \circ (\mu \circ \Delta)^h (a_R) = \sum_R N_{R_1 R_2 \dots R_n}^{R} Z_{h; R} \\
	&=\sum_\cclass \qty(\frac{\abs{G}}{\abs{\cclass}})^{h-1} \chi^{R_1}_\cclass \chi^{R_2}_\cclass \dots \chi^{R_n}_\cclass. \label{eq: fusion genus h n boundaries}
\end{align}

\section{Inverse Vandermonde lemma} \label{apx: inverse vandermonde lemma}
{In equations \eqref{eq: inv vand lemma},\eqref{eq: vander monde used GCTST} and \eqref{eq: vandermonde used RGCTST} we used the following lemma to prove integrality of character sums.}
\setcounter{lemma}{0}
\begin{lemma}
Let $Z$ be a diagonalizable $K \times K$ matrix of non-negative integers with eigenvalues $\lambda_1 \dots, \lambda_K$. Let the set of distinct eigenvalues be $\{z_1, \dots, z_k\}$. Note that $k \leq K$ generally.
Define the $k \times k$ Vandermonde matrix
\begin{equation}
	V_{b,i} = z_i^b.
\end{equation}
In general, the eigenvalues $z_1, \dots, z_k$ are algebraic integers.
But suppose $z_j$ is rational, then the inverse Vandermonde $V^{-1}$ has rational entries in the $j$th row:
\begin{equation}
	(V^{-1})_{j,b} \in \mathbb{Q} \quad \forall \, b \in \{1,\dots,k\}. \label{eq: inv vand lemma apx}
\end{equation}
In this paper we use this lemma for the special case of $z_j$ being an integer.
\end{lemma}

Before we go into the technical details involved in proving this lemma, let us outline the proof.
We will use the fact that the matrix elements $(V^{-1})_{i,b}$ of a Vandermonde matrix can be written in terms of elementary symmetric polynomials (see section 1.2.3 exercise 40 of \cite{TAOCP1})
\begin{equation}
	(V^{-1})_{j,b} = \frac{(-1)^{k-b}e_{k-b}(z_1, \dots, z_{j-1}, z_{j+1}, \dots z_k )}{z_j \prod_{i\neq j} (z_j - z_i)},
\end{equation}
where $e_m$ is the $m$th elementary symmetric polynomial. Given this, it remains to prove that the these elementary symmetric polynomials are rational. The strategy will be to investigate the polynomial
\begin{equation}
	q_{Z,j}(x) = \prod_{i\neq j} (x-z_i) =  \sum_{i=0}^{k-1} (-1)^i e_i(z_1, \dots, z_{j-1},z_{j+1},\dots,z_k)x^{k-1-i}.
\end{equation}
By proving that $q_{Z,j}(x)$ has rational coefficients, it follows that the corresponding elementary symmetric polynomials are rational numbers. The polynomial $q_{Z,j}(x)$ is closely related to the determinant
\begin{equation}
	\det(x-Z) = \prod_{i=1}^K (x-\lambda_i) = \prod_{i=1}^k (x-z_i)^{m_i},
\end{equation}
where $m_i$ is the multiplicity of the eigenvalue $\lambda_i$. The polynomial corresponding to setting $m_i=1$ for all $i$ is called the minimal polynomial of $Z$
\begin{equation}
	p_Z(x) = \prod_{j=1}^k (x-z_j).
\end{equation}
The minimal polynomial differs from $q_{Z,j}(x)$ by a linear factor
\begin{equation}
	p_Z(x) = (x-z_j)q_{Z,j}(x),
\end{equation}
with $z_j \in \mathbb{Q}$. As we will prove in the remainder of this appendix, $p_Z(x)$ has rational coefficients, and comparing the coefficients on both sides gives that $q_{Z_j}(x)$ has rational coefficients as well. It follows that
\begin{equation}
	e_i(z_1, \dots, z_{j-1},z_{j+1},\dots,z_k) \in \mathbb{Q}.
\end{equation}
Lastly,
\begin{equation}
	z_j(z_j - z_1) \dots (z_j - z_{j-1}) (z_j - z_{j+1}) \dots (z_j -z_k) = z_j q_{Z,j}(z_j) \in \mathbb{Q},
\end{equation}
which concludes the proof of the lemma in equation \eqref{eq: inv vand lemma apx}.


We will now prove that the minimal polynomial $p_Z(x)$ has rational coefficients.
The non-negative integers are contained in the field $\mathbb{Q}$ of rationals and therefore $Z$ is an element of the set $M_k(\mathbb{Q})$ of $k\times k$ matrices with rational coefficients.
Let $F$ be a finite field extension of the rationals. The minimal polynomial of a matrix $X$ over a field $F$ is uniquely defined as the monic polynomial $p_X^{(F)}(x) \in F[x]$ of lowest degree such that
\begin{equation}
	p_X^{(F)}(X) = 0.
\end{equation}
The minimal polynomial divides the characteristic polynomial (see \cite[Theorem 3.3.1]{horn2012matrix})
\begin{equation}
	P_X(x) = \det(x - X).
\end{equation}
That is, there exists a monic polynomial $h(x) \in F[x]$ such that
\begin{equation}
	P_X(x) = h(x)p_X^{(F)}(x). \label{eq: min divs char}
\end{equation}
In particular, the characteristic polynomial $P_X(x)$ satisfies 
\be
P_X(X)=0
\ee
and therefore the minimal polynomial of a matrix exists. The minimal polynomial of a matrix is unique. To show this, suppose $p_{X,1}^{(F)}(x)$ and $p_{X,2}^{(F)}(x)$ are minimal polynomials of the matrix $X$. We have
\be
p_{X,2}^{(F)}(x)= q(x) p_{X,1}^{(F)}(x) + r(x)
\ee
for some polynomials $q(x),r(x)\in F[x]$ and the polynomial $r(x)$ has degree less than that of $p_{X,1}^{(F)}(x)$. Evaluating on the matrix $X$, we get
\be
r(X)=0
\ee
Therefore, if $r(X)$ is a non-trivial polynomial, then we have a contradiction with the assumption that $p_{X,1}^{(F)}(x)$ is a polynomial of least degree evaluating to zero on $X$. Therefore, $r(x)=0$. This shows that $p_{X,1}^{(F)}(x)$ divides the polynomial $p_{X,2}^{(F)}(x)$. Since $p_{X,1}^{(F)}(x)$ and $p_{X,2}^{(F)}(x)$ are polynomials of the same degree, $q(x)$ must be a constant polynomial. Since $p_{X,1}^{(F)}(x)$ and $p_{X,2}^{(F)}(x)$ are also monic, we must have $q(x)=1$ and therefore, 
\be
p_{X,2}^{(F)}(x)=p_{X,1}^{(F)}(x)
\ee

In our particular case, we have $X=Z$ is a matrix with integer entries. Let us show that the minimal polynomial $p_{Z}^{(F)}(x)$ has rational coefficients. 
Let
\begin{equation}
	p_Z^{(F)}(x) = x^k + b_{k-1}x^{k-1} + \dots + b_0,
\end{equation}
be the expansion of $p_Z^{(F)}(x)$ with coefficients in $F$.  We will now show that
\begin{equation}
	b_{i} \in \mathbb{Q}~. 
\end{equation}
Let $r_j^{(i)}$ be the $j$th row of $Z^i$, then
\begin{align}
	&p_Z^{(F)}(Z) = Z^k + b_{k-1}Z^{k-1} + \dots + b_0 = 0 
\end{align}
implies that
\begin{align}
	&r_j^{(k)}+\sum_{i=0}^{k-1} b_i r_j^{(i)} = 0 \label{eq: F lindep}
\end{align}
for all $j=1,\dots, K$ where $K$ is the size of the matrix $Z$. That is, the vectors $r_j^{(0)}, \dots, r_j^{(k)}$ are linearly dependent over the field $F$.
Let $f_l$ be a basis for the field $F$ such that
\begin{equation}
	b_i = \sum_l \beta_{i,l}f_l
\end{equation}
where $f_0 = 1$ and $\beta_{i,l} \in \mathbb{Q}$.
Substituting this into \eqref{eq: F lindep} gives
\begin{equation}
	r_j^{(k)}+\sum_{i=0}^{k-1} \sum_{l} \beta_{i,l}f_l r_j^{(i)} = 0.
\end{equation}
Note that $r_j^{(i)}$ are rational numbers. Therefore, the above equation contains a sum over the basis elements $f_l$ with rational coefficients. Since $f_l$ are linearly independent over the rationals, this implies vanishing of all coefficients in front of $f_l$ for every $l$. In particular, using the fact that $f_0 = 1$ we have
\begin{equation}
	r_j^{(k)}+\sum_{i=0}^{k-1} \beta_{i,0}f_0 r_j^{(i)} = \qty(r_j^{(k)}+\sum_{i=0}^{k-1} \beta_{i,0} r_j^{(i)})f_0 = 0,
\end{equation}
and vanishing of the coefficients in front of $f_0$ implies
\begin{align}
	r_j^{(k)}+&\sum_{i=0}^{k-1} \beta_{i,0} r_j^{(i)} = 0, \label{eq: F dep means Q dep}
\end{align}
Since \eqref{eq: F dep means Q dep} holds for all rows $j=1,\dots,K$, we have
\begin{equation}
	Z^k+\beta_{k-1,0} Z^{k-1} + \dots + \beta_{0,0} = 0.
\end{equation}
That is, the monic polynomial $p(x) \in \mathbb{Q}$ defined by
\begin{equation}
	p(x) = x^k+\beta_{k-1,0} x^{k-1} + \dots + \beta_{0,0},
\end{equation}
satisfies $p(Z) = 0$. However, since the minimal polynomial is unique, we must have $p_{X}^{(F)}(x)=p(x)$. Therefore, if the matrix is defined over the rationals, the minimal polynomial has rational coefficients. This justifies the assumption in the above outline of the proof, and therefore we have proven the lemma.

%

\section{Class sizes and characters from $\fusionalg{G}$-CTST}\label{apx: additional formulas}
In this appendix we construct polynomials $F(x)$, from $\fusionalg{G}$-CTST amplitudes, with roots at $x=\Sym C_1, \dots, \Sym C_k$, where $\Sym C = \tfrac{\abs{G}}{\abs{C}}$. Thus, solving the polynomial is a constructive method for finding conjugacy class sizes from CTST. The polynomial $F(x)$ contains contributions from manifolds with more than one genus. The roots of $F(x)$ are directly related to the eigenvalues of the handle creation operator. This is analogous to the $G$-CTST construction \cite{IDFCTS, GCTST2} of dimension of irreducible representations, which uses the fact that the eigenvalues of the $G$-TQFT handle creation operator are $\tfrac{\abs{G}^2}{d_R^2}$.

The handle creation operators can be used to construct level sets of characters with fixed eigenvalues of the handle creation operator, or equivalently, fixed conjugacy class sizes.
Using the previously constructed conjugacy class data, we give explicit formulas for these character sums in terms of elementary symmetric polynomials in $z_i = \Sym C_i$ and $\fusionalg{G}$-CTST amplitudes with one boundary but varying genus. This is dual to the $G$-CTST considerations in \cite{GCTST2}, where the handle creation operator can be used to construct level sets of characters with the same dimension.

By considering surfaces with a single boundary labelled by $R_2$ and $b$ boundaries labelled by $R_1$, we construct level set sums of $\chi^{R_2}_{C}$ with fixed values of $\chi^{R_1}_{C}$. This is the row-column dual fo the construction in \cite{GCTST2}, where surfaces with a single boundary labelled by $C_2$ and $b$ boundaries labelled by $C_1$ gave rise to level sets of normalized characters of $C_2$.

\subsection{Class sizes from $\fusionalg{G}$-CTST}
To construct classes sizes from $\fusionalg{G}$-CTST amplitudes, define
\begin{equation}
	z_\cclass = \frac{\abs{G}}{\abs{\cclass}} = \Sym \cclass
\end{equation}
and the polynomial
\begin{equation}
	F(x) = \prod_{\cclass \in \cclasses{G}} (x-z_{\cclass}).
\end{equation}
The polynomial can be expanded into elementary symmetric functions $e_k$,
\begin{equation}
	F(x) = \sum_k (-1)^{k}x^{K-k} e_k(z_{\cclass_1}, \dots, z_{\cclass_K}), \label{eq: class size root poly}
\end{equation}
where $K = \abs{\cclasses{G}}$.
Elementary symmetric functions evaluated on $z_K$ can be expressed in terms of the partition functions $Z_h^{\fusionalg{G}}$, see \cite[Equation 2.17]{IDFCTS}
\begin{equation}
	e_k(z_{\cclass_1}, \dots, z_{\cclass_K}) = \sum_{p \vdash k} \frac{(-1)^{k-\sum_i p_i}}{\Aut(p)} \prod_{i} (Z_{h=i+1}^{\fusionalg{G}})^{p_i}. \label{eq: esym from Z}
\end{equation}
The sum is over integer partitions $p=(p_1, p_2, \dots, p_k)$ of $k$
\begin{equation}
	\sum_i i p_i = k, \qq{and } \Aut(p) = \prod i^{p_i} p_i!
\end{equation}
Substituting \eqref{eq: esym from Z} into \eqref{eq: class size root poly} gives
\begin{equation}
	F(x) = \sum_k (-1)^{k}x^{K-k} \sum_{p \vdash k} \frac{(-1)^{k-\sum_i p_i}}{\Aut(p)} \prod_{i} (Z_{h=i+1}^{\fusionalg{G}})^{p_i},
\end{equation}
and solving $F(x) = 0$ gives the conjugacy class sizes from $\fusionalg{G}$-CTST data.

\subsection{Reducible characters over level sets of class sizes  from genus $h$ fusion data}
In this section we will probe the irreducible representation $S$ using surfaces labelled by a boundary $S$ with varying genus.
The genus one partition function is
\bea 
&& Z_{ h=1  ; S } = \sum_{ R  } N_{ R , \bar R , S }  = { 1 \over |G| } 
\sum_{ R } N_{ R \bar R S }  \cr 
&& = { 1 \over |G| } \sum_{ g } \chi^\irrep ( g ) \chi^\irrep ( g^{-1} ) \chi^S ( g )   \cr 
&& = \sum_{ \cclass } { |\cclass| \Sym \cclass  \over |G| } \chi^S_\cclass = \sum_{ \cclass} \chi^S_\cclass
\eea
and for higher genus we have 
\bea 
&& Z^{\fusionalg{G}}_{ h; S} = 
\sum_{ R_1 , \cdots , R_{h+1}  } N_{ R_1 \cdots R_{h+1}  , \bar R_1 , \cdots , \bar R_{h+1}  ;  S } \cr 
&& = { 1 \over |G| } \sum_{ g} \sum_{ R_1 \cdots R_{h+1}  } 
\chi^{R_1}  ( g  ) \chi^{ R_2} ( g  ) \cdots \chi^{ R_{h+1 } } ( g  ) 
\chi^{\bar R_1}  ( g  ) \chi^{ \bar R_2} ( g  ) \cdots \chi^{ \bar R_{h+1} } ( g  ) \chi^{ S } ( g )  \cr 
&& = { 1 \over |G| }  \sum_{ \cclass } \abs{\cclass} ( \Sym \cclass )^{ h    } \chi^S_\cclass \cr 
&& =  \sum_{ \cclass } ( \Sym \cclass)^{ h -1   } \chi^S_\cclass  \label{eq: ZhFusion}
\eea
Define the following equivalence relation on $\cclasses{G}$
\begin{equation}
	\cclass_1 \sim \cclass_2 \qq{if $\Sym \cclass_1 = \Sym \cclass_2$}.
\end{equation}
It partitions $\cclasses{G}$ into subsets $\kappa_1, \dots, \kappa_l$ such that
\begin{equation}
	\cclasses{G} =  \kappa_1 \cup \dots \cup \kappa_l.
\end{equation}
We introduce the notation
\begin{equation}
	z_i = \Sym \cclass, \qq{for any $\cclass \in \kappa_i$}.
\end{equation}
Now we can rewrite equation \eqref{eq: ZhFusion} as
\begin{equation}
	Z^{\fusionalg{G}}_{h;S} = \sum_{i=1}^l z_i^{h-1} \sum_{\cclass \in \kappa_i} \chi_\cclass^S.
\end{equation}
Define the $l$-vector
\begin{equation}
	y_h(S) = (Z_{2; S}, \dots, Z_{l+1;S}),
\end{equation}
the $l \times l$ Vandermonde matrix
\begin{equation}
	V_{h, i} = z_i^{h}
\end{equation}
and $l$-vector
\begin{equation}
	x_{i}(S) = \sum_{\cclass \in \kappa_i} \chi^S_{\cclass}.
\end{equation}
Equation \eqref{eq: ZhFusion} reads
\begin{equation}
	y(S) = Vx(S)
\end{equation}
The inverse of the Vandermonde matrix is
\begin{equation}
	(V^{-1})_{i, h} = \frac{(-1)^{l-h} e_{l-h}(z_1 \dots, z_{i-1}, z_{i+1}, \dots, z_l)}{z_i \prod_{j\neq i}(z_i - z_j)},
\end{equation}
This immediately gives a formula for level set sums of characters:
\begin{equation}
	\sum_{\cclass \in \kappa_i} \chi^S_{\cclass} =\frac{\sum_{h=1}^l (-1)^{l-h} e_{l-h}(z_1 \dots, z_{i-1}, z_{i+1}, \dots, z_l) Z_{h+1;S}}{z_i \prod_{j\neq i}(z_i - z_j)},
\end{equation}


\subsection{ Constructing sums of characters of $R_2$ over level sets of characters of $R_1$ }
Similar arguments allow us to obtain sums of characters of level sets of another character. Fix an irreducible representation $\irrep_1$ and consider the partition functions
\bea 
&& Z^{\fusionalg{G}}_{ h =1 ; R_1^b , R_2 } \cr 
&& = \sum_{ R }  N_{ R , \bar R , R_1^b , R_2   }
= { 1 \over |G| } \sum_{ g }\sum_{ R }  \chi^\irrep ( g ) \chi^{\bar R } ( g )  ( \chi^{ R_1} ( g ) )^b \chi^{ R_2} ( g ) \cr 
&& = {   1 \over |G| }   \sum_{ K } |K| . \Sym K . ( \chi^{R_1}_K )^b \chi^{ R_2}_K \cr 
&& = \sum_{ K } ( \chi^{R_1}_K )^b \chi^{ R_2}_K. 
\eea
Define the equivalence relation
\begin{equation}
	\cclass_1 \sim_{R_1} \cclass_2 \qq{if $\chi^{\irrep_1}_{\cclass_1} = \chi^{\irrep_1}_{\cclass_2}$,}
\end{equation}
on $\cclasses{G}$, which partitions conjugacy classes into subsets $\kappa_1, \dots, \kappa_l$. Introduce
\begin{equation}
	z_i^{R_1} = \chi_\cclass^{R_1} \qq{for any $\cclass \in \kappa_i$.}
\end{equation}
We can re-write the partition functions as
\begin{equation}
	Z^{\fusionalg{G}}_{h =1 ; R_1^b ;R_2} = \sum_{i=1}^l ( z^{R_1}_i )^b \sum_{\cclass \in \kappa_i} \chi^{ R_2}_\cclass. \label{eq: char level sums}
\end{equation}
As before, define the $l$-vector
\begin{equation}
	y = (Z_{h=1, R_1^1, R_2}, \dots,Z_{h=1, R_1^l, R_2}),
\end{equation}
the $l \times l$ Vandermonde matrix
\begin{equation}
	V_{b,i}(R_1) = ( z^{R_1}_{i} )^b
\end{equation}
and $l$-vector
\begin{equation}
	x_{i}(R_2) = \sum_{\cclass \in \kappa_i} \chi^{ R_2}_\cclass.
\end{equation}
Then equation \eqref{eq: char level sums} reads
\begin{equation}
	y(R_1, R_2) = V(R_1) x(R_2).
\end{equation}
The inverse of $V(R_1)$ is
\begin{equation}
	(V^{-1})_{i, b} = \frac{(-1)^{l - b} e_{l-b}(z_1^{R_1}, \dots, z_{i-1}^{R_1}, z_{i+1}^{R_1}, \dots, z_l^{R_1})}{z^{R_1}_i \prod_{j\neq i}(z^{R_1}_{i} - z^{R_1}_{j})},
\end{equation}
which gives
\begin{equation}
	\sum_{\cclass \in \kappa_i} \chi^{ R_2}_\cclass = \frac{\sum_{b=1}^l (-1)^{l-b} e_{l-b}(z_1^{R_1}, \dots, z_{i-1}^{R_1}, z_{i+1}^{R_1}, \dots, z_l^{R_1})Z_{h=1;R_1^b;R_2}}{z^{R_1}_i \prod_{j\neq i}(z^{R_1}_{i} - z^{R_1}_{j})}.
\end{equation}

\section{Galois groups, integer rows and columns}

\label{sec:Galrowcoltable}

In section \ref{sec:introwcol}, we showed that if the Galois group of the field containing the character table is cyclic, then the number of integer rows and columns are equal. In this appendix, we will illustrate this result by computing the Galois groups of finite groups of order $\leq 100$ and sporadic groups. We compute their exponent $E$, number of integers rows and columns in the character table, the isomorphism class of the Galois group $\DZ_{E}^{\times}$ and the isomorphism class of the Galois group of the minimal field containing the character table. These computations were done using GAP \cite{GAP4} and SageMath \cite{sagemath}. The isomorphism classes of groups are denoted using the notation in GAP. The highlighted rows correspond to finite groups with different number of integer rows and columns. These groups have non-cyclic Galois groups, which agrees with our results in section \ref{sec:introwcol}.

{
	\tiny
\begin{longtable}{llllll}
		Group & Exp & Int. Rows & Int. Columns & Galois Group &	Faithful Galois Group \\	
		\hline \\
		C1 & $1$ & $1$ & $1$ & C1 &	C1 \\
		C2 & $2$ & $2$ & $2$ & C1 &	C1 \\
		C3 & $3$ & $1$ & $1$ & C2 &	C2 \\
		C4 & $4$ & $2$ & $2$ & C2 &	C2 \\
		C2 x C2 & $2$ & $4$ & $4$ & C1 &	C1 \\
		C5 & $5$ & $1$ & $1$ & C4 &	C4 \\
\end{longtable}
	
\noindent{\normalsize The table below contains the data for sporadic groups. `NR' denotes that we don't have the data for this entry. }
	
	\begin{longtable}{llllll}
		Group & Exp & Int. Rows & Int. Columns & Galois Group &	Faithful Galois Group \\	
		\hline \\
		M11 & $1320$ & $6$ & $6$ & C20 x C2 x C2 x C2 x C2 & C2 x C2 \\
		M12 & $1320$ & $13$ & $13$ & C20 x C2 x C2 x C2 x C2 & C2 \\
		M22 & $9240$ & $8$ & $8$ & C60 x C2 x C2 x C2 x C2 x C2 & C2 x C2 \\
		M23 & $212520$ & $7$ & $7$ & C660 x C2 x C2 x C2 x C2 x C2 x C2 & C2 x C2 x C2 x C2 \\
		M24 & $212520$ & $16$ & $16$ & C660 x C2 x C2 x C2 x C2 x C2 x C2 & C2 x C2 x C2 \\
		HS & $9240$ & $20$ & $20$ & C60 x C2 x C2 x C2 x C2 x C2 & C2 x C2 \\
\end{longtable}
}

\section{Harada's conjecture and row-column duality of TQFTs} 
\label{sec:Harada}

From class algebra 
\bea 
T_{ \cclass_1   } T_{ \cclass_2} = \sum_{ \cclass_3} f_{ \cclass_1 \cclass_2}^{\cclass_3} T_{\cclass_3} 
\eea
Define the matrix $\hat f_{ \cclass } $ by 
\bea 
( f_{ \cclass } )_{ \cclass_1}^{\cclass_2} = f_{\cclass \cclass_1}^{\cclass_3 }. 
\eea
We know 
\bea 
T_{ \cclass } P_R = { \chi^{R } ( T_{ \cclass } ) \over d_R } P_R 
\eea
Thus the eigenvalues of $\hat f_{ \cclass } $ are ${ \chi^{R } ( T_{ \cclass } ) \over d_R }$
The determinant is 
\bea
\det ( \hat f_{ \cclass} ) = \prod_R { \chi^{R } ( T_{ \cclass } ) \over d_R } = \prod_R \frac{\abs{\cclass}}{d_R} \chi^R_\cclass.
\eea
This gives an integer for every column of the character table. The product
\bea 
X = \prod_{ \cclass } \det ( \hat f_{ \cclass } )  = \prod_{ \cclass } \prod_R \frac{\abs{\cclass}}{d_R} \chi^{R }_\cclass,
\eea
of these integers is an integer associated with the whole character table. 

In the fusion algebra theory we have 
\bea 
a_R a_S = N_{ RS }^T a_T 
\eea
Define the matrix $\hat N_R $ for every row by the  equation 
\bea 
( \hat N_R )_S^{~T }   = N_{ RS}^{~~T}  
\eea
We have 
\bea 
a_R A_{ \cclass}  = \chi^\irrep_{ \cclass } A_{ \cclass} 
\eea
This the eigenvalues of $ \hat N_R $ are $ \chi^\irrep_\cclass   $.  
Take the determinant 
\bea 
\det ( \hat N_R ) = \prod_{ \cclass  } \chi^\irrep_{\cclass }.
\eea
This is an integer for every row. Take the product over all $R$ to get 
\bea 
\tilde X = \prod_{ R } \det ( \hat N_R ) = \prod_R \prod_{ \cclass  } \chi^\irrep_{ \cclass}   
\eea
Comparing the two we get 
\bea 
{ X \over \tilde X } =  \prod_\cclass \prod_R  \frac{\abs{\cclass}}{d_R}.
\eea
This ratio is conjectured to be an integer by Harada (see notes at the end of Chapter 4 of \cite{Navarro} and the paper \cite{Harada}).

\newpage
\bibliography{chetdocbib}

\end{document}